\documentclass[12pt,a4paper]{article}
\pdfoutput=1
\usepackage{jheppub-arxiv}
\usepackage[usenames,dvipsnames,table]{xcolor}
 
\usepackage{amssymb} 
\usepackage{amsmath}
\usepackage{mathtools}
\usepackage{amsfonts}    
\usepackage{dsfont}
\usepackage{pdfpages}
\usepackage{verbatim}
\usepackage{stmaryrd}
\usepackage{graphicx}
\usepackage{import}
\usepackage{etoc}
\usepackage{enumitem}
\usepackage{tensor}
\usepackage{mathrsfs}
\usepackage{textgreek} 
\usepackage[mathscr]{euscript}
\usepackage[smalltableaux]{ytableau}
\ytableausetup{centertableaux}
\usepackage{tikz}
\usetikzlibrary{decorations.pathreplacing, decorations.markings,calc,shapes.misc,decorations.pathmorphing,patterns.meta, math,positioning}
\usepackage{tabularx,ragged2e}
\usepackage{physics}

\usepackage{slashed}
\usepackage{datetime}
\usepackage{hyperref}
\hypersetup{
    pdfencoding=unicode,
	colorlinks=true,
	urlcolor=Maroon,
	linkcolor=RoyalBlue,
	citecolor=Maroon,
	pdftitle={Records from the S-Matrix Marathon: Gravitational Physics from Scattering Amplitudes},
	pdfauthor={},
	pdfdisplaydoctitle=true,
	pdfstartview=FitH,
	linktocpage=true
}

\usetikzlibrary{shapes}
\usepackage{enumitem}
\usepackage[export]{adjustbox}
\usepackage{nccmath}
\usepackage{bbm}
\usepackage{braket}
\usepackage{empheq}
\usepackage{cancel}
\usepackage{cleveref}
\usepackage[most]{tcolorbox}

\definecolor{charcoal}{HTML}{343837}
\newcommand{\sumint}{
  \mathop{%
    \mathchoice%
    {\ooalign{$\displaystyle\sum$\cr\hidewidth$\displaystyle\int$\hidewidth\cr}}%
    {\ooalign{$\textstyle\sum$\cr\hidewidth$\textstyle\int$\hidewidth\cr}}%
    {\ooalign{$\scriptstyle\sum$\cr\hidewidth$\scriptstyle\int$\hidewidth\cr}}%
    {\ooalign{$\scriptscriptstyle\sum$\cr\hidewidth$\scriptscriptstyle\int$\hidewidth\cr}}%
  }\displaylimits
}

\usepackage{bm}
\definecolor{yellowish}{rgb}{0.880722,0.611041,0.142051}

\newcommand{\ba}{\begin{align}}

\newcommand{\be}{\begin{equation}}
\newcommand{\ee}{\end{equation}}
\def\bd{\begin{tikzpicture}}
\def\ed{\end{tikzpicture}}

\renewcommand\Re{\mathop{\text{Re}}}

\let\vecc\vec
\renewcommand{\vec}[1]{\mathbf{#1}}
\allowdisplaybreaks

\def\XXint#1#2#3{{\setbox0=\hbox{$#1{#2#3}{\int}$}
     \vcenter{\hbox{$#2#3$}}\kern-.5\wd0}}

\definecolor{light-gray}{gray}{0.75}

\renewcommand\d{\text{d}}

\newcommand{\e}{\mathrm{e}}

\newcommand{\eps}{\varepsilon}

\renewcommand{\geq}{\geqslant}

\newcommand{\rD}{\mathrm{D}}

\def\barv{\bar{v}}
\def\barp{\bar{p}}

\definecolor{dpurple}  {RGB} {189,  147,  249}

\usepackage{ntheorem}
\usepackage[framemethod=tikz,footnoteinside=false,hidealllines=true,topline=true,bottomline=true,linecolor=black!10!white,linewidth=1pt,innerleftmargin=0em,innerrightmargin=0em,frametitlerule=true,ntheorem]{mdframed}
\theorembodyfont{\upshape}

\newmdtheoremenv{mtheorem}{Theorem}[section]
\newmdtheoremenv[]{mdexample}{Example}[section]
\newmdtheoremenv{mdremark}{Remark}[section]
\newmdtheoremenv{mddefinition}{Definition}[section]
\newmdtheoremenv{mdcorollary}{Corollary}[section]
\newmdtheoremenv{mdproposition}{Proposition}[section]
\newmdtheoremenv{QA}{Audience question}[section]
\newcommand{\question}[1]{%
    #1\\[2ex] 
    \textit{Answer:} 
}

\title{{\Large\normalfont Records from the S-Matrix Marathon:}\\ Gravitational Physics from Scattering Amplitudes}

\author{{\normalfont Lecturers:}}
\author[1]{Miguel~Correia,}
\author[2]{Holmfridur~S.~Hannesdottir,}
\author[3]{Giulia~Isabella,}
\author[3]{Anna~M.~Wolz,}
\author[4]{Zihan~Zhou}

\author{\\ {\normalfont Notes edited by:}}
\author[1]{Mathieu~Giroux,}
\author[2]{Holmfridur~S.~Hannesdottir,}
\author[2,4,5]{Sebastian~Mizera,}
\author[1]{Celina~Pasiecznik}

\affiliation[1]{Department of Physics, McGill University, 3600 Rue University,\\ Montr\'eal, H3A 2T8, QC Canada}
\affiliation[2]{Institute for Advanced Study, Princeton, NJ 08540, USA}
\affiliation[3]{Mani L. Bhaumik Institute for Theoretical Physics, Department of Physics\\ and Astronomy, University of California Los Angeles, Los Angeles, CA 90095, USA}
\affiliation[4]{Department of Physics, Princeton University, Princeton, NJ 08544, USA}
\affiliation[5]{Princeton Center for Theoretical Science,\\ Princeton University, Princeton, NJ 08544, USA}

\abstract{
These lecture notes explain how classical gravitational physics emerges from scattering amplitudes. We emphasize the role of different kinematic regimes in probing various aspects of bound and unbound problems, as illustrated by the Hydrogen atom example. Classical predictions of General Relativity, such as the Shapiro time delay and perihelion precession, emerge from these considerations. We also explain a number of recent approaches to probing black hole physics from the perspective of amplitudes, including applications of worldline effective field theory in astrophysics, predictions of gravitational waveforms, and the hierarchical three-body problem.

These notes are based on a series of lectures held during the S-Matrix Marathon workshop at the Institute for Advanced Study on 11--22 March 2024.
}

\begin{document}

\maketitle
\setcounter{page}{1}

\setcounter{tocdepth}{4}
\setcounter{secnumdepth}{4}

\makeatletter
\g@addto@macro\bfseries{\boldmath}
\makeatother

\newpage
\section*{Preface}

This article is a chapter from the \emph{Records from the S-Matrix Marathon}, a series of lecture notes covering selected topics on scattering amplitudes~\cite{RecordsBook}. They are based on lectures delivered during a workshop on 11--22 March 2024 at the Institute for Advanced Study in Princeton, NJ. We hope that they can serve as a pedagogical introduction to the topics surrounding the S-matrix theory.

These lecture notes were prepared by the lectures in collaboration with the above-mentioned editors.

\vfill
\section*{Acknowledgments}

M.C. is supported by the Simons Collaboration on the Nonperturbative Bootstrap. M.G.’s and C.P.'s work is supported in parts by the National Science and Engineering Council of Canada (NSERC) and the Canada Research
Chair program, reference number CRC-2022-00421. Additionally, C.P. is supported by the Walter C. Sumner Memorial Fellowship.
H.S.H. gratefully acknowledges funding provided by the J. Robert Oppenheimer Endowed Fund of the Institute for Advanced Study. G.I. is supported by the US Department of Energy under award number DE-SC0024224, the Sloan Foundation and the Mani L. Bhaumik Institute for Theoretical Physics.
S.M.
gratefully acknowledges funding provided by the Sivian Fund and the Roger Dashen Member Fund at the Institute for Advanced Study. 
This material is based upon work supported by the U.S. Department of Energy, Office of Science, Office of High Energy Physics under Award Number DE-SC0009988.
A.M.W. is supported by the US Department of Energy under award number DE-SC0024224, Alfred P. Sloan Foundation, and the NSF Graduate Research Fellowship under Grant No. DGE-2034835, and would like to thank the Mani L. Bhaumik Institute for Theoretical Physics.

The S-Matrix Marathon workshop was sponsored by the Institute for Advanced Study and the Carl P. Feinberg Program in Cross-Disciplinary Innovation.

\newpage

\usetikzlibrary{angles,quotes,calc}

\section*{\label{ch:IsabellaEtAl}Gravitational Physics from Scattering Amplitudes\\
\normalfont{\textit{Miguel Correia, Hofie Hannesdottir, Giulia Isabella,\\ Anna M. Wolz, Zihan Zhou}}}

\setcounter{section}{0}

\noindent\rule{\textwidth}{0.25pt}
\vspace{-0.8em}
\etocsettocstyle{\noindent\textbf{Contents}\vskip0pt}{}
\localtableofcontents
\vspace{0.5em}
\noindent\rule{\textwidth}{0.25pt}
\vspace{1em}

These lectures will focus on how to use the classical limit of quantum field theory (QFT) to compute observables, in particular in electromagnetism and general relativity (GR). In other contributions to \cite{RecordsBook}, we have learned about the beautiful analytic structure of the S-matrix in QFT. Here, we will examine to what extent this structure survives in the classical limit, and what insight it offers for classical physics.

The question of how to take the classical limit of QFT is very old, but it is particularly relevant today. Since gravitational waves were detected at LIGO~\cite{LIGOScientific:2016aoc}, the application of scattering-amplitude techniques to the computation of gravitational observables has become an active area of research. Indeed, people have realized that perturbative techniques in QFT, developed over decades in the context of making predictions for the Large Hadron Collider at CERN, can be directly applied to compute quantities that are otherwise hard to obtain directly from General Relativity (GR).

Interestingly, many physical observables in GR (scattering angle, perihelion shifts, etc.) must be computed non-perturbatively, and thus provide an interesting case study for resummation of perturbative expansions within QFT. A simple dimensional-analysis argument shows why we need to go beyond perturbation theory: since the gravitational constant $G_N=1/M_{\text{Pl}}^2$ is dimensionful (where $M_{\text{Pl}}$ is the Planck mass), it must be multiplied by some kinematic quantity, which can be taken to be the center-of-mass energy, in order to become a dimensionless perturbative coupling. The relevant coupling is therefore the center-of-mass energy squared in units of the Planck mass: $E_{\text{cm}}^2/M_{\text{Pl}}^2$. Since the Planck mass is of order $M_{\text{Pl}}\sim 10^{-8}$ kg, this coupling constant becomes $\mathcal{O}(1)$ even for the scattering of mosquitoes. Extracting classical observables thus requires a sufficiently thorough understanding of scattering-amplitude techniques to make \emph{all-orders} statements. In these lectures, we will review some of the things that are known to hold at all orders in perturbation theory.

Before going on, let us clarify what we mean by the classical limit. We have the following kinematic scales in our problem (dynamics will come later): 
\begin{itemize}
    \item Compton wavelength, $\lambda_{\text{C}}=\frac{\hbar}{m}$, where $m$ is the lightest mass participating in the scattering event. This scale is associated with particle production and QFT effects (precisely the ones we want to eliminate in a classical setup).
    \item De Broglie wavelength (for massive particles), $\lambda_{\text{dB}}=\frac{\hbar}{|p|}$, associated with quantum-mechanical effects, where the wave nature of the particle becomes important.
    \item The impact parameter $b$, which is the typical separation length between the scattered bodies.
\end{itemize}

In order to get rid of all the particle production/QFT effects, we will always consider 
\begin{equation}
    b\gg \lambda_{\text{C}}\, .
\end{equation} We can then distinguish between four interesting regimes: 
\be\label{eq:regimes}
\includegraphics[scale=0.4,valign=c
]{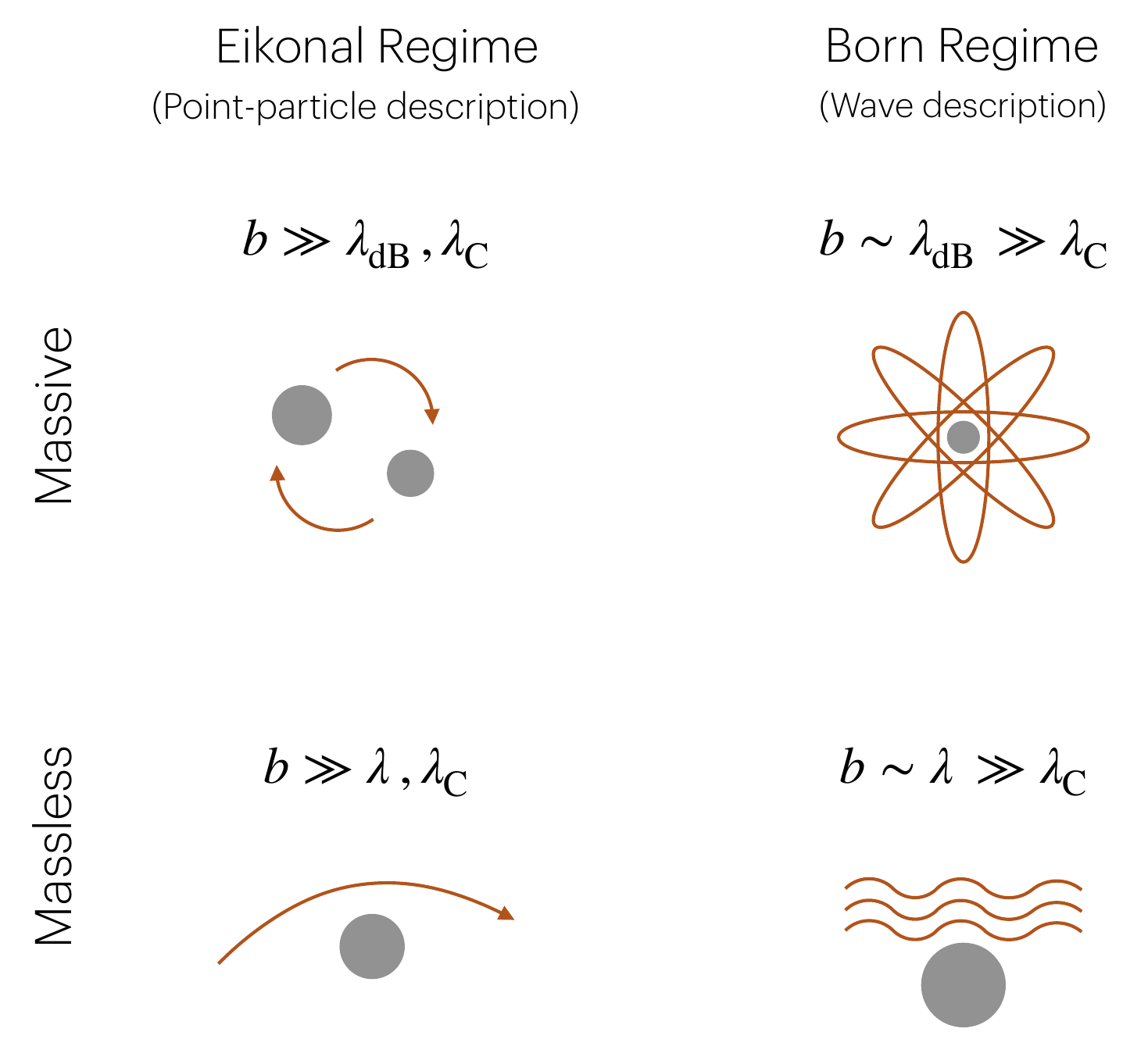}
\ee
The first one involves a massive body in a point-particle approximation (top left). This corresponds to the regime $b \gg \lambda_{\text{dB}}, \lambda_{\text{C}}$. Similarly, we can remain in the point-particle setup, but consider a massless particle traveling along a null geodesic (bottom left). The regime relevant to this event is $b \gg \lambda, \lambda_{\text{C}}$, where $\lambda$ is the wavelength of the massless particle. There are also wave counterparts to both regimes. If the de Broglie wavelength of a massive particle is comparable to the size of the experiment (top right), we have $b \sim \lambda_{\text{dB}} \gg \lambda_{\text{C}}$. Notice that this inequality implies a non-relativistic setup where $|p|\ll m$. We will discuss this regime at length later in the context of the non-relativistic hydrogen atom.  Likewise, if the wavelength of a massless particle is of a size comparable to the impact parameter (bottom right), we are in the $b \sim \lambda \gg \lambda_{\text{C}}$ regime. In this case, this is a probe limit, where the frequency of the wave is much smaller than any of the masses present in the problem. 

\begin{figure}[t!]
    \centering
    \includegraphics[scale=1,valign=c]{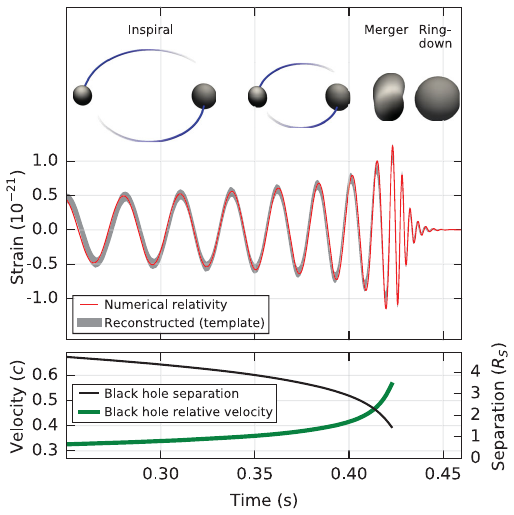}
    \caption{The gravitational-wave signal from the GW150914 event, illustrating the three phases of a binary black hole coalescence. Figure reprinted from \cite{LIGOScientific:2016aoc}.}
    \label{fig:ligomerger}
\end{figure}

For our purpose of describing gravitational scattering, the most important situations will be the first and the last: either massive scattering in a point-particle approximation, or a massless wave scattering in the probe limit.
Indeed, the actual observable we measure in gravitational wave physics is a signal that looks something like Fig.~\ref{fig:ligomerger}.
The signal can be divided into three different regions in time. The first one is the inspiral phase, where two black holes or other heavy objects orbit around each other and slowly coalesce. This phase corresponds to the massive point-particle regime. The middle phase is the merger. It is a strongly coupled regime where the two black holes collapse, and numerical GR techniques are needed. The final phase is the ringdown, where the system settles into a single black hole and can be thought of as a perturbation over the Schwarzschild metric. It corresponds to the massless wave regime explained above.

The outline of this chapter is as follows. First, we will explore in detail the case of the non-relativistic hydrogen atom. 
It turns out that QED in the non-relativistic limit is a wonderful case study, which despite being computationally much simpler, allows us to introduce all the main ingredients that will be needed later. Then, once we have a solid understanding in QED, we will apply an analogous setup to gravity.
Finally, the latter sections focus on absorption effects, radiation, and the three-body problem, hopefully offering a more complete and broad view on this active and fascinating subject.

\section[Feynman diagrams, hydrogen atom, and classical limit]{Feynman diagrams, hydrogen atom, and classical limit\\
\normalfont{\textit{Giulia Isabella}}}\label{sec:Isabella}

Disclaimer: The first two sections of this chapter are the result of some work and study of the literature over the last months. This is of course a huge and active subject with thousands of papers, and the content of these sections will be a personal biased view of the authors, but hopefully a coherent picture of the subject.
The references that we will mostly use are:
\begin{itemize}
    \item ``Quasipotential equation corresponding to the relativistic eikonal approximation'' by Todorov (1970) \cite{Todorov:1970gr}.
    \item ``Analyticity in the complex angular momentum plane of the coulomb scattering amplitude'' by Singh (1961) \cite{Singh:1962qgg}.
    \item ``From scattering amplitudes to classical potentials in the post-Minkowskian expansion'' by Cheung, Rothstein, Solon (2018) \cite{Cheung:2018wkq}.
    \item ``Post-Minkowskian effective field theory for conservative binary dynamics'' by K\"alin, Porto (2020) \cite{Kalin:2020mvi}.
    \item ``Scalar QED as a toy model for higher-order effects in classical gravitational scattering'' by Bern, Gatica, Herrmann, Luna, Zeng (2021) \cite{Bern:2021xze}.
    \item ``Classical vs quantum eikonal scattering and its causal structure'' by Bellazzini, GI, Riva (2022) \cite{Bellazzini:2022wzv}.
    \item ``The Born regime of gravitational amplitudes" by Correia, GI (2024) \cite{Correia:2024jgr}.
\end{itemize}

\subsection{Classical observables for the hydrogen atom}
In this section, we will model the hydrogen atom from scattering amplitudes and study its various kinematic limits. Our first goal is to show how to compute observables in the limit of the top left corner of \eqref{eq:regimes}, which in momentum space corresponds to taking the limit of transferred momentum to be smaller than the masses and momenta for any individual particle.  We could think of this regime as throwing some macroscopic charged balls that pass next to each other, where the electromagnetic attraction/repulsion will affect their trajectories.  

The $2\to 2$ scattering configuration that we have in mind in impact parameter space is as follows:
\be\label{eq:scattering-kinematics}
\raisebox{-1.5cm}{
\begin{tikzpicture}[scale=1.2, every node/.style={scale=1}]

\coordinate (P1) at (-0.5,0);
\coordinate (P2) at (5,-0.75);
\coordinate (P3) at (5,-2);
\coordinate (P4) at (-0.5,-1.5);
\coordinate (up) at (2.5,0);
\coordinate (down) at (2,-2);

\draw[Maroon, very thick, ->] (P1) node[above left,black] {$p_1$} -- (2.4,0) ;
\draw[RoyalBlue, very thick, ->] (down) -- (P4) node[above left,black] {$p_4$};

\draw[Maroon, very thick, ->] (up) -- (P2) node[below right,black] {$p_3$};
\draw[RoyalBlue, very thick, ->] (P3) node[below right,black] {$p_2$} -- (2.1,-2);

\draw[dashed] (2.4,0) -- (5,0);

\draw[dashed] (2.25,0) -- (2.25,-2) node[midway,right] {$b$};

\node at (4,-0.2) {\footnotesize$\theta$};

\end{tikzpicture}
}
\ee
Let us call the mass, momentum and charge of the particles $m_i$, $p_i$ and $Q_i$ respectively, for $i=1,2$.
The kinematics can be described in terms of the scattering angle $\theta$, the center-of-mass energy squared $s = \vec{p}^2 =  (p_1 + p_2)^2 > 0$, and the momentum transfer squared $t = -\vec{q}^2 = (p_1 + p_3)^2 < 0$. They satisfy the constraint
\begin{equation}
    \cos\theta=1-t/2\vec{p}^2\, .
\end{equation}
We use mostly-minus metric signature. Bold faced quantities denote 3-vectors, such that $p_i^2 = (p_i^0)^2 - \vec{p}_i^2 = m_i^2$. 

As mentioned earlier, the point particle regime we are interested in is given by $|\vec{q}|\ll m_i, |\vec{p}|$. Additionally, we are going to study this scattering process in the simplest possible setup: non-relativistically, in the probe limit ($m_1 \gg m_2$), and ignoring spin effects by considering scalar QED for simplicity. 

Let us start by drawing and computing diagrams for this process. At tree level, we simply have the contribution from the photon exchange diagram, which in this limit evaluates to 
\be
\mathcal{M}_0 = 
\raisebox{-1.5em}{
\begin{tikzpicture}[line width=1.1, scale=0.7,
    electron/.style={postaction={decorate}, decoration={
        markings,
        mark=at position #1 with {\arrow{latex}}
    }},
    electron/.default=0.5,  
    positron/.style={postaction={decorate}, decoration={
        markings,
        mark=at position #1 with {\arrow{latex}}
    }},
    positron/.default=0.5  
]
    \coordinate (left1) at (-1,-1);
    \coordinate (left2) at (-1, 1);
    \coordinate (right1) at (1, -1);
    \coordinate (right2) at (1,1);
    \coordinate (up1) at (0,1);
    \coordinate (low1) at (0,-1);

    \draw[decorate, decoration=snake] (up1) -- (low1);

    \draw[RoyalBlue] (left1) -- (right1);
    \draw[Maroon] (left2) -- (right2);

\end{tikzpicture}
}
= -4 m_1 m_2\frac{Q_1 Q_2}{\vec{q}^2}\, .
\ee
Recall that in this notation $\vec{q}$ is the (spatial) momentum transfer.
We are interested in the amplitude in the $b$ space, where the impact parameter $b$ is the separation between the two bodies in the plane transverse to the scattering as illustrated in \eqref{eq:scattering-kinematics}. It turns out that at small momentum transfer, the vector $\vec{q}$ is essentially two-dimensional. So, the next step is to simply (2D) Fourier transform the above result to the $b$ space. The result is
\begin{equation}
    \text{FT}[\mathcal{M}_0] = -  Q_1 Q_2\log{b/b_{\text{IR}}}\, .
\end{equation}
which is IR divergent and we have regularized it with a sharp IR cutoff $b_{\text{IR}}$. 

We will continue the computation at the one-loop level. It turns out that in this limit the only diagram that contributes to leading order is the planar box.
Its contribution can be readily computed and gives
\begin{equation}
     \mathcal{M}_1=
     \raisebox{-1.5em}{
\begin{tikzpicture}[line width=1.1, scale=0.7,
    electron/.style={postaction={decorate}, decoration={
        markings,
        mark=at position #1 with {\arrow{latex}}
    }},
    electron/.default=0.5,  
    positron/.style={postaction={decorate}, decoration={
        markings,
        mark=at position #1 with {\arrow{latex}}
    }},
    positron/.default=0.5  
]
    \coordinate (left1) at (-1,-1);
    \coordinate (left2) at (-1, 1);
    \coordinate (right1) at (2, -1);
    \coordinate (right2) at (2,1);
    \coordinate (up1) at (0,1);
    \coordinate (low1) at (0,-1);
    \coordinate (up2) at (1,1);
    \coordinate (low2) at (1,-1);

    \draw[decorate, decoration=snake] (up1) -- (low1);
    \draw[decorate, decoration=snake] (up2) -- (low2);

    \draw[RoyalBlue] (left1) -- (right1);
    \draw[Maroon] (left2) -- (right2);

\end{tikzpicture}
}    
     = i  \frac{Q_1^2 Q_2^2 m_1 m_2}{\pi \vec{q}^2 \sqrt{2E/m_2}}\log{\vec{q}^2} \, .
\end{equation}
Here, $E=|\vec{p}|^2/2 m_2$ is the non-relativistic kinetic energy of the probe particle. Its Fourier transform is 
\be
\text{FT}[\mathcal{M}_1] = i \frac{(Q_1 Q_2 \log{b/b_{\text{IR}}})^2}{\sqrt{E/m_2}}\, .
\ee

This pattern continues. If you do a two-loop computation, you get a term proportional to $(\log b)^3$ from the double box, etc. Taking care of all the factors of $2$ and overall normalizations (we will fix them later), it is easy to see that the computation using the diagrams
\be
\raisebox{-1.5em}{
\begin{tikzpicture}[line width=1.1, scale=0.7,
    electron/.style={postaction={decorate}, decoration={
        markings,
        mark=at position #1 with {\arrow{latex}}
    }},
    electron/.default=0.5,  
    positron/.style={postaction={decorate}, decoration={
        markings,
        mark=at position #1 with {\arrow{latex}}
    }},
    positron/.default=0.5  
]
    \coordinate (left1) at (-1,-1);
    \coordinate (left2) at (-1, 1);
    \coordinate (right1) at (0, -1);
    \coordinate (right2) at (0,1);
    
    \draw[RoyalBlue] (left1) -- (right1);
    \draw[Maroon] (left2) -- (right2);

\end{tikzpicture}
}
+
\raisebox{-1.5em}{
\begin{tikzpicture}[line width=1.1, scale=0.7,
    electron/.style={postaction={decorate}, decoration={
        markings,
        mark=at position #1 with {\arrow{latex}}
    }},
    electron/.default=0.5,  
    positron/.style={postaction={decorate}, decoration={
        markings,
        mark=at position #1 with {\arrow{latex}}
    }},
    positron/.default=0.5  
]
]
    \coordinate (left1) at (-1,-1);
    \coordinate (left2) at (-1, 1);
    \coordinate (right1) at (1, -1);
    \coordinate (right2) at (1,1);
    \coordinate (up1) at (0,1);
    \coordinate (low1) at (0,-1);

    \draw[decorate, decoration=snake] (up1) -- (low1);

    \draw[RoyalBlue] (left1) -- (right1);
    \draw[Maroon] (left2) -- (right2);

\end{tikzpicture}
}
+
\raisebox{-1.5em}{
\begin{tikzpicture}[line width=1.1, scale=0.7,
    electron/.style={postaction={decorate}, decoration={
        markings,
        mark=at position #1 with {\arrow{latex}}
    }},
    electron/.default=0.5,  
    positron/.style={postaction={decorate}, decoration={
        markings,
        mark=at position #1 with {\arrow{latex}}
    }},
    positron/.default=0.5  
]
    \coordinate (left1) at (-1,-1);
    \coordinate (left2) at (-1, 1);
    \coordinate (right1) at (2, -1);
    \coordinate (right2) at (2,1);
    \coordinate (up1) at (0,1);
    \coordinate (low1) at (0,-1);
    \coordinate (up2) at (1,1);
    \coordinate (low2) at (1,-1);

    \draw[decorate, decoration=snake] (up1) -- (low1);
    \draw[decorate, decoration=snake] (up2) -- (low2);

    \draw[RoyalBlue] (left1) -- (right1);
    \draw[Maroon] (left2) -- (right2);

\end{tikzpicture}
}
+
\raisebox{-1.5em}{
\begin{tikzpicture}[line width=1.1, scale=0.7,
    electron/.style={postaction={decorate}, decoration={
        markings,
        mark=at position #1 with {\arrow{latex}}
    }},
    electron/.default=0.5,  
    positron/.style={postaction={decorate}, decoration={
        markings,
        mark=at position #1 with {\arrow{latex}}
    }},
    positron/.default=0.5  
]
    \coordinate (left1) at (-1,-1);
    \coordinate (left2) at (-1, 1);
    \coordinate (right1) at (3, -1);
    \coordinate (right2) at (3,1);
    \coordinate (up1) at (0,1);
    \coordinate (low1) at (0,-1);
    \coordinate (up2) at (1,1);
    \coordinate (low2) at (1,-1);
    \coordinate (up3) at (2,1);
    \coordinate (low3) at (2,-1);

    \draw[decorate, decoration=snake] (up1) -- (low1);
    \draw[decorate, decoration=snake] (up2) -- (low2);
    \draw[decorate, decoration=snake] (up3) -- (low3);

    \draw[RoyalBlue] (left1) -- (right1);
    \draw[Maroon] (left2) -- (right2);

\end{tikzpicture}
}
+ \ldots
\label{eq:ladder_sum}
\ee
exponentiates to
\be
S(E,b) = \e^{2i\delta(E,b)} = \e^{-i \frac{2Q_1 Q_2}{\sqrt{2E/m_2}} \log{b/b_{\text{IR}}}} = \e^{-2i\alpha_e \log{b/b_{\text{IR}}}}\,,
\ee
in the impact parameter space. Note that we included the disconnected piece, which gives the $1$ in a small-$\alpha_e$ expansion of the exponential (the same $1$ as the matrix element of $\mathbbm{1}$ in $S = \mathbbm{1} + iT$). Here
\be\label{eq:alpha-e}
\alpha_e=\frac{Q_1 Q_2}{\sqrt{2E/m_2}}\,,
\ee
is the effective QED coupling. We will refer to $\delta(E,b)$ as the \emph{phase shift}. Notice that restoring the $\hbar$ gives you the phase shift $\delta(E,b) = -\tfrac{1}{\hbar} \alpha_e \log b/b_{\text{IR}}$ so the loop diagrams are in fact more classical, i.e., more superleading to the tree diagrams as $\hbar \to 0$. These contributions are usually referred to as superclassical. Later on, we will see that an analogous exponentiation takes place for gravity: it is basically a resummation of the Newtonian potential.

We are now interested in extracting some classical observables from this expression. We know that the amplitude is a function of the transferred momentum $\vec{q}$ and the energy of the probe particle $E$. We already saw that the transferred momentum is conjugate to the impact parameter $b$. Since time translation acts on the amplitude as a phase $\e^{iH T}$, where $H$ is the Hamiltonian, it is easy to see that the energy is conjugate to some time delay. If we assume that we scatter wave packets sharply peaked at a given energy, then the inverse transform (from $b$ to $\vec{q}$) will look like\footnote{Note that we are using a slight abuse of notation between 2D and 3D vectors. For simplicity we can think of $\vec{b}$ as a 3D vector with non-zero components only in the $xy$ plane.} 
\begin{equation}
    \int \d E \int \d^2 \vec{b}\, \e^{ 2i\delta(E,b)+i \vec{q}\cdot \vec{b}+i T E}\, ,
\end{equation}
where $b = |\vec{b}|$.
This is a highly-oscillatory integral.
It is dominated by two saddle points. One appearing when integrating over $E$, which leads to 
\begin{equation}
    T=\frac{2\partial\text{Re}\,\delta(E,b)}{\partial E}=\frac{-Q_1 Q_2}{\sqrt{2}m_2 (E/m_2)^{3/2}}\log{b}\,,
\end{equation}
and a second one emerging from the $\vec{b}$ integration 
\be
\theta = \frac{2}{|\vec{p}|}\frac{\partial\text{Re}\delta(E,b)}{\partial b}= - \frac{Q_1 Q_2}{E b}\, ,
\ee
where we used $ \theta \sim \frac{|\vec{q}|}{|\vec{p}|}$. 
Hence, this calculation recovers the classical scattering angle and time delay!

Notice that in order to develop a saddle point, we need a large oscillating phase that happens when we scatter objects with macroscopic charges. In Sec.~\ref{sec:Correia}, we will do a similar analysis and discuss what regime gives measurable observables for gravity (the same exercise for gravity leads to the Shapiro time delay and the leading scattering angle $R_s/b$). 

\subsection{Large angular momentum limit of partial waves}

Up to this point, everything worked rather nicely at the leading order, but we have been somewhat heuristic. Let us try to do this computation more carefully.

When the impact parameter $b$ is very large, the classical angular momentum $J$ also becomes large. Therefore, it is easy to imagine that we can access this regime by projecting the amplitude in partial waves and taking the large $J$ limit.

We can carry this construction out very explicitly for the hydrogen atom. It turns out that the resummation can be performed directly in the $\vec{q}$ space for this simple case. The result is
\begin{equation}\label{eq:resumHA}
\mathcal{M}\sim \frac{Q_1 Q_2}{t}\frac{\Gamma(1-i\alpha_e)}{\Gamma(1+i\alpha_e)}\left( \frac{-t}{\mu^2}\right)^{i\alpha_e}\, ,
\end{equation}
where $\mu$ is the scale from dimensional regularization that cuts off the IR divergences. Notice the appearance of an infinite series of poles coming from the Gamma function in the numerator. Of course, these are associated with the energy levels of the hydrogen atom. One might wonder why we are observing energy levels, which are quantum mechanical properties of the wave regime, in this classical point-particle limit. This is a special feature of the hydrogen atom: in momentum space (and non-relativistic limit), the amplitude is the same for both the wave and point particle regime. This is an accident that does not hold in gravity.

It is more standard to see how this plays out in the $J$ space.
We can then compute the partial waves which (up to an overall IR divergent phase) are 
\begin{equation}\label{eq:MJE}
S_J(E)=\frac{\Gamma(1+J-i\alpha_e)}{\Gamma(1+J+i\alpha_e)}\, ,
\end{equation}
where once again we have included the disconnected piece. Let us now take the large $J$ limit of this expression. Using the Stirling approximation, it is easy to see that the leading contribution is indeed
\begin{equation}
    \e^{-2i \alpha_e \log{J}}\, ,
\end{equation}
where $J=|\vec{p}|b=\sqrt{2m_2 E}b$ is the classical angular momentum and the IR divergence is just the overall phase. This expression thus matches the naive Fourier transform that we did earlier! It is clearly the right direction. 

Of course, in general, we will not be able to access the full partial-waves solution. Therefore, let us study this limit directly at the level of the transform.

For completeness, let us do this derivation with spinning external states. Partial waves are representations of the little group in the center-of-mass frame that is $\text{SO}(3)$. They can be computed as
\begin{equation}
    {\mathcal{M}_J}^{\lambda_3 \lambda_4}_{\lambda_1\lambda_2}(s) = \mathcal{N}^{-1}\int_{-1}^{1} \d\!\cos\theta \,\, d^{J}_{\lambda_{12} \lambda_{34}}(\theta) \, \mathcal{M}_{\lambda_1 \lambda_2}^{\lambda_3\lambda_4}(p_i)\, ,
\end{equation}
where $\mathcal{N}$ is a normalization factor, $\lambda_{ij} = \lambda_i - \lambda_j$, and the Wigner $d$-matrix is
\begin{equation}\label{eq:wigner-d}
    d^{J}_{\lambda \lambda^\prime}(\theta) =\langle \lambda J| \e^{-i\theta J_2}|\lambda^\prime J\rangle\, .
\end{equation}
The bras and kets label the elements of the spherical basis of $\mathrm{SO}(3)$.
Let us study the large $J$ limit of this Wigner $d$-matrix.

We will understand this limit at the group level. In particular, the raising and lowering operator act on a state as
\begin{equation}
    J_3|J \lambda\rangle=\lambda |J \lambda\rangle\,, \qquad J_{\pm}| J \lambda \rangle= \sqrt{\mathcal{J}^2 -\lambda(\lambda\pm1)}| J\lambda\pm 1\rangle\, ,
\end{equation}
 where $\mathcal{J}^2=J(J+1)$ is the Casimir operator, and the tower contains $2J+1$ states within the irrep. In the limit $\lambda/J\ll 1$, we get
\begin{equation}
    J_{\pm}| J \lambda \rangle= \sqrt{\mathcal{J}^2}| J\lambda\pm 1\rangle+\mathcal{O}(\lambda/J)\, ,
\end{equation}
where we see that the raising and lowering operators now commute:
\be
[J_+,J_-]=0\,,
\ee
and the irreps become infinite dimensional. What we are observing is the contraction of the little group $\text{SO}(3)$ to $\text{ISO}(2)$, which is a non-compact group and as such admits only infinite dimensional irreps. Hence, at large angular momentum we recover the ``flat-earth limit'', where the isometries of a sphere reduce to those of a plane:
\be
\adjustbox{valign=c,scale={0.9}{0.9}}{\tikzset{every picture/.style={line width=0.75pt}} 
\begin{tikzpicture}[x=0.75pt,y=0.75pt,yscale=-1,xscale=1]
\draw   (94,157.5) .. controls (94,131.82) and (114.82,111) .. (140.5,111) .. controls (166.18,111) and (187,131.82) .. (187,157.5) .. controls (187,183.18) and (166.18,204) .. (140.5,204) .. controls (114.82,204) and (94,183.18) .. (94,157.5) -- cycle ;
\draw  [draw opacity=0] (187,160.13) .. controls (173.05,162.44) and (157.06,163.76) .. (140.06,163.77) .. controls (123.76,163.77) and (108.38,162.58) .. (94.84,160.45) -- (140.05,133.77) -- cycle ; \draw   (187,160.13) .. controls (173.05,162.44) and (157.06,163.76) .. (140.06,163.77) .. controls (123.76,163.77) and (108.38,162.58) .. (94.84,160.45) ;  
\draw  [draw opacity=0][dash pattern={on 4.5pt off 4.5pt}] (94.84,160.4) .. controls (108.79,158.1) and (124.78,156.79) .. (141.78,156.79) .. controls (158.08,156.79) and (173.46,157.99) .. (187,160.13) -- (141.78,186.79) -- cycle ; \draw  [dash pattern={on 4.5pt off 4.5pt}] (94.84,160.4) .. controls (108.79,158.1) and (124.78,156.79) .. (141.78,156.79) .. controls (158.08,156.79) and (173.46,157.99) .. (187,160.13) ;  
\draw  [color={rgb, 255:red, 74; green, 144; blue, 226 }  ,draw opacity=1 ][fill={rgb, 255:red, 74; green, 144; blue, 226 }  ,fill opacity=0.18 ] (158.72,103.46) -- (229.07,169.34) -- (159.28,158.79) -- (88.93,92.91) -- cycle ;
\draw  [draw opacity=0][fill={rgb, 255:red, 0; green, 0; blue, 0 }  ,fill opacity=1 ] (156.38,131.13) .. controls (156.38,129.68) and (157.55,128.5) .. (159,128.5) .. controls (160.45,128.5) and (161.63,129.68) .. (161.63,131.13) .. controls (161.63,132.57) and (160.45,133.75) .. (159,133.75) .. controls (157.55,133.75) and (156.38,132.57) .. (156.38,131.13) -- cycle ;
\draw  [draw opacity=0] (365.63,114.77) .. controls (371.74,106.21) and (387.6,100.04) .. (406.25,99.91) .. controls (425.3,99.78) and (441.53,106) .. (447.45,114.78) -- (406.4,122.35) -- cycle ; \draw[->]   (365.63,114.77) .. controls (371.74,106.21) and (387.6,100.04) .. (406.25,99.91) .. controls (425.3,99.78) and (441.53,106) .. (447.45,114.78) node[above,midway,xshift=-20,yshift=5]{$J\to \infty$} ;  
\draw (281,105.4) node [anchor=north west][inner sep=0.75pt]    {$\text{SO(3)}$};
\draw (247,129) node [anchor=north west][inner sep=0.75pt]   [align=left] {\begin{minipage}[lt]{85.33pt}\setlength\topsep{0pt}
Compact group:
\begin{center}
finite dim. irrep
\end{center}
\end{minipage}};
\draw (487,99.4) node [anchor=north west][inner sep=0.75pt]    {$\text{ISO(2)}$};
\draw (454,123) node [anchor=north west][inner sep=0.75pt]   [align=left] {\begin{minipage}[lt]{105.19pt}\setlength\topsep{0pt}
\begin{center}
Non-compact group:\\continuous irrep
\end{center}
\end{minipage}};
\draw (465,165) node [anchor=north west][inner sep=0.75pt]   [align=left] {\begin{minipage}[lt]{82.49pt}\setlength\topsep{0pt}
\begin{center}
(Emergence of \\continuous $\displaystyle J$)
\end{center}
\end{minipage}};
\end{tikzpicture}}
\ee

Let us return to the Wigner $d$-matrix. It would be ideal to find a basis given by the eigenstates of $J_+$ and $J_-$, which would drastically simplify the evaluation of the Wigner $d$-matrix elements \eqref{eq:wigner-d}. It turns out that such a basis exists and is the continuous-spin basis:
\begin{equation}
j_{\pm}|\varphi\rangle  =  \e^{\mp i\varphi } |\varphi\rangle\, ,
\end{equation}
where $\varphi$ is an angle. The continuous-spin basis and the $|\lambda\rangle$ basis are connected via a Fourier series, namely 
\begin{equation}
\label{continSpinBasis}
| \varphi \rangle \equiv \sum_{\lambda \in \text{(half-)integers}} \e^{i\varphi\lambda} |\lambda \rangle \quad \longleftrightarrow\quad  |\lambda\rangle =\int_{0}^{2\pi}\frac{\d\varphi}{2\pi} \e^{-i\lambda\varphi}|\varphi\rangle\,.
\end{equation}

In summary, in the large $J$ limit, the $|J \lambda \rangle$ states can be decomposed into a suitable basis of $\text{ISO}(2)$ irreducible representations for which $J_2$ matrix elements are diagonal.    This procedure allows us to recover the Wigner $d$-matrix $d^J(\theta)$ in the large angular momentum limit, as follows:
\begin{align}
\label{dlimitExplicit}
d^J_{\lambda^\prime \lambda}(\theta)   \xrightarrow[J\rightarrow \infty ]{\theta\rightarrow 0}    \int_{0}^{2\pi} \frac{\d\varphi}{2\pi} \e^{i(\lambda^\prime-
\lambda)\varphi} \e^{i\theta \mathcal{J} \sin\varphi } =  J_{\lambda-\lambda^\prime}(\mathcal{J} \theta) \ ,
\end{align}
where $\mathcal{J}\equiv \sqrt{J(J+1)}$ as before.
This is just an integral representation of the Bessel $J_\nu(x)$ function.

The Fourier transform emerges naturally from this picture. For example, the integral over the angles $\theta$ can be translated into that over $q$. The Wigner $d$-matrix becomes the integral over $\varphi$ as above. In the large $J$ limit, the amplitude exponentiates with the phase shift. More precisely, we have :
\be
\underbracket[0.4pt]{{\mathcal{M}_J}^{\lambda_3 \lambda_4}_{\lambda_1\lambda_2}(s)}_{\frac{\e^{2i \delta_J} - 1}{i}} = \mathcal{N}^{-1} \underbracket[0.4pt]{\int_{-1}^{1} \d\!\cos\theta}_{\sim \int_0^\infty \d q\, q} \!\!\!\!\!\!\!\!\!\!\!\!\!\!\!\!\!\!\!\! \underbracket[0.4pt]{d^{J}_{\lambda_{12} \lambda_{34}}(\theta)}_{\qquad\qquad \int \frac{\d \varphi}{2\pi} \e^{i(\lambda_{12} - \lambda_{34})\varphi} \e^{i\theta J \sin \varphi}} \!\!\!\!\!\!\!\!\!\!\!\!\!\!\!\!\!\!\!\! \mathcal{M}_{\lambda_1 \lambda_2}^{\lambda_3\lambda_4}(p_i)\, ,
\ee

This procedure gives a systematic way to extract the phase shift order by order in $\theta$. The regime of the eikonal approximation is
\begin{equation}
\theta = \frac{\alpha_e}{J}\ll 1\, .
\end{equation}
Up to $\mathcal{O}(\alpha_e^2)$, it is a simple Fourier transform, but at subleading orders there will be some corrections to this limit of the Wigner $d$-matrix, which is in principle known to all orders. For example, if we wanted to extract the $\mathcal{O}(\alpha_e^3)$ contribution from the amplitude, we need to slightly modify the 2D Fourier transform to make it consistent with the large $J$ limit of the partial waves.

Notice that the saddle point is dominated by the scaling $J\theta \sim 1$, so at large angular momentum we are automatically in the small angle $\theta \sim t/s\sim \alpha_e /J$ regime. 
This is the reason why people often directly expand the amplitude at small $t/s$ and refer to this as the ``classical limit''.
In gravity those $\alpha_g/J$ perturbative corrections to $\log{b}$ are precisely the Post-Minkowskian (PM) corrections, as will be illustrated in the next section.

\subsection{\label{sec:resumming}Resumming at \texorpdfstring{$\alpha_e\sim J$}{alphaeJ} and bound orbits}

We can now ask what happens when the ratio $\alpha_e/J$ becomes $\mathcal{O}(1)$. It is easy to see that this will correspond to a larger scattering angle, which (spoiler alert) will lead us to bound orbits.

Since in the case of the hydrogen atom we actually know the full solution, we can study that limit explicitly. Recall from the introduction that we are interested in the point-particle limit $b\gg \lambda_{\text{dB}}=\frac{1}{|\vec{p}|}$, which implies $J=|\vec{p}|b\gg 1$. Therefore, this regime must have the following hierarchy: $J\sim \alpha_e\gg 1$ (alternatively, this can be seen by taking the naive $\hbar \rightarrow 0$ limit).

We first observe that (up to a $J$ independent phase), the amplitude \eqref{eq:MJE} exponentiates into the following object
\begin{equation}
S_J \sim \e^{2\frac{i}{\hbar} I(E,J)}\,,
\end{equation}
with
\begin{equation}
    I(E,J) = \alpha_e \left[-\frac{1}{2} \log{(J^2 + \alpha_e^2)} + 1 - \frac{J}{\alpha_e} \arctan (\frac{\alpha_e}{J})\right]\, .
\end{equation}
We wrote down explicitly the $\hbar$ dependence to highlight the fact that it becomes a ``classical action''. 
This is a fully classical object usually referred to as the \emph{radial action}.

It can be obtained by solving equations of motion for a point particle in an effective potential. Here, we are in the probe limit, so the potential is literally the Coulomb $1/r$ potential generated by the heavy object, but it can be generalized away from the probe limit by constructing an effective potential for one effective body with reduced mass. In Sec.~\ref{sec:Correia}, we will see more details about this computation in the gravitational case. 
We will return to the above effective potential and how it can be extracted from the amplitude in a moment.

Before that, let us examine the analytic structure of the amplitude in partial waves $S_J(E)$ from \eqref{eq:MJE}, which looks as follows:
\be
\adjustbox{valign=c,scale={1}{1}}{\DeclareRobustCommand{\svdots}{
  \vbox{%
    \baselineskip=0.33333\normalbaselineskip
    \lineskiplimit=0pt
    \hbox{.}\hbox{.}\hbox{.}%
    \kern-0.2\baselineskip
  }%
}
\tikzset{every picture/.style={line width=0.75pt}} 
\begin{tikzpicture}[x=0.75pt,y=0.75pt,yscale=-1,xscale=1]
\draw [color=ForestGreen  ,draw opacity=1 ]   (360,169) -- (377,169) ;
\draw [color=ForestGreen  ,draw opacity=1 ]   (357,162) -- (378,162) ;
\draw [color=ForestGreen  ,draw opacity=1 ]   (356,157) -- (380,157) ;
\draw [color=ForestGreen  ,draw opacity=1 ]   (356,154) -- (381,154) ;
\draw[<-]    (139.5,91) -- (139.5,194) ;
\draw[->]    (74,142.5) -- (205,142.5) ;
\draw[<-]    (339.5,91) -- (339.5,194) node[pos=0,left] {$V_{\text{eff}}$};
\draw   (204,104) -- (190,104) node[above,xshift=6,yshift=-2]{$E$} -- (190,90);
\draw  [draw opacity=0][fill=Orange  ,fill opacity=1 ] (128,142.75) .. controls (128,141.78) and (128.78,141) .. (129.75,141) .. controls (130.72,141) and (131.5,141.78) .. (131.5,142.75) .. controls (131.5,143.72) and (130.72,144.5) .. (129.75,144.5) .. controls (128.78,144.5) and (128,143.72) .. (128,142.75) -- cycle ;
\draw  [draw opacity=0][fill=Orange  ,fill opacity=1 ] (107,142.75) .. controls (107,141.78) and (107.78,141) .. (108.75,141) .. controls (109.72,141) and (110.5,141.78) .. (110.5,142.75) .. controls (110.5,143.72) and (109.72,144.5) .. (108.75,144.5) .. controls (107.78,144.5) and (107,143.72) .. (107,142.75) -- cycle ;
\draw  [draw opacity=0][fill=Orange  ,fill opacity=1 ] (92,142.75) .. controls (92,141.78) and (92.78,141) .. (93.75,141) .. controls (94.72,141) and (95.5,141.78) .. (95.5,142.75) .. controls (95.5,143.72) and (94.72,144.5) .. (93.75,144.5) .. controls (92.78,144.5) and (92,143.72) .. (92,142.75) -- cycle ;
\draw  [draw opacity=0][fill=Orange  ,fill opacity=1 ] (79,142.75) .. controls (79,141.78) and (79.78,141) .. (80.75,141) .. controls (81.72,141) and (82.5,141.78) .. (82.5,142.75) .. controls (82.5,143.72) and (81.72,144.5) .. (80.75,144.5) .. controls (79.78,144.5) and (79,143.72) .. (79,142.75) -- cycle ;
\draw  [draw opacity=0][fill=Orange  ,fill opacity=1 ] (131.5,142.75) .. controls (131.5,141.78) and (132.28,141) .. (133.25,141) .. controls (134.22,141) and (135,141.78) .. (135,142.75) .. controls (135,143.72) and (134.22,144.5) .. (133.25,144.5) .. controls (132.28,144.5) and (131.5,143.72) .. (131.5,142.75) -- cycle ;
\draw  [draw opacity=0][fill=Orange  ,fill opacity=1 ] (122.5,142.75) .. controls (122.5,141.78) and (123.28,141) .. (124.25,141) .. controls (125.22,141) and (126,141.78) .. (126,142.75) .. controls (126,143.72) and (125.22,144.5) .. (124.25,144.5) .. controls (123.28,144.5) and (122.5,143.72) .. (122.5,142.75) -- cycle ;
\draw  [draw opacity=0][fill=Orange  ,fill opacity=1 ] (118,142.75) .. controls (118,141.78) and (118.78,141) .. (119.75,141) .. controls (120.72,141) and (121.5,141.78) .. (121.5,142.75) .. controls (121.5,143.72) and (120.72,144.5) .. (119.75,144.5) .. controls (118.78,144.5) and (118,143.72) .. (118,142.75) -- cycle ;
\draw [color={rgb, 255:red, 208; green, 2; blue, 27 }  ,draw opacity=1 ]   (139.5,142.5) .. controls (141.17,140.83) and (142.83,140.83) .. (144.5,142.5) .. controls (146.17,144.17) and (147.83,144.17) .. (149.5,142.5) .. controls (151.17,140.83) and (152.83,140.83) .. (154.5,142.5) .. controls (156.17,144.17) and (157.83,144.17) .. (159.5,142.5) .. controls (161.17,140.83) and (162.83,140.83) .. (164.5,142.5) .. controls (166.17,144.17) and (167.83,144.17) .. (169.5,142.5) .. controls (171.17,140.83) and (172.83,140.83) .. (174.5,142.5) .. controls (176.17,144.17) and (177.83,144.17) .. (179.5,142.5) .. controls (181.17,140.83) and (182.83,140.83) .. (184.5,142.5) .. controls (186.17,144.17) and (187.83,144.17) .. (189.5,142.5) .. controls (191.17,140.83) and (192.83,140.83) .. (194.5,142.5) .. controls (196.17,144.17) and (197.83,144.17) .. (199.5,142.5) .. controls (201.17,140.83) and (202.83,140.83) .. (204.5,142.5) -- (205,142.5) -- (205,142.5) ;
\draw    (353,142.5) -- (405,142.5) ;
\draw[->]   (274,142.5) -- (405,142.5) node[right]{$r$};
\draw    (346,98) .. controls (347.1,126.57) and (364,204.5) .. (373,179.5) .. controls (382,154.5) and (378,142.5) .. (405,142.5) ;

\draw  [draw opacity=0][fill=ForestGreen  ,fill opacity=1 ] (367,184.75) .. controls (367,183.78) and (367.78,183) .. (368.75,183) .. controls (369.72,183) and (370.5,183.78) .. (370.5,184.75) .. controls (370.5,185.72) and (369.72,186.5) .. (368.75,186.5) .. controls (367.78,186.5) and (367,185.72) .. (367,184.75) -- cycle ;

\draw (46,149.4) node [anchor=north west][inner sep=0.75pt]  [font=\small,color=Orange  ,opacity=1 ]  {$\sqrt{\frac{m}{2}}\frac{Q_{1} Q_2}{( 1+J)}$};
\draw (365.25,143) node [anchor=north west][inner sep=0.75pt]  [font=\tiny,color=ForestGreen  ,opacity=1 ]  {$\svdots$};
\end{tikzpicture}}
\ee
It is clear that at negative energy the gamma functions develop some poles, which are the energy levels of the hydrogen atom bound states. In the $\hbar \rightarrow 0$ limit, the poles get closer and create a continuum:
\be
\adjustbox{valign=c,scale={1}{1}}{\tikzset{every picture/.style={line width=0.75pt}} 
\begin{tikzpicture}[x=0.75pt,y=0.75pt,yscale=-1,xscale=1]
\draw[<-]    (119.5,71) -- (119.5,174) ;
\draw[->]     (54,122.5) -- (185,122.5) ;
\draw[<-]     (319.5,71) -- (319.5,174) node[pos=0,left] {$V_{\text{eff}}$};
\draw   (184,84) -- (170,84) node[above,xshift=6,yshift=-2]{$E$} -- (170,70);
\draw  [draw opacity=0][fill=Orange ,fill opacity=1 ] (59,122.75) .. controls (59,121.78) and (59.78,121) .. (60.75,121) .. controls (61.72,121) and (62.5,121.78) .. (62.5,122.75) .. controls (62.5,123.72) and (61.72,124.5) .. (60.75,124.5) .. controls (59.78,124.5) and (59,123.72) .. (59,122.75) -- cycle ;
\draw [color=Maroon ,draw opacity=1 ]   (119.5,122.5) .. controls (121.17,120.83) and (122.83,120.83) .. (124.5,122.5) .. controls (126.17,124.17) and (127.83,124.17) .. (129.5,122.5) .. controls (131.17,120.83) and (132.83,120.83) .. (134.5,122.5) .. controls (136.17,124.17) and (137.83,124.17) .. (139.5,122.5) .. controls (141.17,120.83) and (142.83,120.83) .. (144.5,122.5) .. controls (146.17,124.17) and (147.83,124.17) .. (149.5,122.5) .. controls (151.17,120.83) and (152.83,120.83) .. (154.5,122.5) .. controls (156.17,124.17) and (157.83,124.17) .. (159.5,122.5) .. controls (161.17,120.83) and (162.83,120.83) .. (164.5,122.5) .. controls (166.17,124.17) and (167.83,124.17) .. (169.5,122.5) .. controls (171.17,120.83) and (172.83,120.83) .. (174.5,122.5) .. controls (176.17,124.17) and (177.83,124.17) .. (179.5,122.5) .. controls (181.17,120.83) and (182.83,120.83) .. (184.5,122.5) -- (185,122.5) -- (185,122.5) ;
\draw    (333,122.5) -- (385,122.5) ;
\draw [draw opacity=0][fill=ForestGreen ,fill opacity=0.2 ]   (333,122.5) .. controls (339,150.5) and (348,176.5) .. (353,159.5) .. controls (358,142.5) and (358,122.5) .. (385,122.5) ;
\draw[->]    (254,122.5) -- (385,122.5) node[right]{$r$};
\draw    (326,78) .. controls (327.1,106.57) and (344,184.5) .. (353,159.5) .. controls (362,134.5) and (358,122.5) .. (385,122.5) ;

\draw  [draw opacity=0][fill=ForestGreen ,fill opacity=1 ] (347,164.75) .. controls (347,163.78) and (347.78,163) .. (348.75,163) .. controls (349.72,163) and (350.5,163.78) .. (350.5,164.75) .. controls (350.5,165.72) and (349.72,166.5) .. (348.75,166.5) .. controls (347.78,166.5) and (347,165.72) .. (347,164.75) -- cycle ;
\draw [color=Orange ,draw opacity=1 ]   (60.75,122.75) .. controls (62.41,121.08) and (64.08,121.07) .. (65.75,122.73) .. controls (67.42,124.39) and (69.09,124.38) .. (70.75,122.71) .. controls (72.41,121.04) and (74.08,121.03) .. (75.75,122.69) .. controls (77.42,124.34) and (79.09,124.33) .. (80.75,122.66) .. controls (82.41,120.99) and (84.08,120.98) .. (85.75,122.64) .. controls (87.42,124.3) and (89.09,124.29) .. (90.75,122.62) .. controls (92.41,120.95) and (94.08,120.94) .. (95.75,122.6) .. controls (97.42,124.26) and (99.09,124.25) .. (100.75,122.58) .. controls (102.41,120.91) and (104.08,120.9) .. (105.75,122.56) .. controls (107.42,124.22) and (109.09,124.21) .. (110.75,122.54) .. controls (112.41,120.87) and (114.08,120.86) .. (115.75,122.52) -- (119.5,122.5) -- (119.5,122.5) ;
\draw  [draw opacity=0] (160.12,107.89) .. controls (156.26,108.43) and (152.21,107) .. (149.57,103.75) .. controls (147.26,100.91) and (146.54,97.3) .. (147.28,93.98) -- (158.5,96.5) -- cycle ; \draw  [color=Maroon ,draw opacity=1 ] (160.12,107.89) .. controls (156.26,108.43) and (152.21,107) .. (149.57,103.75) .. controls (147.26,100.91) and (146.54,97.3) .. (147.28,93.98) ;  
\draw  [draw opacity=0] (134.39,106.76) .. controls (138.04,105.37) and (142.3,105.87) .. (145.6,108.45) .. controls (148.48,110.71) and (149.99,114.07) .. (150,117.47) -- (138.5,117.5) -- cycle ; \draw  [color=Maroon ,draw opacity=1 ] (134.39,106.76) .. controls (138.04,105.37) and (142.3,105.87) .. (145.6,108.45) .. controls (148.48,110.71) and (149.99,114.07) .. (150,117.47) ;  
\draw  [color=Orange ,draw opacity=1 ] (96,113) .. controls (96,111.9) and (100.25,111) .. (105.5,111) .. controls (110.75,111) and (115,111.9) .. (115,113) .. controls (115,114.1) and (110.75,115) .. (105.5,115) .. controls (100.25,115) and (96,114.1) .. (96,113) -- cycle ;
\draw  [color=Orange ,draw opacity=1 ] (71,112.5) .. controls (71,110.01) and (75.25,108) .. (80.5,108) .. controls (85.75,108) and (90,110.01) .. (90,112.5) .. controls (90,114.99) and (85.75,117) .. (80.5,117) .. controls (75.25,117) and (71,114.99) .. (71,112.5) -- cycle ;
\draw  [color=Orange ,draw opacity=1 ] (45,109) .. controls (45,102.92) and (49.92,98) .. (56,98) .. controls (62.08,98) and (67,102.92) .. (67,109) .. controls (67,115.08) and (62.08,120) .. (56,120) .. controls (49.92,120) and (45,115.08) .. (45,109) -- cycle ;

\draw (26,129.4) node [anchor=north west][inner sep=0.75pt]  [font=\small,color=Orange ,opacity=1 ]  {$\sqrt{\frac{m}{2}}\frac{Q_{1} Q_2}{( 1+J)}$};
\end{tikzpicture}
}
\ee
From the picture of the potential it is easy to see that this should correspond to classical bound orbits.
This effect can be seen in the radial action, which develops an imaginary part at negative energies.

Now, let us try to do the analytic continuation explicitly. For this purpose, we write
\begin{equation}
   \sqrt{E} = i \sqrt{-E}\,,
\end{equation}
with Feynman $i \varepsilon$ prescription, $E \to E + i 0^+$ to choose the correct branch. The real part of the action is given by 
\begin{equation}
  \mathrm{Re}\, I(E,J) = 
   \left[ \frac{\pi Q_1 Q_2 }{2\sqrt{-2E/m} } + \frac{\pi J}{2} \right]  \Theta\big(- \frac{mQ_1^2 Q_2^2}{2 J^2} < E < 0\big) \, .
\end{equation}
This allows us to compute the orbital period and the angle swept after one orbit by taking derivatives with respect to the real part. We find 
\be
   T = 2 {\partial \, \mathrm{Re} I \over \partial E} = \frac{\partial \alpha_e}{\partial E}
={\pi Q_1 Q_2 m^{1/2}\over 2\sqrt{2} (-E)^{3/2}}\,,
\ee
and
\be
\Delta \phi = 2 {\partial \, \mathrm{Re} I \over \partial J}+\pi = 2 \pi.
\ee
The extra factor of $\pi$ is there in order to go from the deflection angle (that we were interested about in the scattering problem) to the angle swept by the particle.
Adding subleading corrections can be done using this procedure. We will see later how to derive the perihelion shift in gravity from such considerations.

\begin{QA}
\question{
Why is the time delay supposed to compute the period?}
The statement is that the time delay and the period are analytic continuations of each other. This should not have been obvious a priori. Understanding this type of continuation, in general, remains an open problem.
\end{QA}

A fascinating aspect of this derivation is the fact that from a purely classical perspective there is no reason why the bound and unbound regimes are an analytic continuation of each other or why the choice of $\varepsilon$ prescription is the correct one. These features are purely derived from the underlying quantum theory.

\subsection{Potential, Born series, and Schr\"odinger equation}

We obviously do not always have the luxury of easily resumming the amplitudes in momentum space, so how can we access the $\alpha_e \sim J$ regime by computing only a finite number of diagrams in perturbation theory? This section is going to be a bit vague and is meant to give only a general idea behind this problem. This will be made more precise in the next section.

As mentioned earlier, the object we need is the potential $V$. It is already easy to see that the 3D Fourier transform of the tree-level diagram gives 
\begin{equation}
\raisebox{-1.5em}{
\begin{tikzpicture}[line width=1.1, scale=0.7,
    electron/.style={postaction={decorate}, decoration={
        markings,
        mark=at position #1 with {\arrow{latex}}
    }},
    electron/.default=0.5,  
    positron/.style={postaction={decorate}, decoration={
        markings,
        mark=at position #1 with {\arrow{latex}}
    }},
    positron/.default=0.5  
]
    \coordinate (left1) at (-1,-1);
    \coordinate (left2) at (-1, 1);
    \coordinate (right1) at (1, -1);
    \coordinate (right2) at (1,1);
    \coordinate (up1) at (0,1);
    \coordinate (low1) at (0,-1);

    \draw[decorate, decoration=snake] (up1) -- (low1);

    \draw[RoyalBlue] (left1) -- (right1);
    \draw[Maroon] (left2) -- (right2);

\end{tikzpicture}
} =  - m_1 m_2  \int \d^3 \vec{q}\, \e^{iqr}\frac{Q_1 Q_2}{\vec{q}^2}\, ,
\end{equation}
where $q = |\vec{q}|$. Up to a normalization (which we will ignore right now), it is just the Coulomb potential:
\begin{equation}
    V(r)=\frac{Q_1 Q_2}{r}\, .
\end{equation}
Let us now examine the form of the one-loop integrand in the probe non-relativistic limit. The first step is to solve explicitly for the first component $k^0$ of the loop momentum $k = (k^0, \vec{k})$. Then, expanding the result in the large mass limit, we only get a contribution from the ``matter pole'' of the form
\begin{align}
\raisebox{-1.5em}{
\begin{tikzpicture}[line width=1.1, scale=0.7,
    electron/.style={postaction={decorate}, decoration={
        markings,
        mark=at position #1 with {\arrow{latex}}
    }},
    electron/.default=0.5, 
    positron/.style={postaction={decorate}, decoration={
        markings,
        mark=at position #1 with {\arrow{latex}}
    }},
    positron/.default=0.5 
]
    \coordinate (left1) at (-1,-1);
    \coordinate (left2) at (-1, 1);
    \coordinate (right1) at (2, -1);
    \coordinate (right2) at (2,1);
    \coordinate (up1) at (0,1);
    \coordinate (low1) at (0,-1);
    \coordinate (up2) at (1,1);
    \coordinate (low2) at (1,-1);
    \draw[decorate, decoration=snake] (up1) -- (low1);
    \draw[decorate, decoration=snake] (up2) -- (low2);
    \draw[RoyalBlue] (left1) -- (right1);
    \draw[Maroon] (left2) -- (right2);
\end{tikzpicture}
} & = \int \frac{\d k_0\, \d^3 \vec{k}}{(2\pi)^4} \frac{1}{(k-p_1)^2+i\varepsilon} \frac{1}{(k-p_3)^2+i\varepsilon} \\
&\qquad\qquad\quad\times\frac{1}{k^2-m_1^2+i\varepsilon} \frac{1}{(p_1+p_2-k)^2-m_2^2+i \varepsilon} \nonumber\\
 &\sim   \frac{1}{4m_1}\int \frac{\d^3 \vec{k}}{(2\pi)^3}\frac{Q_1 Q_2}{(p-k)^2}\frac{1}{{\vec{k}^2 \over 2m_2}-E-i\varepsilon}\frac{Q_1 Q_2}{(p+q-k)^2} \nonumber\\
 &\sim \frac{1}{4m_1}\int \frac{\d^3 \vec{k}}{(2\pi)^3} V((p-k)^2)\frac{1}{{\vec{k}^2 \over 2m_2}-E-i\varepsilon} V((p+q-k)^2)\, ,\nonumber
\end{align}
which is some sort of iteration of the potential $V$, confirming that in this limit all the dynamics is hidden in the Coulomb potential. At two loops we get an extra iteration of $V$, at three loops we get two extra iterations of $V$, etc.

What is the kernel that controls this iteration? By Fourier transforming the kernel in position space we get 
\begin{equation}
    G_+(x)\sim \frac{\e^{ip|x-x^\prime|}}{|x-x^\prime|}\, ,
\end{equation}
which is the retarded Green's function of the wave equation. This should not come as a surprise since in the classical limit the scattering is only influenced by events happening in the past lightcone.

The structure of the amplitude we are observing is nothing but a Born series, which is a perturbative way of solving a wave equation in the small potential regime. We are not going into details here, as this is standard quantum mechanics textbook material (see, for instance, \cite{Sakurai:2011zz}). What is the wave equation whose solution is the non-relativistic hydrogen atom? Obviously, the answer is the Schr\"odinger equation 
\begin{equation}
    \left(\frac{\nabla^2}{2m_2} + E \right)\psi=V(r)\, \psi\,,
\end{equation}
with $V(r)=Q_1Q_2/r$. 

As a last exercise before moving on to gravity, we can ask how does the radial action emerge from this formulation. Since the radial action is written in the angular momentum space $J$, we start by rewriting the Schr\"odinger equation in spherical harmonics: 
\begin{equation}
-\frac{1}{2m_2 r^2}\frac{\d}{\d r}\left(r^2 \frac{\d}{\d r}\psi_J(r)\right) = \left(-E +V(r)+\frac{J(J+1)}{2m_2r^2}\right)\psi_J(r) \, .
\end{equation}

What is the semi-classical limit of the Schr\"odinger equation? That is precisely the regime of the WKB approximation $\psi_J=e^{\frac{i}{\hbar}I_J(r)}$. Plugging this into the Schr\"odinger equation, taking the $\hbar \rightarrow 0$ limit, and solving for $I_J$ yields
\begin{equation}\label{eq:IJ}
    I_J(E)=\int \d r \sqrt{2m_2\left(E-V(r)\right)-\frac{J^2}{r^2}}\, ,
\end{equation}
which is precisely the radial action that we constructed from the amplitude earlier. 

At this stage, let us summarize this section with a graphic representing all the connections we have established:
\be
\adjustbox{valign=c,scale={1.1}{1.1}}{\tikzset{ma/.style={decoration={markings,mark=at position 0.5 with {\arrow[scale=0.7]{>}}},postaction={decorate}}}
\tikzset{ma2/.style={decoration={markings,mark=at position 0.3 with {\arrow[scale=0.7,Maroon!70!black]{>}}},postaction={decorate}}}
\tikzset{ma3/.style={decoration={markings,mark=at position 0.7 with {\arrow[scale=0.7,RoyalBlue!80!black]{>}}},postaction={decorate}}}
\tikzset{mar/.style={decoration={markings,mark=at position 0.5 with {\arrowreversed[scale=0.7]{>}}},postaction={decorate}}}
\tikzset{every picture/.style={line width=0.75pt}}
\begin{tikzpicture}[x=0.75pt,y=0.75pt,yscale=-1,xscale=1]
\draw  [color=Maroon ,draw opacity=1 ][fill=Maroon ,fill opacity=0.1 ] (244.5,16) -- (472,16) -- (374.5,56) -- (147,56) -- cycle ;
\draw  [color=RoyalBlue  ,draw opacity=1 ][fill=RoyalBlue  ,fill opacity=0.1 ] (242.5,175.5) -- (470,175.5) -- (372.5,215.5) -- (145,215.5) -- cycle ;
\draw    (230,44) -- (281,25) ;
\draw    (231,202) -- (351,197) ;
\draw  [dash pattern={on 0.84pt off 2.51pt}]  (230,44) -- (231,202) ;
\draw  [dash pattern={on 0.84pt off 2.51pt}]  (281,25) -- (282,183) ;
\draw    (231,202) -- (282,183) ;
\draw    (282,183) -- (351,197) ;
\draw[color=ForestGreen,mar]  [dash pattern={on 4.5pt off 4.5pt}]  (230,44) -- (351,197) ;
\draw  [draw opacity=0][fill=black  ,fill opacity=1 ] (227.75,44) .. controls (227.75,42.76) and (228.76,41.75) .. (230,41.75) .. controls (231.24,41.75) and (232.25,42.76) .. (232.25,44) .. controls (232.25,45.24) and (231.24,46.25) .. (230,46.25) .. controls (228.76,46.25) and (227.75,45.24) .. (227.75,44) -- cycle ;
\draw  [draw opacity=0][fill=black  ,fill opacity=1 ] (278.75,25) .. controls (278.75,23.76) and (279.76,22.75) .. (281,22.75) .. controls (282.24,22.75) and (283.25,23.76) .. (283.25,25) .. controls (283.25,26.24) and (282.24,27.25) .. (281,27.25) .. controls (279.76,27.25) and (278.75,26.24) .. (278.75,25) -- cycle ;
\draw  [draw opacity=0][fill=black  ,fill opacity=1 ] (348.75,197) .. controls (348.75,195.76) and (349.76,194.75) .. (351,194.75) .. controls (352.24,194.75) and (353.25,195.76) .. (353.25,197) .. controls (353.25,198.24) and (352.24,199.25) .. (351,199.25) .. controls (349.76,199.25) and (348.75,198.24) .. (348.75,197) -- cycle ;
\draw  [draw opacity=0][fill=black  ,fill opacity=1 ] (228.75,202) .. controls (228.75,200.76) and (229.76,199.75) .. (231,199.75) .. controls (232.24,199.75) and (233.25,200.76) .. (233.25,202) .. controls (233.25,203.24) and (232.24,204.25) .. (231,204.25) .. controls (229.76,204.25) and (228.75,203.24) .. (228.75,202) -- cycle ;
\draw  [draw opacity=0][fill=black  ,fill opacity=1 ] (279.75,183) .. controls (279.75,181.76) and (280.76,180.75) .. (282,180.75) .. controls (283.24,180.75) and (284.25,181.76) .. (284.25,183) .. controls (284.25,184.24) and (283.24,185.25) .. (282,185.25) .. controls (280.76,185.25) and (279.75,184.24) .. (279.75,183) -- cycle ;
\draw[->]    (262,146) -- (277,178) ;

\draw (476,13) node [anchor=north west][inner sep=0.75pt]  [font=\scriptsize] [align=left] {\begin{minipage}[lt]{42.07pt}\setlength\topsep{0pt}
\begin{center}
\textcolor{Maroon}{Point-like}
\textcolor{Maroon}{particle}
\end{center}

\end{minipage}};
\draw (472,178.5) node [anchor=north west][inner sep=0.75pt]  [font=\scriptsize] [align=left] {\begin{minipage}[lt]{24.54pt}\setlength\topsep{0pt}
\begin{center}
\textcolor{RoyalBlue}{Wave}
\textcolor{RoyalBlue}{regime}
\end{center}

\end{minipage}};
\draw (197,117.4) node [anchor=north west][inner sep=0.75pt]  [font=\scriptsize]  {$J\gg1$};
\draw (282,67.4) node [anchor=north west][inner sep=0.75pt]  [font=\scriptsize]  {$\text{WKB}$};
\draw (229,180.53) node [anchor=north west][inner sep=0.75pt]  [font=\tiny,rotate=-338.3] [align=left] {Solve\\ WE};
\draw (262.6,203.31) node [anchor=north west][inner sep=0.75pt]  [font=\tiny,rotate=-357.84] [align=left] {Partial wave};
\draw (290.87,174.51) node [anchor=north west][inner sep=0.75pt]  [font=\tiny,rotate=-11.91] [align=left] {Extract pot.};
\draw (239.37,29.76) node [anchor=north west][inner sep=0.75pt]  [font=\tiny,rotate=-338.12] [align=left] {Solve};
\draw (188.32,36.01) node [anchor=north west][inner sep=0.75pt]  [font=\scriptsize]  {$\e^{iI( b,s)}$};
\draw (195.92,198.23) node [anchor=north west][inner sep=0.75pt]  [font=\scriptsize,rotate=-359]  {$A_{J}( s)$};
\draw (357.06,191.11) node [anchor=north west][inner sep=0.75pt]  [font=\scriptsize,rotate=-359]  {$A(s,t)$};
\draw (224.51,117) node [anchor=north west][inner sep=0.75pt]  [font=\tiny,rotate=0] [align=left] {\begin{minipage}[lt]{45.87pt}\setlength\topsep{0pt}
\begin{center}
Spherical \\harmonics WE
\end{center}
\end{minipage}};
\draw[color=ForestGreen] (319,126.4) node [anchor=north west][inner sep=0.75pt]  {$\int \e^{ib\cdot q}$};
\end{tikzpicture}
}
\ee

At the level of the wave regime (lower plane on the diagram) we are effectively describing the non-relativistic hydrogen atom (neglecting corrections due to spin, relativistic effects, finite size, etc.). We were able to resum all ladder diagrams into a compact expression \eqref{eq:resumHA}. By projecting into partial waves, we obtained \eqref{eq:MJE}, whose poles at negative energies were identified as energy levels of the hydrogen atom. This solution can be obtained directly by solving the Schr\"odinger equation in spherical harmonics. The dynamics of this problem is fully fixed by the Coulomb potential $V=Q_1 Q_2/r$ which can be readily extracted from the tree-level amplitude. We saw that in fact ladder diagrams in this regime organise themselves into a Born series of iterations of the Coulomb potential.

From this regime we can take the point particle limit (upper plane of the diagram). At the level of the amplitude this is equivalent to taking the limit $q \ll |\vec{p}|$. In this particular example, the form of the amplitude does not change (this is not the case in gravity as we will see next). Similarly, the point particle limit of the amplitude in angular momentum space is reached by taking $J\gg1$. In this limit the amplitude naturally exponentiates into the radial action. From this object we can extract classical observables (such as scattering angles and time delays) by performing a saddle point analysis. Since in general we do not have access to the full $S_J(E)$ solution, we studied how it can be extracted perturbatively from amplitude calculations, by taking the $J\gg 1$ limit directly at the level of the transform, which up to $\mathcal{O}(\alpha_e^2)$ reduces to a 2D Fourier transform. The radial action can be obtained alternatively from the Schr\"odinger equation by considering the WKB approximation.

\section[Gravity and relativistic Born series]{Gravity and relativistic Born series\\
\normalfont{\textit{Miguel Correia}}}\label{sec:Correia}

In this section, we are going to generalize what we have seen before to the case of gravity. Recall that we are interested in extracting classical observables from scattering amplitudes. To illustrate how to do this, we are going to review the appearance of some of the classical aspects of GR: bending of light, Shapiro time delay, precession of the perihelion of Mercury, the effects of horizon dissipation (in Sec.~\ref{sec:Zhou}), emission of gravitational waves (in Sec.~\ref{sec:Hannesdottir}), and non-linear three-body dynamics (in Sec.~\ref{sec:Wolz}). The purpose of this section is to cover the first three effects.

\subsection{Gravitational amplitudes}

We want to compute scattering amplitudes in the presence of gravity. Let us start with reviewing the basic setup for approaching this computation, as if we lived in the 1950's. The first step is to write down the action, which we can split into:
\begin{equation}
    S = S_\text{gravity} + S_\text{matter} + S_\text{gauge fixing} + S_\text{ghost}\, .
\end{equation}
Let us consider the Einstein--Hilbert action for GR with two minimally coupled scalars $\Phi_1$ and $\Phi_2$:
\begin{equation}
    S_\text{gravity} = {1 \over 16 \pi G} \int \d^4 x \sqrt{-g} \, R\,,
\end{equation}
and
\begin{equation}
S_\text{matter} = {1 \over 2} \int \d^4 x \sqrt{-g} \left[ g^{\mu \nu} \partial_\mu \Phi_1 \partial_\nu \Phi_1 -  m_1^2 \Phi_1^2  + g^{\mu \nu} \partial_\mu \Phi_2 \partial_\nu \Phi_2 - m_2^2 \Phi_2^2  \right]\, .
\end{equation}

To construct the Feynman rules, we expand the metric around flat space
\begin{equation}
    g_{\mu \nu} = \eta_{\mu \nu} + \sqrt{32 \pi G} \, h_{\mu \nu}\,,
\end{equation}
and consider $h_{\mu\nu}$ to be the field we want to scatter.
This procedure generates higher-point interactions. Schematically, the action looks like
\begin{equation}
    S_\text{gravity} = \int \d^4 x \left[ (\partial h)^2 + \sqrt{G} \partial^2 h^3 + \mathcal{O}(G) \right]\,.
\end{equation}
and
\begin{equation}
    S_\text{matter} = \int \d^4 x \left[ (\partial \Phi_1)^2 + (\partial \Phi_2)^2 + \sqrt{G} h_{\mu \nu} (T^{\mu \nu}[\Phi_1] + T^{\mu \nu}[\Phi_2])  + \mathcal{O}(G) \right]\, .
\end{equation}
The action has an infinite number of terms, which becomes quite cumbersome at higher orders. This is where the on-shell revolution plays a big role. Essentially, we have learned that we do not need any of this formalism.\footnote{At this stage, Miguel crossed out all the above equations.}

We are going to follow the ``bootstrap'' philosophy and use general principles to constrain the amplitudes instead. It will suffice to know that the graviton is a spin-$2$ massless particle and use a few of the modern tools such as:
\begin{itemize}
    \item Unitarity and factorization on poles
    \item Spinor-helicity formalism
    \item Double copy relations (gravity $\sim  \text{Yang--Mills}^2$)
    \item Generalized unitarity 
    \item Integration by parts and differential equations
\end{itemize}
In this lecture we are not going to review any of the above methods, instead we are mostly interested in how to extract physical observables from scattering amplitudes.  

Analogously to the hydrogen atom case discussed previously, we start with tree-level scattering of $m_1 m_2 \to m_1 m_2$ in gravity, which reads:
\begin{equation}
    i \mathcal{M}_0 = 
    \raisebox{-1.5em}{
\begin{tikzpicture}[line width=1.1, scale=0.7,
    electron/.style={postaction={decorate}, decoration={
        markings,
        mark=at position #1 with {\arrow{latex}}
    }},
    electron/.default=0.5,  
    positron/.style={postaction={decorate}, decoration={
        markings,
        mark=at position #1 with {\arrow{latex}}
    }},
    positron/.default=0.5  
]
    \coordinate (left1) at (-1,-1);
    \coordinate (left2) at (-1, 1);
    \coordinate (right1) at (1, -1);
    \coordinate (right2) at (1,1);
    \coordinate (up1) at (0,1);
    \coordinate (low1) at (0,-1);

    \draw[decorate, decoration=snake] (up1) -- (low1);
    \draw[decorate, decoration=snake] (0.15,1) -- (0.15,-1);
    
    \draw[RoyalBlue] (left1) -- (right1);
    \draw[Maroon] (left2) -- (right2);

\end{tikzpicture}
}
= 16 \pi G { (2(p_1\cdot p_2)^2-m_1^2 m_2^2)\over t} \equiv 16 \pi G{ m_1^2 m_2^2 (2\sigma^2-1)\over t}\,,
\end{equation}
in the small $t$ limit. We define $\sigma \equiv \frac{p_1 \cdot p_2}{m_1 m_2}$.
The eikonal phase at the leading order in $G$ is then
\begin{equation}
    \delta(s,b) = - \alpha_G(s) \log(b/b_{IR})\,,
\end{equation}
with 
\begin{equation}
    \alpha_G(s) =  G  {2(p_1 \cdot p_2)^2 - m_1^2 m_2^2 \over \sqrt{s} |\mathbf{p}|} ={ G m_1 m_2 (2\sigma^2-1) \over\sqrt{\sigma^2-1} } \, .
\end{equation}
In the non-relativistic limit (say one of the particles is very heavy, $m_1 = M \gg m = m_2$) we have $p_1 \cdot p_2 = M m$ and $\sqrt{s} = M$. This gives us
\begin{equation}
    \alpha_G = {2 G M m^2 \over |\mathbf{p}|} ={2 G M m \over \sqrt{2  E/m}}\, ,
\end{equation}
where $E \equiv {|\mathbf{p}|^2 \over 2 m}$ coming from $E \equiv \sqrt{s} - M - m$. Notice that structurally, this is the same effective coupling we got before in \eqref{eq:alpha-e} for QED.

Let us now consider a classical test of GR: bending of light. When one of the particles is massless, we have $\sqrt{s} = M + \omega$ and $p_1 \cdot p_2 = M \omega$. This yields
\begin{equation}
    \alpha_G = 2 G M \omega\, .
\end{equation}
The deflection angle is given by $\theta = 2 {\partial \delta(s,b) \over \partial J}$, where $b = J/\omega$. This gives us
\begin{equation}
    \theta = {4 G M \omega  \over J} =  {4 G M \over b}\, ,
\end{equation}
which is the classic deflection angle formula in GR.
Notice that the deflection angle does not depend on the frequency $\omega$, just on the impact parameter $b$.
\begin{QA}
Is the fact that $\omega$ cancels out a consequence of the equivalence principle?\\

\noindent\emph{Answer:} Yes, exactly. Higher orders including quantum corrections will give dependence on the frequency.
\end{QA}

We can also consider the Shapiro time delay.
Recall that the time delay is given by $\Delta t = 2 {\partial \delta(s,b) \over \partial \sqrt{s}}$. In the case of a null geodesic, for which $\sqrt{s} = \omega + M$, we get
\begin{equation}
    \Delta t = -4 G M \log(b / b_{IR}) > 0\, ,
\end{equation}
which again is the correct answer in GR.
As expected, gravity slows you down.

Let us now do something slightly more interesting and ask what would happen as we go to higher orders in the expansion. The amplitude admits an expansion with more and more loops:
\be
i \mathcal{M} = \raisebox{-1.5em}{
\begin{tikzpicture}[line width=1.1, scale=0.7,
    electron/.style={postaction={decorate}, decoration={
        markings,
        mark=at position #1 with {\arrow{latex}}
    }},
    electron/.default=0.5,  
    positron/.style={postaction={decorate}, decoration={
        markings,
        mark=at position #1 with {\arrow{latex}}
    }},
    positron/.default=0.5  
]
    \coordinate (left1) at (-1,-1);
    \coordinate (left2) at (-1, 1);
    \coordinate (right1) at (1, -1);
    \coordinate (right2) at (1,1);
    \coordinate (up1) at (0,1);
    \coordinate (low1) at (0,-1);

    \draw[decorate, decoration=snake] (up1) -- (low1);
    \draw[decorate, decoration=snake] (0.15,1) -- (0.15,-1);
    
    \draw[RoyalBlue] (left1) -- (right1);
    \draw[Maroon] (left2) -- (right2);

\end{tikzpicture}
}
+
\raisebox{-1.5em}{
\begin{tikzpicture}[line width=1.1, scale=0.7,
    electron/.style={postaction={decorate}, decoration={
        markings,
        mark=at position #1 with {\arrow{latex}}
    }},
    electron/.default=0.5,  
    positron/.style={postaction={decorate}, decoration={
        markings,
        mark=at position #1 with {\arrow{latex}}
    }},
    positron/.default=0.5  
]
    \coordinate (left1) at (-1,-1);
    \coordinate (left2) at (-1, 1);
    \coordinate (right1) at (2, -1);
    \coordinate (right2) at (2,1);
    \coordinate (up1) at (0,1);
    \coordinate (low1) at (0,-1);
    \coordinate (up2) at (1,1);
    \coordinate (low2) at (1,-1);

    \draw[decorate, decoration=snake] (up1) -- (low1);
    \draw[decorate, decoration=snake] (0.15,1) -- (0.15,-1);
    \draw[decorate, decoration=snake] (up2) -- (low2);
    \draw[decorate, decoration=snake] (1.15,1) -- (1.15,-1);

    \draw[RoyalBlue] (left1) -- (right1);
    \draw[Maroon] (left2) -- (right2);
\end{tikzpicture}
}
+
\raisebox{-1.5em}{
\begin{tikzpicture}[line width=1.1, scale=0.7,
    electron/.style={postaction={decorate}, decoration={
        markings,
        mark=at position #1 with {\arrow{latex}}
    }},
    electron/.default=0.5,  
    positron/.style={postaction={decorate}, decoration={
        markings,
        mark=at position #1 with {\arrow{latex}}
    }},
    positron/.default=0.5  
]
    \coordinate (left1) at (-1,-1);
    \coordinate (left2) at (-1, 1);
    \coordinate (right1) at (2, -1);
    \coordinate (right2) at (2,1);
    \coordinate (up1) at (0,1);
    \coordinate (low1) at (0,-1);
    \coordinate (up2) at (1,1);
    \coordinate (low2) at (1,-1);

    \draw[decorate, decoration=snake] (up1) -- (low2);
    \draw[decorate, decoration=snake] (0.15,1) -- (1.15,-1);
    \filldraw[white] (0.5,0) circle (10pt);
    \draw[decorate, decoration=snake] (up2) -- (low1);
    \draw[decorate, decoration=snake] (1.15,1) -- (0.15,-1);

    \draw[RoyalBlue] (left1) -- (right1);
    \draw[Maroon] (left2) -- (right2);
\end{tikzpicture}
}
+
\raisebox{-1.5em}{
\begin{tikzpicture}[line width=1.1, scale=0.7,
    electron/.style={postaction={decorate}, decoration={
        markings,
        mark=at position #1 with {\arrow{latex}}
    }},
    electron/.default=0.5,  
    positron/.style={postaction={decorate}, decoration={
        markings,
        mark=at position #1 with {\arrow{latex}}
    }},
    positron/.default=0.5  
]
    \coordinate (left1) at (-1,-1);
    \coordinate (left2) at (-1, 1);
    \coordinate (right1) at (2, -1);
    \coordinate (right2) at (2,1);
    \coordinate (up1) at (0,1);
    \coordinate (low1) at (0,-1);
    \coordinate (up2) at (1,1);
    \coordinate (low2) at (1,-1);

    \draw[decorate, decoration=snake] (0.5,1) -- (0.5,0);
    \draw[decorate, decoration=snake] (0.65,1) -- (0.65,0);
    
    \draw[decorate, decoration=snake] (0.5,0) -- (low2);
    \draw[decorate, decoration=snake] (0.65,0) -- (1.15,-1);

    \draw[decorate, decoration=snake] (0.5,0) -- (low1);
    \draw[decorate, decoration=snake] (0.65,0) -- (0.15,-1);

    \draw[RoyalBlue] (left1) -- (right1);
    \draw[Maroon] (left2) -- (right2);
\end{tikzpicture}
}
+
\raisebox{-1.5em}{
\begin{tikzpicture}[line width=1.1, scale=0.7,
    electron/.style={postaction={decorate}, decoration={
        markings,
        mark=at position #1 with {\arrow{latex}}
    }},
    electron/.default=0.5,  
    positron/.style={postaction={decorate}, decoration={
        markings,
        mark=at position #1 with {\arrow{latex}}
    }},
    positron/.default=0.5  
]
    \coordinate (left1) at (-1,-1);
    \coordinate (left2) at (-1, 1);
    \coordinate (right1) at (2, -1);
    \coordinate (right2) at (2,1);
    \coordinate (up1) at (0,1);
    \coordinate (low1) at (0,-1);
    \coordinate (up2) at (1,1);
    \coordinate (low2) at (1,-1);

    \draw[decorate, decoration=snake] (0.5,-1) -- (0.5,0);
    \draw[decorate, decoration=snake] (0.65,-1) -- (0.65,0);
    
    \draw[decorate, decoration=snake] (0.5,0) -- (up2);
    \draw[decorate, decoration=snake] (0.65,0) -- (1.15,1);

    \draw[decorate, decoration=snake] (0.5,0) -- (up1);
    \draw[decorate, decoration=snake] (0.65,0) -- (0.15,1);

    \draw[RoyalBlue] (left1) -- (right1);
    \draw[Maroon] (left2) -- (right2);
\end{tikzpicture}
}
+ \ldots
\ee
In the hydrogen atom, the tree-level diagram contains the Coulomb potential and the ladder diagrams are iterations of it. However, in GR we have mixing between different terms. To distinguish which terms contribute to the `potential' and which terms are iterations, we will use the connection to an ``effective-one-body" Lippmann--Schwinger equation, which resums all such contributions. We will follow the exposition of Todorov \cite{Todorov:1970gr}, who did his work in 1970 while at IAS (see also \cite{Correia:2024jgr} for a modern review).

\subsection{Relativistic Born series}

In the center of mass frame we have the ingoing momenta
\begin{align}
p_1^\mu &= (E_1, \mathbf{p})\, , & E_1 = \sqrt{|\mathbf{p}|^2 + m_1^2}\, , \\
p_2^\mu &= (E_2, -\mathbf{p})\, , & E_2 = \sqrt{|\mathbf{p}|^2 + m_2^2}\, , 
\end{align}
with the outgoing momenta given by
\begin{align}
    p_3^\mu &= (E_1, \mathbf{p}') = (E_1, \mathbf{p}+ \mathbf{q})\, ,  \\
p_4^\mu &= (E_2, -\mathbf{p}') =  (E_2, -\mathbf{p} - \mathbf{q})\, .
\end{align}
Here, $\mathbf{q} = \mathbf{p}' - \mathbf{p}$ is the momentum exchange. Let $T(s,t)$ be the scattering amplitude for the scattering of two scalars of mass $m_1$ and $m_2$ where the Mandelstam invariants read
\begin{equation}
    s = (p_1 + p_2)^2, \qquad t = (p_1 - p_3)^2\, .
\end{equation}
These relate to $\mathbf{p}$ and $\mathbf{q}$ in the center of mass frame via
\begin{equation}
\label{eq:st}
    |\mathbf{p}|^2 = {\big[s - (m_1+m_2)^2\big]\big[s - (m_1-m_2)^2\big] \over 4 s}, \qquad  t =  -|\mathbf{p}' - \mathbf{p}|^2 = -|\mathbf{q}|^2 .
\end{equation}
The vectors $\mathbf{p}$ and $\mathbf{q}$ are further constrained by the condition $\mathbf{p} \cdot \mathbf{q} = - |\mathbf{q}|^2 / 2$ that stems from energy conservation $|\mathbf{p}| = |\mathbf{p}'|$. We will use the notation $T(\mathbf{p}, \mathbf{p}')$ where it is understood that the dependence on $s$ and $t$ is given in terms of $\mathbf{p}$ and $\mathbf{p}'$ in Eq.~\eqref{eq:st}.
\par
We define the \emph{potential} $V(\mathbf{p}, \mathbf{q})$ in terms of the Lippmann--Schwinger equation:
\begin{equation}\label{eq:LS}
    T(\mathbf{p},\mathbf{p}^\prime)=V(\mathbf{p},\mathbf{p}^\prime)+\int \d^3\mathbf{k}\,T(\mathbf{p},\mathbf{k})G(\mathbf{p},\mathbf{k})V(\mathbf{k},\mathbf{p}^\prime)\, ,
\end{equation}
where $G(\mathbf{p},\mathbf{k})$ is a Green's function. It is not unique, but it will have to satisfy some consistency conditions and we will fix it shortly. This equation is easier to understand at the level of pictures:
\be
\includegraphics[scale=0.23,valign=c]{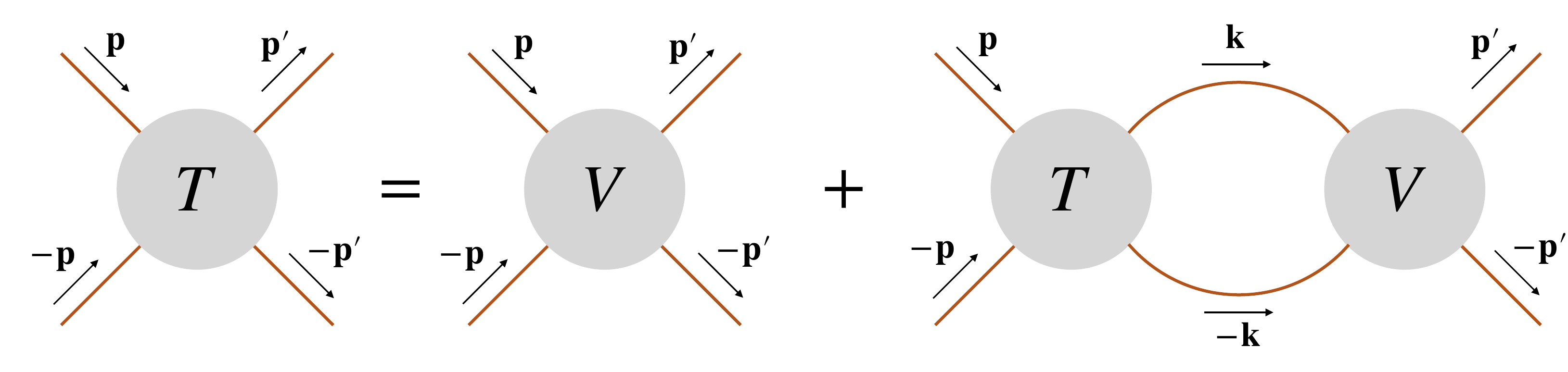}
\ee
Note that in the last term, the momentum $\mathbf{k}$ is off-shell and we integrate over it.

The fact we chose this to be a 3D integral makes it connect directly to the standard Lippmann--Schwinger equation of one-body quantum mechanics. In fact, in the eikonal or WKB limit one finds a relation to the effective one-body formalism by Buonanno and Damour \cite{Buonanno:1998gg, Fiziev:2000mh}.\footnote{Miguel was not sure about the spelling of Alessandra Buonanno's family name. The Italians in the audience advised that it should contain three instances of the letter n.} 

We now require the potential $V(\mathbf{p},\mathbf{p}') = V(s,t)$ to be a real function in the scattering regime $s \geq (m_1 + m_2)^2$, i.e.,
\begin{equation}
\mathrm{Im} V(\mathbf{p},\mathbf{p'}) = 0 \qquad \text{ for } \qquad s \geq (m_1 + m_2)^2\, .
\end{equation}
We will now also require consistency of the Lippmann--Schwinger equation with elastic unitarity in order to fix the form of $G(\mathbf{p},\mathbf{k})$. Elastic unitarity, which is the optical theorem applied below the three-particle threshold, reads
\begin{align}
\mathrm{Im} \, T(\mathbf{p},\mathbf{p}')&= {1 \over 8 \pi^2} \int {\d^4 k} \; \delta^+\!\big(k^2 - m_1^2\big) \delta^+\!\big((k-p_1 - p_2)^2 - m_2^2 \big) T(p,k) \, T^*(k,p') \notag \\
&= {1 \over 16 \pi^2 \sqrt{s}} \int {\d^3 \mathbf{k} } \, \delta(|\mathbf{k}|^2 - |\mathbf{p}|^2) \, T(\mathbf{p},\mathbf{k}) \, T^*(\mathbf{k},\mathbf{p}')\, ,
\end{align}
where $k$ is the 4-momentum and $\delta^+$ denote putting the corresponding particle on-shell and imposing the positive-energy condition. In the second line, we recast it as a 3D integral in order to resemble the Lippmann--Schwinger equation, by getting rid of one of the delta functions.
Since we imposed that $V(\mathbf{p}, \mathbf{q})$ is real across the 2-particle cut, the discontinuity of the amplitude should be captured by the Green's function $G(\mathbf{p}, \mathbf{k})$. 

To arrive at the constraint on $G$ as quickly as possible, it will pay off to work with a short-hand notation. Schematically, the relativistic Born series can be written as
\begin{equation}\label{eq:T-Born}
    T = V + V G V + V G V G V + \dots = V {1 \over 1 - G V}\, ,
\end{equation}
where multiplication denotes integrating over the intermediate momentum, as above, and we suppress all the constants and kinematic dependence.
Imposing the reality condition, $V^* = V$, gives
\begin{equation}\label{eq:T-star}
    T^* = V {1 \over 1 - G^* V}\, .
\end{equation}
In the same notation, Elastic unitarity takes the form
\begin{equation}
    T - T^* = T T^* \, .
\end{equation}
Using \eqref{eq:T-Born} and \eqref{eq:T-star}, the right-hand side can be written as
\begin{equation}
    T T^* = V {1 \over 1 - G V} V {1 \over 1 - G^* V} \, ,
\end{equation}
whereas the left-hand side is
\begin{equation}
    T - T^* = V \left[ {1 \over 1 - G V} - {1 \over 1 - G^* V} \right] = V {1 \over 1 - G V} (G - G^*) V {1 \over 1 - G^* V} \, .
\end{equation}
Unitarity then fixes $G - G^* = 1$, i.e., the imaginary part of the Green's function is a delta function. After carefully working out all the factors, in terms of equations this constraint reads\begin{equation}
    \mathrm{Im} \, G(\mathbf{p}, \mathbf{k}) ={\delta(|\mathbf{k}|^2 - |\mathbf{p}|^2) \over 16 \pi^2 \sqrt{s}}  \, .
    \label{eq:Im_G}
\end{equation}
The simplest choice is to write the solution as
\begin{equation}
    G(\mathbf{p},\mathbf{k}) = {1 \over (2 \pi)^3 } {1 \over 2 \sqrt{s}}  {1 \over |\mathbf{k}|^2 - |\mathbf{p}|^2 - i \epsilon}\, ,
    \label{eq:G}
\end{equation}
This solution is written up to analytic (real) terms which can be absorbed into the potential.

\begin{QA}
So what exactly is the freedom in choosing $G$?\\

\noindent\emph{Answer:} You can add any analytic function to the above solution, as long as it is compatible with the constraint \eqref{eq:Im_G}. Different choices will lead to other effective potentials $V$ differing by off-shell pieces. In coordinate space these choices correspond to different coordinate systems \cite{Correia:2024jgr}. We observe that the above choice selects isotropic coordinates (see below).
\end{QA}

Let us take the non-relativistic probe limit as a cross-check. This amounts to setting:
\begin{equation}
    \sqrt{s} = M + m + E, \qquad \mathbf{|p|}^2 = 2 \mu E = {2m M \over m + M} E\, ,
\end{equation}
where we made use of Eq.~\eqref{eq:st} and $\mu = \frac{m_1 m_2}{m_1 + m_2}$ is the reduced mass of the system. In this limit, we get
\be
G \sim \frac{1}{\frac{|\mathbf{k}|^2}{2\mu} - E - i\eps}\, .
\ee
This is precisely the non-relativistic propagator from the Schr\"odinger equation, as expected.

Let us examine this Green's function closer. The $|\mathbf{k}|^2$ dependence is the same in the relativistic and non-relativistic cases. It means that after a Fourier transform, we still get quadratic derivatives in the spatial directions. The difference is the dependence on $E$, which is linear in the non-relativistic case. This is in contrast to the relativistic case, in which we had
\be
|\vec{p}|^2 = \frac{[s - (m_1 + m_2)^2][s - (m_1 - m_2)^2]}{4s}\,,
\label{eq:psquare}
\ee
and the energy dependence is very non-linear. To summarize, we end up with an effective Schr\"odinger equation with
\be
|\vec{k}|^2 \to -\nabla^2\, , \qquad
|\vec{p}|^2 = f(E)\,,
\ee
for $f(E)$ given by eq. \eqref{eq:psquare} where $s = E^2$. This is called the relativistic effective one-body Schr\"odinger equation:
\be
f(E) \Psi = (\nabla^2 + V ) \Psi\, .
\ee
Similar manipulations can of course also be done for the hydrogen atom. By drawing more and more complicated diagrams, one can compute things like the QED corrections to the spectrum, vacuum polarization, anomalous magnetic moment of the electron, positronium lifetime, etc \cite{Todorov:1970gr}.

The summary so far is as follows. We first wrote down the Lippmann--Schwinger equation. Then we found that the only way it can be consistent with elastic unitarity with $V$ being real is that $G$ has the form given above. After this is done, we can use a perturbative expansion in terms of Feynman diagrams to fix the potential $V$ and compute classical observables.

\subsection{Potential at \texorpdfstring{$\mathcal{O}(G^2)$}{OG2}, perihelion precession, and higher orders}

We may now go ahead and compute loop corrections to the potential by matching the Feynman diagram expansion with the Born series of the Lippmann--Schwinger equation in Eq.~\eqref{eq:LS}. 

For example, at one-loop we have:
\be
\raisebox{-1.5em}{
\begin{tikzpicture}[line width=1.1, scale=0.7,
    electron/.style={postaction={decorate}, decoration={
        markings,
        mark=at position #1 with {\arrow{latex}}
    }},
    electron/.default=0.5,  
    positron/.style={postaction={decorate}, decoration={
        markings,
        mark=at position #1 with {\arrow{latex}}
    }},
    positron/.default=0.5  
]
    \coordinate (left1) at (-1,-1);
    \coordinate (left2) at (-1, 1);
    \coordinate (right1) at (2, -1);
    \coordinate (right2) at (2,1);
    \coordinate (up1) at (0,1);
    \coordinate (low1) at (0,-1);
    \coordinate (up2) at (1,1);
    \coordinate (low2) at (1,-1);

    \draw[decorate, decoration=snake] (up1) -- (low1);
    \draw[decorate, decoration=snake] (0.15,1) -- (0.15,-1);
    \draw[decorate, decoration=snake] (up2) -- (low2);
    \draw[decorate, decoration=snake] (1.15,1) -- (1.15,-1);

    \draw[RoyalBlue] (left1) -- (right1);
    \draw[Maroon] (left2) -- (right2);
\end{tikzpicture}
}
\qquad
\raisebox{-1.5em}{
\begin{tikzpicture}[line width=1.1, scale=0.7,
    electron/.style={postaction={decorate}, decoration={
        markings,
        mark=at position #1 with {\arrow{latex}}
    }},
    electron/.default=0.5,  
    positron/.style={postaction={decorate}, decoration={
        markings,
        mark=at position #1 with {\arrow{latex}}
    }},
    positron/.default=0.5  
]
    \coordinate (left1) at (-1,-1);
    \coordinate (left2) at (-1, 1);
    \coordinate (right1) at (2, -1);
    \coordinate (right2) at (2,1);
    \coordinate (up1) at (0,1);
    \coordinate (low1) at (0,-1);
    \coordinate (up2) at (1,1);
    \coordinate (low2) at (1,-1);

    \draw[decorate, decoration=snake] (up1) -- (low2);
    \draw[decorate, decoration=snake] (0.15,1) -- (1.15,-1);
    \filldraw[white] (0.5,0) circle (10pt);
    \draw[decorate, decoration=snake] (up2) -- (low1);
    \draw[decorate, decoration=snake] (1.15,1) -- (0.15,-1);

    \draw[RoyalBlue] (left1) -- (right1);
    \draw[Maroon] (left2) -- (right2);
\end{tikzpicture}
}
\qquad
\raisebox{-1.5em}{
\begin{tikzpicture}[line width=1.1, scale=0.7,
    electron/.style={postaction={decorate}, decoration={
        markings,
        mark=at position #1 with {\arrow{latex}}
    }},
    electron/.default=0.5,  
    positron/.style={postaction={decorate}, decoration={
        markings,
        mark=at position #1 with {\arrow{latex}}
    }},
    positron/.default=0.5  
]
    \coordinate (left1) at (-1,-1);
    \coordinate (left2) at (-1, 1);
    \coordinate (right1) at (2, -1);
    \coordinate (right2) at (2,1);
    \coordinate (up1) at (0,1);
    \coordinate (low1) at (0,-1);
    \coordinate (up2) at (1,1);
    \coordinate (low2) at (1,-1);

    \draw[decorate, decoration=snake] (0.5,1) -- (0.5,0);
    \draw[decorate, decoration=snake] (0.65,1) -- (0.65,0);
    
    \draw[decorate, decoration=snake] (0.5,0) -- (low2);
    \draw[decorate, decoration=snake] (0.65,0) -- (1.15,-1);

    \draw[decorate, decoration=snake] (0.5,0) -- (low1);
    \draw[decorate, decoration=snake] (0.65,0) -- (0.15,-1);

    \draw[RoyalBlue] (left1) -- (right1);
    \draw[Maroon] (left2) -- (right2);
\end{tikzpicture}
}
\qquad
\raisebox{-1.5em}{
\begin{tikzpicture}[line width=1.1, scale=0.7,
    electron/.style={postaction={decorate}, decoration={
        markings,
        mark=at position #1 with {\arrow{latex}}
    }},
    electron/.default=0.5,  
    positron/.style={postaction={decorate}, decoration={
        markings,
        mark=at position #1 with {\arrow{latex}}
    }},
    positron/.default=0.5  
]
    \coordinate (left1) at (-1,-1);
    \coordinate (left2) at (-1, 1);
    \coordinate (right1) at (2, -1);
    \coordinate (right2) at (2,1);
    \coordinate (up1) at (0,1);
    \coordinate (low1) at (0,-1);
    \coordinate (up2) at (1,1);
    \coordinate (low2) at (1,-1);

    \draw[decorate, decoration=snake] (0.5,-1) -- (0.5,0);
    \draw[decorate, decoration=snake] (0.65,-1) -- (0.65,0);
    
    \draw[decorate, decoration=snake] (0.5,0) -- (up2);
    \draw[decorate, decoration=snake] (0.65,0) -- (1.15,1);

    \draw[decorate, decoration=snake] (0.5,0) -- (up1);
    \draw[decorate, decoration=snake] (0.65,0) -- (0.15,1);

    \draw[RoyalBlue] (left1) -- (right1);
    \draw[Maroon] (left2) -- (right2);
\end{tikzpicture}
}
\ee
In this case, the cross-box (second diagram) contributes at leading order in the classical, or small $q$, limit and cancels part of the box (first diagram). This is a crucial difference between defining the potential via the Lippmann--Schwinger eq. \eqref{eq:LS}. In the latter case only the box contributes as an ``iteration" while the crossed-box gets absorbed in the potential, which makes the potential ill-defined.\footnote{%
In his work \cite{Todorov:1970gr}, Todorov writes: ``I wish to thank Professor F. Dyson for an
enlightening discussion prior to this work, and especially for his refusal to be satisfied with any two-particle equation which does not lead [in the probe limit $m_1 \ll m_2$] to the Klein--Gordon (or Dirac) equation in an external field.''}

Performing the full computations at one-loop order gives use the $\mathcal{O}(G^2)$ terms: 
\begin{equation}
\hspace{-0.3cm}
    \mathcal{M}_2 = {6 \pi^2 G^2(m_1{+}m_2)(5(p_1{\cdot}p_2)^2{-}m_1^2 m_2^2)  \over|\mathbf{q}|} +  \int \d^3\mathbf{k} \, \mathcal{M}_1(\mathbf{p}, \mathbf{k})  G(\mathbf{p}, \mathbf{k}) \mathcal{M}_1(\mathbf{k}, \mathbf{p}')\, 
\end{equation}
where
\begin{equation}
    \mathcal{M}_1(\mathbf{p}, \mathbf{k}) = -{8 \pi G \big(m_1^2 m_2^2-2(p_1\cdot p_2)^2\big) \over |\mathbf{q}|^2}\,,
\end{equation}
is the tree-level amplitude.
\par
Matching it to the Born series leads to the effective potential
\begin{equation}
\begin{split}
        V(\mathbf{p},\mathbf{q}) = &-{8 \pi G \big(m_1^2 m_2^2-2(p_1\cdot p_2)^2\big) \over |\mathbf{q}|^2} \\&- {6 \pi^2 G^2(m_1+m_2)(m_1^2 m_2^2-5(p_1\cdot p_2)^2)  \over|\mathbf{q}|} + \mathcal{O}(G^3)\, .
\end{split}
\end{equation}
The Fourier transform to position space gives
\begin{equation}
V(r) = {G \over r} {m_1^2 m_2^2-2(p_1\cdot p_2)^2 \over  \sqrt{s}} - {G^2 \over r^2}  {3  m_1^2 m_2^2 (m_1 + m_2) \over 2 \sqrt{s}} (1 - 5 \sigma^2) + \mathcal{O}(G^3)\, ,
\end{equation}
where we remind that
\be
\sigma \equiv \frac{p_1 \cdot p_2}{m_1 m_2}\, .
\ee
We can then compute $\mathcal{O}(G^2)$ corrections to the deflection angle. In fact, the state of the art is currently $\mathcal{O}(G^4)$, or equivalently 3 loops. 
For the purposes of this section, we will be satisfied with $\mathcal{O}(G^2)$. 

Let us compute one more classical test of GR: the precession of the perihelion of Mercury. It is reproduced by plugging the above potential into the radial action \eqref{eq:IJ} and taking its $J$ derivative:
\begin{equation}
    \Delta \Phi = \pi + {\partial I_J(E) \over \partial J} = G^2 {3 \pi m_1^2 m_2^2 (m_1 + m_2) \over 2 J^2 \sqrt{s}} (5 \sigma^2 - 1)\, .
\end{equation}
In the non-relativistic limit we have $\sigma \to 1$ and $\sqrt{s} \to m_1 + m_2$. If in addition, we use the probe limit $m = m_2 \ll m_1 = M$ and make use of the identity $J^2 = m^2 G M (1 - e^2) a$, we find the famous result of the Mercury perihelion shift 
\begin{equation}
    \Delta \Phi_\text{Mercury} = {6 \pi G M \over (1 - e^2) a }\, .
\end{equation}

Now, let us say some words about higher orders in this perturbative expansion in $G$. What is exciting about this expansion is that there is some new physical effect at every order. At $\mathcal{O}(G^2)$ we have the perihelion precession just discussed. At order $\mathcal{O}(G^3)$ there are radiation effects. For radiative effects one can use the so-called Kosower--Maybee--O'Connell (KMOC) formalism \cite{Kosower:2018adc}. The big open problem in this area is how to analytically continue the result of the scattering problem to the bound-state problem. The problem occurs at $\mathcal{O}(G^4)$, where you get diagrams of this form:
\be
\raisebox{-1.3em}{
\begin{tikzpicture}[line width=1.1, scale=0.7,
    electron/.style={postaction={decorate}, decoration={
        markings,
        mark=at position #1 with {\arrow{latex}}
    }},
    electron/.default=0.5,  
    positron/.style={postaction={decorate}, decoration={
        markings,
        mark=at position #1 with {\arrow{latex}}
    }},
    positron/.default=0.5  
]
    \coordinate (left1) at (-1,-1);
    \coordinate (left2) at (-1, 1);
    \coordinate (right1) at (3, -1);
    \coordinate (right2) at (3,1);
    \coordinate (up1) at (0,1);
    \coordinate (low1) at (0,-1);
    \coordinate (up2) at (2,1);
    \coordinate (low2) at (2,-1);

    \draw[decorate, decoration=snake] (up1) -- (low1);
    \draw[decorate, decoration=snake] (0.15,1) -- (0.15,-1);
    
    \draw[decorate, decoration=snake] (up2) -- (low2);
    \draw[decorate, decoration=snake] (2.15,1) -- (2.15,-1);

    \draw[decorate, decoration=snake] (0.15,0) -- (2,0);
    \draw[decorate, decoration=snake] (0.15,-0.15) -- (2,-0.15);

    \draw[decorate, decoration=snake] (1,0) -- (1,1);
    \draw[decorate, decoration=snake] (1.15,0) -- (1.15,1);

    \draw[RoyalBlue] (left1) -- (right1);
    \draw[Maroon] (left2) -- (right2);
\end{tikzpicture}
}
\qquad
\raisebox{-1.3em}{
\begin{tikzpicture}[line width=1.1, scale=0.7,
    electron/.style={postaction={decorate}, decoration={
        markings,
        mark=at position #1 with {\arrow{latex}}
    }},
    electron/.default=0.5,  
    positron/.style={postaction={decorate}, decoration={
        markings,
        mark=at position #1 with {\arrow{latex}}
    }},
    positron/.default=0.5  
]
    \coordinate (left1) at (-1,-1);
    \coordinate (left2) at (-1, 1);
    \coordinate (right1) at (3, -1);
    \coordinate (right2) at (3,1);
    \coordinate (up1) at (0,1);
    \coordinate (low1) at (0,-1);
    \coordinate (up2) at (2,1);
    \coordinate (low2) at (2,-1);

    \draw[decorate, decoration=snake] (up1) -- (low1);
    \draw[decorate, decoration=snake] (0.15,1) -- (0.15,-1);
    
    \draw[decorate, decoration=snake] (up2) -- (low2);
    \draw[decorate, decoration=snake] (2.15,1) -- (2.15,-1);

    \draw[decorate, decoration=snake] (-0.5,1) -- (1,2);
    \draw[decorate, decoration=snake] (-0.6,1) -- (1,2.1);

    \draw[decorate, decoration=snake] (2.5,1) -- (1,2);
    \draw[decorate, decoration=snake] (2.6,1) -- (1,2.1);

    \draw[decorate, decoration=snake] (1,2) -- (1,1);
    \draw[decorate, decoration=snake] (1.15,2) -- (1.15,1);

    \draw[RoyalBlue] (left1) -- (right1);
    \draw[Maroon] (left2) -- (right2);
\end{tikzpicture}
}
\label{eq:taileffects}
\ee
These are responsible for the \emph{tail effects}. The radiated graviton feels the attraction of the two-body system it was emitted from (this does not happen in QED due to lack of photon self-interactions). When trying to compute the effective potential with these diagrams, we get terms of the form $G^4 \rho(E)$, where $\rho(E)$ is a distribution, not a function, of the energy. It does not admit a good analytic continuation. The way to deal with it is a big open problem.

\subsection{Wave regime and Regge--Wheeler equation}

Let us now consider the wave regime mentioned in Sec.~\ref{sec:Isabella}, as presented in the bottom right corner of equation \eqref{eq:regimes}, which is the relevant regime for the ringdown. In this case, we have one massless particle with frequency $\omega$ scattering off a massive particle with mass $M$. Hence, the energy is $\sqrt{s} = M + \omega$ with $\omega \ll M$. 
We are in the regime in which the impact parameter $b \sim \lambda \sim {1 \over \omega} \gg \lambda_{\text{C}}$, where $\lambda_{\text{C}}$ is the Compton wavelength.

In this case, we can compute the potential once again, but this time the triangle diagram gives an extra contribution. After the dust settles, we get:
\begin{equation}\label{eq:2PMVmom}
    V(\mathbf{p}, \mathbf{q}) = \frac{16\pi G M^2 \omega^2}{|\mathbf{q}|^2}+\frac{15\pi^2 G^2M^3\omega^2}{|\mathbf{q}|}+\frac{G^2M^2\pi^2}{2}\frac{\mathbf{p} \cdot \mathbf{q}}{|\mathbf{q}|} + \ldots
\end{equation}
The first term is the tree-level contribution. The second term is precisely what comes from the triangle diagram. The third and later terms are subdominant in the point-particle limit.

What does this potential give after the Fourier transform? The answer is
\begin{equation}
  V(r)=\left(\frac{4 MG \omega^2}{r}+\frac{15G^2 M^2 \omega^2 }{2 r^2} \right) + \frac{G^2 M^2}{2 r^3}\hbar^2 \partial_r + \ldots
\end{equation}
Only the first two terms are relevant.
The corresponding wave operator is 
\begin{equation}
    |\mathbf{p}|^2 - |\mathbf{k}|^2 \to \omega^2 - \nabla^2\,,
\end{equation}
and the associated Schr\"odinger equation reads
\begin{equation}\label{eq:WEmassless}
    (\omega^2 - \nabla^2) \phi = V \phi\, .
\end{equation}
What is this equation? It turns out to be the
Regge--Wheeler equation in isotropic coordinates. Let us confirm this.

We are going to start with the action for a scalar field on a gravitational background, which reads
\begin{equation}
S = \int \d^4 x \sqrt{-g} \, g^{\mu \nu} \partial_\mu \phi \,\partial_\nu \phi\, .
\end{equation}
The wave equation follows from
\begin{equation}
{\delta S \over \delta \phi} = 0 \qquad \implies \qquad \partial_\mu(\sqrt{-g} \, g^{\mu \nu} \partial_\nu \phi) = 0.
\end{equation}
The Schwarzschild metric, in isotropic coordinates, is written as
\begin{equation}
    g_{\mu \nu} = \mathrm{diag}\big(A(r), - B(r), - r^2 B(r), - r^2 \sin^2 \theta  B(r) \big)\,,
\end{equation}
or equivalently
\begin{equation}
    \d s^2 = A(r) \d t^2 - B(r)\big[ \d r^2 + r^2 \d\theta^2 + r^2 \sin^2 \theta \,\d\varphi^2 \big]\,,
\end{equation}
with
\begin{equation}\label{eq:ABdef}
    A(r) = {\big( 1 - {G M \over 2 r}\big)^2 \over \big( 1 + {G M \over 2 r}\big)^2 }, \qquad B(r) = \Big( 1 + {G M \over 2 r}\Big)^4.
\end{equation}
Noting that $\sqrt{-g} =  \sqrt{A B^3} \, r^2 \sin \theta$, the wave equation on this background is given by
\begin{equation}
    {1 \over A} \partial_t^2 \phi - {1 \over B} \nabla^2 \phi - {1 \over \sqrt{AB^3}} \partial_r \left( {\sqrt{AB}} \right)\partial_r \phi  = 0\, ,
\end{equation}
where $\nabla^2 = \eta^{ i j} \partial_i \partial_j$ is the Euclidean Laplacian. Rearranging terms we obtain
\begin{equation}
      \nabla^2 \phi = \left({B \over A} \partial_t^2 - {\partial_r (AB) \over 2 A B} \partial_r  \right)\phi\, .
\end{equation}
Next, we write it in terms of the operators $\hat{E} = - i \hbar \partial_t$ and $\hat{\mathbf{p}} = - i \hbar \vecc{\partial}$, and replace $A$ and $B$ from \eqref{eq:ABdef}. Lastly, by matching to the wave equation \eqref{eq:WEmassless}, we can extract the potential at all orders in $G$
\begin{equation}\label{eq:allordV}
 V =  \left(\frac{\left(1+\frac{ G M}{2r}\right)^6}{\left(1-\frac{ G M}{2r}\right)^2} -1\right)\partial_t^2  + \frac{2G^2M^2}{4r^3-G^2M^2r}\hbar^2\partial_r\, .
\end{equation}
Expanding the potential up to $\mathcal{O}(G^2)$ we obtain
\begin{equation}\label{eq:V2PM}
  V =  \left(\frac{4 MG}{r}+\frac{15G^2 M^2 }{2 r^2} \right) \omega^2 + \frac{G^2 M^2}{2 r^3}\hbar^2 \partial_r  + \mathcal{O}(G^3)\, .
\end{equation}
The Fourier Transform of this expression precisely matches \eqref{eq:2PMVmom}.

Notice that the first bracket dominates at high frequencies, which corresponds to the Eikonal regime described earlier. We can indeed recognise the corrections up to $\mathcal{O}(G^2)$ to the light bending problem that enter in the relevant radial action and allow one to extract the classical scattering angle $\theta=\frac{2GM}{b}+\frac{15\pi G^2M^2}{8b^2}$. 

We  may use this understanding to compute gravitational amplitudes for massless particles scattering off black hole backgrounds via the Born series and Fourier transforms, instead of standard Feynman integrals. We envision two possible applications:
\begin{itemize}
    \item \textbf{Scattering off a Kerr black hole.} Current interest in the community is to compute the tree-level $\mathcal{O}(G)$ scattering amplitude off the Kerr black hole in the classical wave regime. We believe that this Kerr--Compton amplitude is simply the Born amplitude, which can be computed by a Fourier transform of the Kerr potential \cite{Correia:2024jgr}.
    \item \textbf{Black hole tidal Love numbers.} The next section will describe how to account for dissipative and tidal effects in terms of Feynman diagrams. We envision that the matching calculation between effective field theory (EFT) and the exact solution from GR can be done at the level of the Born series more efficiently than via the use of Feynman diagrams. We expect that tidal Love numbers are encoded as `delta function' contributions in the potential.
\end{itemize}

\section[Worldline EFT, astrophysics applications, and gravitational Raman scattering]{Worldline EFT, astrophysics applications, and gravitational Raman scattering\\
\normalfont{\textit{Zihan Zhou}}}
\label{sec:Zhou}

\subsection{Worldline EFT and astrophysics applications}

In the previous sections, we calculated the scattering amplitude of binary dynamics within the point particle approximation. While this is a good approximation when the two compact objects are far away, it often fails to accurately model the internal structures such as the tidal response of neutron stars and black holes. To address these limitations and incorporate more complex physical effects, we turn to the worldline EFT, which utilizes a multipole expansion approach analogous to that used in electromagnetism \cite{Jackson:1998nia}. The multipole expansion in worldline EFT is structured as a series where higher order terms capture more detailed angular information about the system. The simplest term is the monopole, represented by $m$. To account for the object's structure more comprehensively, we introduce higher-order terms: dipole $P_i$, quadrupole $Q_{ij}$, octupole $Q_{ijk}$, and potentially higher multipoles. These terms allow us to model the detailed structure of compact objects beyond the basic point particle description.

Based on this decomposition, we can write down the effective action within a given gravitational background \cite{Goldberger:2004jt,Rothstein:2014sra,Porto:2016pyg,Goldberger:2022rqf}
\be
S_{\rm pp} = - \int \d \tau \left[ m + Q_{L}^E E^{L} + \ldots \right] + (E \to B)\, ,
\label{eq:worldline_action}
\ee
where $L = i_1i_2\cdots i_\ell$ shows the multipole indices. Similar to what we have learned in electromagnetism, in gravitational contexts $E^L$ and $B^L$ represent the electric and magnetic components of the tidal fields, respectively. Each component exhibits distinct parity transformation properties. These fields are defined as follows:
\be
E_L = \begin{dcases}
\partial_{\langle L\rangle} \phi \quad&\text{if}\quad s = 0\, ,\\
\partial_{\langle L-1\rangle} E_i \quad&\text{if}\quad s = 1\, ,\\
\partial_{\langle L-2\rangle} E_{ij} \quad&\text{if}\quad s = 2\, .
\end{dcases}
\quad 
B_L = \begin{dcases}
\partial_{\langle L-1\rangle} B_i \quad&\text{if}\quad s = 1\, ,\\
\partial_{\langle L-2\rangle} B_{ij} \quad&\text{if}\quad s = 2\, ,
\end{dcases}
\ee
where $s$ denotes the spin of the external field.
Let us use the notation $\langle \cdots \rangle$ to represent the symmetric-trace-free component of tensors. In our discussion on gravitational interactions (spin-2 fields), we consider a point particle with its four-velocity, $u^\mu$, and tetrads $e_i^\mu$. The definitions of the electric and magnetic tidal fields are given by
\be
E_{ij} = u^\mu e_i^\nu u^\rho e_j^\sigma C_{\mu\nu\rho\sigma}\, ~, \quad B_{ij} = u^\mu e_i^\nu u^\rho e_j^\sigma \star C_{\mu\nu\rho\sigma} \, ,
\ee
where $C_{\mu\nu\rho\sigma}$ is the Weyl tensor and $\star C_{\mu\nu\rho\sigma}$ stands for its dual. 
One can readily verify that the effective action presented in Eq.~\eqref{eq:worldline_action} is both gauge-invariant and re-parametrization invariant.

From the EFT perspective, we are going to treat $Q^{E/B}_L$ as composite operators that depend on some microscopic variables $X$
\be
S_{\rm pp} \to S_{\rm pp} + \int \d \tau \mathcal{L}_Q[Q^{E/B}_L(X), \dot Q^{E/B}_L(X)]\, .
\label{eq:worldline_dynamics}
\ee
For a fluid star, the microscopic variable $X$ corresponds to the Lagrangian displacement $\xi$ of the fluid element, $X=\xi$ \cite{poisson2014gravity}. With this basic knowledge in mind, let us try to apply this theoretical framework in some astrophysical context.

\begin{QA}
\question{How much of an approximation is treating stars as fluid stars? For example, is the Sun a fluid star?}
It is typically a very good approximation. Our Sun can be treated as a fluid star with polytropic index $n=3$.
\end{QA}

\subsubsection[Stellar oscillation, tidal encounters and tidal disruption events]{Application 1: Stellar oscillation, tidal encounters and tidal disruption events}

The first application is to use worldline EFT to model stellar oscillations, specifically to analyze tidal encounters and tidal disruption events. For simplicity, we will focus on the Newtonian limit and assume a linear tides approximation. Part of the discussion in this section comes from ongoing work with J. Li, I. Martínez-Rodríguez.

By varying the action in Eq.~\eqref{eq:worldline_dynamics} with respect to the composite variable $Q_L^E$, we can get the following Euler-Lagrangian equation that governs the evolution of the multipole moments
\begin{equation}
    \frac{\d}{\d\tau} \Big(\frac{\partial \mathcal{L}_Q}{\partial \dot Q^E_L}\Big) - \frac{\partial \mathcal{L}_Q}{\partial Q^E_L} = - E^L ~.
\end{equation}
Although the equation presented above may seem quite abstract at this stage, we can still employ the method of retarded Green's functions
\begin{equation}
    \langle Q_L^E(\tau) Q_{L'}^E(\tau') \rangle_{\rm ret} = i \langle [Q_L^E(\tau), Q_{L'}^{E}(\tau')] \rangle \Theta(\tau-\tau') \equiv \delta_{LL'} F_{\ell}(\tau-\tau') ~,
\end{equation}
to solve for the evolution of $Q_L^E$
\begin{equation}
    \langle Q_L^E (\tau)\rangle = - \int_{-\infty}^\tau \d\tau' F_\ell(\tau-\tau') E_L(\tau') ~.
\end{equation}
To better understand the structure of this retarded Green's function, we will use the spectral representation, which decomposes the full response function into a summation of eigenmodes $\omega_n$ accompanied by their corresponding overlap integrals $\langle n | Q_{L}|0\rangle$
\be
     {\rm Re} F_{\ell} (\omega) =  -\sum_n |\left\langle n\left|Q_{L}\right| 0\right\rangle|^2 {\rm PV} \frac{2\omega_n}{\omega^2 - \omega_n^2} ~,
\ee
where PV here stands for the principle value. Now, once we have determined the eigenmode frequencies and the overlapping integral, we can obtain the evolution of the quadrupole moments. However, the calculation of eigenmodes falls outside the scope of the EFT; instead, these must be obtained through matching the EFT with an UV theory. For our purposes, this UV theory is the stellar perturbation theory. In Newtonian gravity, linear fluid perturbations are well-understood, and the corresponding response function is given by Eq.~(3.29) in \cite{Pitre:2023xsr}.

Now, turning to astrophysical applications, let us examine the scenario of a star and black-hole encounter. Consider a star on a parabolic orbit around a supermassive black hole, with a pericenter distance $r_p$ much greater than the Schwarzschild radius $r_s$, allowing us to disregard relativistic effects. We want to ask the question: how much energy does the star gain during this process? Based on our intuition from the driven harmonic oscillators, the answer is just the work done by the external tidal field
\begin{equation}
    \langle  \mathcal{E}_{Q} \rangle (\tau)  = - \int_{-\infty}^\tau \d\tau' \langle \dot  Q_L^E(\tau') \rangle E^L(\tau') \,.
\end{equation}
We call $\mathcal{E}_Q$ the tidal energy resulting from the star and black-hole encounter. In most cases, this tidal energy is less than the star's binding energy, leading to the star experiencing only minor oscillations after the encounter. However, there are exceptions where, for some unlucky stars, the tidal energy during the encounter greatly exceeds their binding energy $U$, i.e.
\begin{equation}
    \mathcal{E}_Q \gg |U| ~.
\end{equation}
In this case, the star cannot be stable anymore, and will break apart in the end. This is known as the tidal disruption event (TDE). Currently, we have observed around 100 such events \cite{van2011optical,Mockler:2018xne,Hammerstein:2022wia}.

\subsubsection[Gravitational wave data analysis]{Application 2: Gravitational wave data analysis}

The second application is to put constraints on the tidal deformation and dissipation coefficients from the current LIGO--Virgo--KAGRA data. The candidates in the current data catalog are mostly quasi-circular binary systems. During the inspiral phase, the wavelength of the gravitational waves is significantly larger than the radius of the black hole. This allows us to apply a low-frequency expansion to the retarded response function. Specifically, we will concentrate on the non-spinning quadrupole sector $\ell=2$:
\be
\langle Q_{ij} Q_{kl} \rangle_{\text{ret}}(\omega) = - M (GM)^4 \left[ \Lambda {+} i (GM\omega)H_\omega {+} (iGM\omega)^2 \Lambda_{\omega^2} {+} \ldots \right] \delta_{\langle ij \rangle,\langle kl \rangle} ~.
\ee
In the above expansion, $\Lambda$ is the static Love number $\sim (\tfrac{R}{GM})^5$ ($\Lambda \sim \Big(\frac{R}{G M}\Big)^{2\ell+1}$ for general multipole sectors). $H_\omega$ is known as the dissipation number because it relates to the time-reversal odd component of the response function. In astrophysical literature, this is often referred to as tidal heating, as the orbital binding energy is converted into heat within the star. $\Lambda_{\omega^2}$ is the dynamical Love number, reflecting the frequency-dependent tidal response. While the impact of static Love numbers on waveforms has been extensively studied for both black holes and neutron stars, the aspect of dissipation has not received much attention. Here, we will briefly summarize our latest results from \cite{Chia:2024bwc}.

\begin{figure}[t!]
    \centering
    \includegraphics[scale=0.4,valign=c]{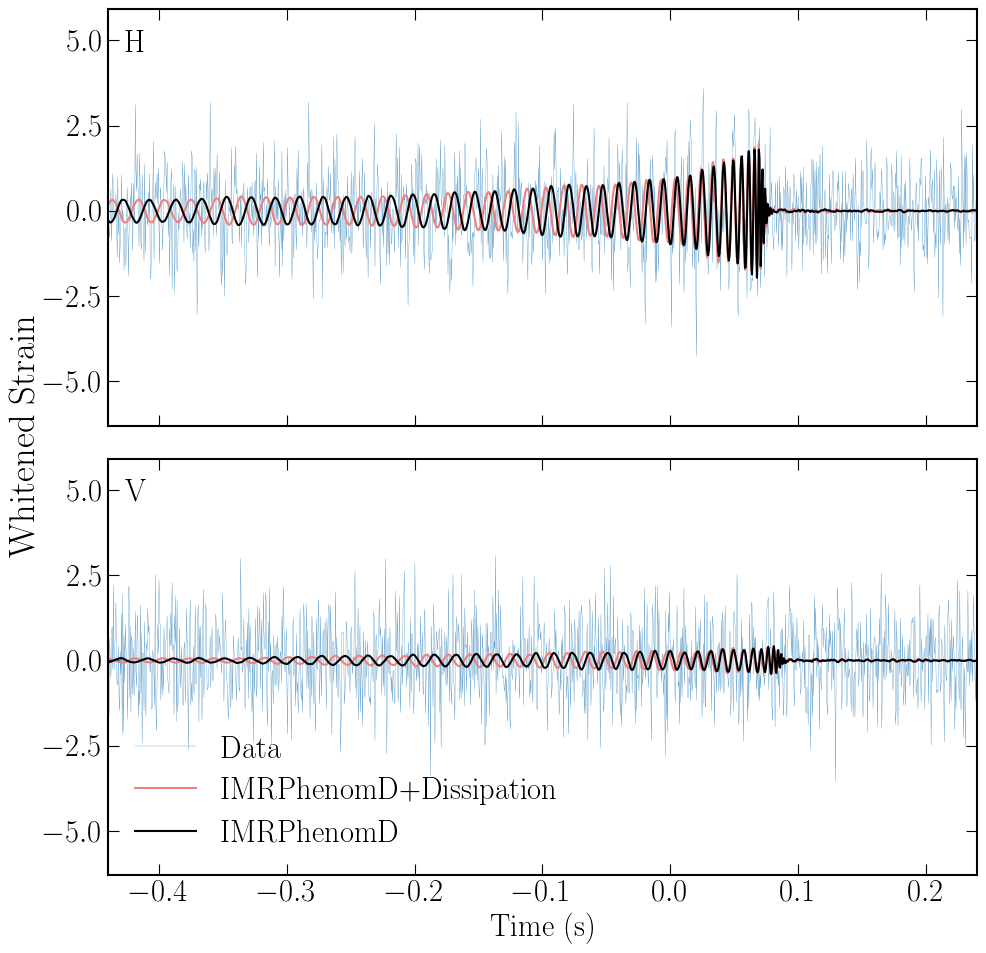}
    \caption{GW strains of the IMRPhenomD and IMRPhenomD+Dissipation waveforms for a GW191216\_213338 event in the Hanford and Virgo data. We choose the individual 
    dissipation
    parameters $H_{1\omega} = H_{2\omega}=10$ to clearly illustrate dephasing of the waveform due to tidal heating. Figure reprinted from \cite{Chia:2024bwc}.}
    \label{fig:dephasing}
\end{figure}

At the level of gravitational waveforms, introducing dissipation leads to dephasing effects. Specifically, as shown in Fig.~\ref{fig:dephasing}, we provide a direct comparison between the standard IMRPhenomD waveform (which describes the inspiral-merger-ringdown waveform for aligned-spin, quasi-circular binary black holes) and the IMRPhenomD waveform modified to include dissipation (IMRPhenomD+dissipation). While these two waveforms are synchronized at the merger point, they begin to deviate from each other as we trace them back to the earlier inspiral phase.
After conducting a detailed Bayesian analysis, we have established constraints on the mass-weighted dissipation number $\mathcal{H}_0 \equiv \frac{1}{M^4}  \left( m_1^4 H_{1 \omega} + m_2^4 H_{2\omega} \right)$ which range from $-13 < \mathcal{H}_0 < 20$ at $90\%$ confidence level (CL). Additionally, we have constrained the ratio of the energy lost due to tidal dissipation $\Delta E_{\rm H}$, to the radiative energy observed at infinity $\Delta E_{\infty}$, finding that $-0.0026 < \Delta E_{\rm H} / \Delta E_{\infty} <0.0025$ at $90\%$ CL. Our results are consistent with GR predictions, which suggest that $\mathcal{H}_0=2/45$ for equal mass binary black holes and $\Delta E_{\rm H} / \Delta E_{\infty} \simeq (10^{-5}, 10^{-4})$. 

\subsection{Gravitational Raman scattering}

As discussed, worldline EFT proves highly valuable in studying tidal effects within astrophysical contexts. Tidal Love numbers and dissipation numbers can, in principle, be constrained using observational data. However, the analysis becomes significantly more complicated for relativistic compact objects, as it involves systematic consideration of relativistic corrections. To achieve a high precision study of tidal effects and relativistic corrections, let us consider the following massless wave scattering on the stellar background, which we call gravitational Raman scattering \cite{Ivanov:2024sds}:
\be
\begin{gathered}
    \begin{tikzpicture}[line width=1,photon/.style={decorate, decoration={snake, amplitude=1pt, segment length=6pt}}]
    \draw[line width = 1, photon] (0,0) -- (1,1);
    \draw[line width = 1, photon] (0,0) -- (1,-1);
    \draw[line width = 0.8, ->] (0.85,0.6) -- (1.05,0.8);
    \draw[line width = 0.8, ->] (1.05,-0.8) -- (0.85,-0.6);
    \filldraw[fill=gray!5, line width=1.2,yshift=2](0,0) circle (0.6) node {\small star};
    \end{tikzpicture}
    \end{gathered}
\ee
This type of scattering amplitude is particularly interesting because it allows for a perturbative computation from the perspective of EFT, while from the UV side, the amplitude is encoded in the solutions to the linear perturbations of stars. Specifically, for 4D black holes, the perturbation equation, known as the Teukolsky equation \cite{Teukolsky:1973ha}, can be solved quasi-exactly \cite{Bonelli:2021uvf,Bonelli:2022ten,Bautista:2023sdf}. Solving this equation provides a way to uncover various properties of black holes.

Essentially, there are two regimes in which one can study this problem. In the eikonal limit where $G M \omega \gg 1, \omega b \gg 1$ as discussed in Section~\ref{sec:Correia}, the analysis approximates null geodesics. However, this regime cannot probe the internal structure of compact objects. Conversely, in the wave limit, where $G M \omega \ll 1$ the finite size effects of the compact object will be important. We will focus on the latter regime. Moreover, the gravitational Raman amplitude can be separated into two parts. The first part captures the scattering against the background metric
\be
\begin{aligned}
\hspace{-0.7cm}
\left[
\begin{gathered}
    \begin{tikzpicture}[line width=1,photon/.style={decorate, decoration={snake, amplitude=1pt, segment length=6pt}}]
    \draw[line width = 1, photon] (0,0) -- (1,1);
    \draw[line width = 1, photon] (0,0) -- (1,-1);
    \filldraw[fill=gray!5, line width=1.2,yshift=2](0,0) circle (0.6) node {\small star};
    \end{tikzpicture}
\end{gathered}
\right]_{\rm BG}
& = 
\begin{gathered}
    \begin{tikzpicture}[line width=1,photon/.style={decorate, decoration={snake, amplitude=1pt, segment length=6pt}}]
    \draw[line width = 1, photon] (0,0) -- (1,1);
    \draw[line width = 1, photon] (0,0) -- (1,-1);
    \draw[line width = 1, dashed] (-1,0) node[midway, above, xshift=-10] {\footnotesize $G M /r$} -- (0,0);
    \filldraw[fill=gray!5, line width=1.2](-1,0) circle (0.15) node {$\times$};
    \end{tikzpicture}
\end{gathered}
+
\begin{gathered}
    \begin{tikzpicture}[line width=1,photon/.style={decorate, decoration={snake, amplitude=1pt, segment length=6pt}}]
    \draw[line width = 1, photon] (0,0.5) -- (1,1.5);
    \draw[line width = 1, photon] (0,-0.5) -- (1,-1.5);
    \draw[line width = 1, photon] (0,-0.5) -- (0,0.5);
    \draw[line width = 1, dashed] (-1,-0.5) -- (0,-0.5);
    \draw[line width = 1, dashed] (-1,0.5) -- (0,0.5);
    \filldraw[fill=gray!5, line width=1.2](-1,-0.5) circle (0.15) node {$\times$};
    \filldraw[fill=gray!5, line width=1.2](-1,0.5) circle (0.15) node {$\times$};
    \end{tikzpicture}
\end{gathered}
+
\begin{gathered}
    \begin{tikzpicture}[line width=1,photon/.style={decorate, decoration={snake, amplitude=1pt, segment length=6pt}}]
    \draw[line width = 1, photon] (0,0.0) -- (1,1.0);
    \draw[line width = 1, photon] (0,0.0) -- (1,-1.0);
    \draw[line width = 1, dashed] (-1,-0.5) -- (0,0.0);
    \draw[line width = 1, dashed] (-1,0.5) -- (0,0.0);
    \filldraw[fill=gray!5, line width=1.2](-1,-0.5) circle (0.15) node {$\times$};
    \filldraw[fill=gray!5, line width=1.2](-1,0.5) circle (0.15) node {$\times$};
    \end{tikzpicture}
\end{gathered}
+
\begin{gathered}
    \begin{tikzpicture}[line width=1,photon/.style={decorate, decoration={snake, amplitude=1pt, segment length=6pt}}]
    \draw[line width = 1, photon] (0,1.0) -- (1,2.0);
    \draw[line width = 1, photon] (0,-1.0) -- (1,-2.0);
    \draw[line width = 1, photon] (0,-1.0) -- (0.0,0.0);
    \draw[line width = 1, photon] (0.0,0.0) -- (0.0,1.0);
    \draw[line width = 1, dashed] (-1,-1.0) -- (0,-1.0);
    \draw[line width = 1, dashed] (-1,1.0) -- (0,1.0);
    \draw[line width = 1, dashed] (-1,0.0) -- (0,0.0);
    \filldraw[fill=gray!5, line width=1.2](-1,-1.0) circle (0.15) node {$\times$};
    \filldraw[fill=gray!5, line width=1.2](-1,1.0) circle (0.15) node {$\times$};
    \filldraw[fill=gray!5, line width=1.2](-1,0.0) circle (0.15) node {$\times$};
    \end{tikzpicture}
\end{gathered}
\\
& \quad +
\begin{gathered}
    \begin{tikzpicture}[line width=1,photon/.style={decorate, decoration={snake, amplitude=1pt, segment length=6pt}}]
    \draw[line width = 1, photon] (0,0.0) -- (1,1.0);
    \draw[line width = 1, photon] (0,-1.0) -- (1,-2.0);
    \draw[line width = 1, photon] (0.0,0.0) -- (0.0,-1.0);
    \draw[line width = 1, dashed] (-1,-1.0) -- (0,-1.0);
    \draw[line width = 1, dashed] (-1,0.5) -- (0,0.0);
    \draw[line width = 1, dashed] (-1,-0.5) -- (0,0.0);
    \filldraw[fill=gray!5, line width=1.2](-1,-1.0) circle (0.15) node {$\times$};
    \filldraw[fill=gray!5, line width=1.2](-1,0.5) circle (0.15) node {$\times$};
    \filldraw[fill=gray!5, line width=1.2](-1,-0.5) circle (0.15) node {$\times$};
    \end{tikzpicture}
\end{gathered}
+
\begin{gathered}
    \begin{tikzpicture}[line width=1,photon/.style={decorate, decoration={snake, amplitude=1pt, segment length=6pt}}]
    \draw[line width = 1, photon] (0,0.0) -- (1,1.0);
    \draw[line width = 1, photon] (0,-1.0) -- (1,-2.0);
    \draw[line width = 1, photon] (0.0,0.0) -- (0.0,-1.0);
    \draw[line width = 1, dashed] (-1,-1.5) -- (0,-1.0);
    \draw[line width = 1, dashed] (-1,0.0) -- (0,0.0);
    \draw[line width = 1, dashed] (-1,-0.5) -- (0,-1.0);
    \filldraw[fill=gray!5, line width=1.2](-1,-1.5) circle (0.15) node {$\times$};
    \filldraw[fill=gray!5, line width=1.2](-1,0.0) circle (0.15) node {$\times$};
    \filldraw[fill=gray!5, line width=1.2](-1,-0.5) circle (0.15) node {$\times$};
    \end{tikzpicture}
\end{gathered}
+
\begin{gathered}
    \begin{tikzpicture}[line width=1,photon/.style={decorate, decoration={snake, amplitude=1pt, segment length=6pt}}]
    \draw[line width = 1, photon] (0,0.0) -- (1,1.0);
    \draw[line width = 1, photon] (0,0.0) -- (1,-1.0);
    \draw[line width = 1, dashed] (-1,-1.0) -- (0,0.0);
    \draw[line width = 1, dashed] (-1,1.0) -- (0,0.0);
    \draw[line width = 1, dashed] (-1,0.0) -- (0,0.0);
    \filldraw[fill=gray!5, line width=1.2](-1,-1.0) circle (0.15) node {$\times$};
    \filldraw[fill=gray!5, line width=1.2](-1,1.0) circle (0.15) node {$\times$};
    \filldraw[fill=gray!5, line width=1.2](-1,0.0) circle (0.15) node {$\times$};
    \end{tikzpicture}
\end{gathered}
+\cdots ~,
\end{aligned}
\ee
which is made up of scattering with various mass monopole insertions on the worldline. The second part is the scattering against the star surface, which includes the tidal response
\be
\begin{aligned}
\left[
\begin{gathered}
    \begin{tikzpicture}[line width=1,photon/.style={decorate, decoration={snake, amplitude=1pt, segment length=6pt}}]
    \draw[line width = 1, photon] (0,0) -- (1,1);
    \draw[line width = 1, photon] (0,0) -- (1,-1);
    \filldraw[fill=gray!5, line width=1.2,yshift=2](0,0) circle (0.6) node {\small star};
    \end{tikzpicture}
\end{gathered}
\right]_{\rm tides}
& =
\begin{gathered}
    \begin{tikzpicture}[line width=1,photon/.style={decorate, decoration={snake, amplitude=1pt, segment length=6pt}}]
    \draw[line width = 1, photon] (0,0.5) -- (1,1.5);
    \draw[line width = 1, photon] (0,-0.5) -- (1,-1.5);
    \draw[line width = 1, dashed, double] (0,-0.5) -- (0,0.5);
    \filldraw[fill=black, line width=1.2](0,0.5) circle (0.15) node[left]{\small$Q\;$};
    \filldraw[fill=black, line width=1.2](0,-0.5) circle (0.15) node[left]{\small$Q\;$};
    \end{tikzpicture}
\end{gathered}
+
\begin{gathered}
    \begin{tikzpicture}[line width=1,photon/.style={decorate, decoration={snake, amplitude=1pt, segment length=6pt}}]
    \draw[line width = 1, photon] (0,0.5) -- (1,1.5);
    \draw[line width = 1, photon] (0,-0.5) -- (1,-1.5);
    \draw[line width = 1, dashed, double] (0,-0.5) -- (0,0.5);
    \draw[line width = 1, dashed] (0,1.2) -- (0.5,1.0);
    \filldraw[fill=black, line width=1.2](0,0.5) circle (0.15);
    \filldraw[fill=black, line width=1.2](0,-0.5) circle (0.15);
    \filldraw[fill=gray!5, line width=1.2](0,1.2) circle (0.15) node {$\times$};
    \end{tikzpicture}
\end{gathered}
+
\begin{gathered}
    \begin{tikzpicture}[line width=1,photon/.style={decorate, decoration={snake, amplitude=1pt, segment length=6pt}}]
    \draw[line width = 1, photon] (0,0.5) -- (1,1.5);
    \draw[line width = 1, photon] (0,-0.5) -- (1,-1.5);
    \draw[line width = 1, dashed, double] (0,-0.5) -- (0,0.5);
    \draw[line width = 1, dashed] (0,-1.2) -- (0.5,-1.0);
    \filldraw[fill=black, line width=1.2](0,0.5) circle (0.15);
    \filldraw[fill=black, line width=1.2](0,-0.5) circle (0.15);
    \filldraw[fill=gray!5, line width=1.2](0,-1.2) circle (0.15) node {$\times$};
    \end{tikzpicture}
\end{gathered}
+
\begin{gathered}
    \begin{tikzpicture}[line width=1,photon/.style={decorate, decoration={snake, amplitude=1pt, segment length=6pt}}]
    \draw[line width = 1, photon] (0,0.5) -- (1,1.5);
    \draw[line width = 1, photon] (0,-0.5) -- (1,-1.5);
    \draw[line width = 1, dashed, double] (0,-0.5) -- (0,0.5);
    \draw[line width = 1, dashed] (0,1.2) -- (0.5,1.0);
    \draw[line width = 1, dashed] (0,-1.2) -- (0.5,-1.0);
    \filldraw[fill=black, line width=1.2](0,0.5) circle (0.15);
    \filldraw[fill=black, line width=1.2](0,-0.5) circle (0.15);
    \filldraw[fill=gray!5, line width=1.2](0,-1.2) circle (0.15) node {$\times$};
    \filldraw[fill=gray!5, line width=1.2](0,1.2) circle (0.15) node {$\times$};
    \end{tikzpicture}
\end{gathered}
\\
& \quad +
\begin{gathered}
    \begin{tikzpicture}[line width=1,photon/.style={decorate, decoration={snake, amplitude=1pt, segment length=6pt}}]
    \draw[line width = 1, photon] (0,0.5) -- (1,1.5);
    \draw[line width = 1, photon] (0,-0.5) -- (1,-1.5);
    \draw[line width = 1, dashed, double] (0,-0.5) -- (0,0.5);
    \draw[line width = 1, dashed] (0,1.0) -- (0.5,1.0);
    \draw[line width = 1, dashed] (0,1.7) -- (0.5,1.0);
    \filldraw[fill=black, line width=1.2](0,0.5) circle (0.15);
    \filldraw[fill=black, line width=1.2](0,-0.5) circle (0.15);
    \filldraw[fill=gray!5, line width=1.2](0,1.0) circle (0.15) node {$\times$};
    \filldraw[fill=gray!5, line width=1.2](0,1.7) circle (0.15) node {$\times$};
    \end{tikzpicture}
\end{gathered}
+
\begin{gathered}
    \begin{tikzpicture}[line width=1,photon/.style={decorate, decoration={snake, amplitude=1pt, segment length=6pt}}]
    \draw[line width = 1, photon] (0,0.5) -- (1,1.5);
    \draw[line width = 1, photon] (0,-0.5) -- (1,-1.5);
    \draw[line width = 1, dashed, double] (0,-0.5) -- (0,0.5);
    \draw[line width = 1, dashed] (0,-1.0) -- (0.5,-1.0);
    \draw[line width = 1, dashed] (0,-1.7) -- (0.5,-1.0);
    \filldraw[fill=black, line width=1.2](0,0.5) circle (0.15);
    \filldraw[fill=black, line width=1.2](0,-0.5) circle (0.15);
    \filldraw[fill=gray!5, line width=1.2](0,-1.0) circle (0.15) node {$\times$};
    \filldraw[fill=gray!5, line width=1.2](0,-1.7) circle (0.15) node {$\times$};
    \end{tikzpicture}
\end{gathered}
+
\begin{gathered}
    \begin{tikzpicture}[line width=1,photon/.style={decorate, decoration={snake, amplitude=1pt, segment length=6pt}}]
    \draw[line width = 1, photon] (0,0.5) -- (1,1.5);
    \draw[line width = 1, photon] (0,-0.5) -- (1,-1.5);
    \draw[line width = 1, dashed, double] (0,-0.5) -- (0,0.5);
    \draw[line width = 1, dashed] (0,1.2) -- (0.5,1.0);
    \draw[line width = 1, dashed] (0,1.7) -- (0.8,1.3);
    \filldraw[fill=black, line width=1.2](0,0.5) circle (0.15);
    \filldraw[fill=black, line width=1.2](0,-0.5) circle (0.15);
    \filldraw[fill=gray!5, line width=1.2](0,1.2) circle (0.15) node {$\times$};
    \filldraw[fill=gray!5, line width=1.2](0,1.7) circle (0.15) node {$\times$};
    \end{tikzpicture}
\end{gathered}
+
\begin{gathered}
    \begin{tikzpicture}[line width=1,photon/.style={decorate, decoration={snake, amplitude=1pt, segment length=6pt}}]
    \draw[line width = 1, photon] (0,0.5) -- (1,1.5);
    \draw[line width = 1, photon] (0,-0.5) -- (1,-1.5);
    \draw[line width = 1, dashed, double] (0,-0.5) -- (0,0.5);
    \draw[line width = 1, dashed] (0,-1.2) -- (0.5,-1.0);
    \draw[line width = 1, dashed] (0,-1.7) -- (0.8,-1.3);
    \filldraw[fill=black, line width=1.2](0,0.5) circle (0.15);
    \filldraw[fill=black, line width=1.2](0,-0.5) circle (0.15);
    \filldraw[fill=gray!5, line width=1.2](0,-1.2) circle (0.15) node {$\times$};
    \filldraw[fill=gray!5, line width=1.2](0,-1.7) circle (0.15) node {$\times$};
    \end{tikzpicture}
\end{gathered}
+\cdots ~.
\end{aligned}
\ee
We parametrize our tidal response function by performing the low-frequency expansion
\be
\langle Q_{L_1} Q_{L_2} \rangle_{\text{ret}}(\omega) = \delta_{\langle L_1 \rangle, \langle L_2 \rangle} \left( C_{\ell,0} + i\omega C_{\ell,1} + (i\omega)^2 C_{\ell,2} + \ldots \right)\, .
\label{eq:QQ_expand}
\ee

Detailed calculations reveal that this amplitude exhibits both IR and UV divergences. IR divergences arise from the long-range nature of the Newtonian potential $G M/r$, analogous to Coulomb interactions discussed in Section~\ref{sec:Isabella}. These divergences will exponentiate when we resum all ladder-like diagrams as shown in Eq.~\eqref{eq:ladder_sum}. Regarding the UV divergences, they lead to renormalization group (RG) runnings in the tidal response function. Specifically focusing on scalar perturbations, the first type of RG running is induced by tidal effects. At the two-loop level, we derive the following RG equation:
\be
\frac{\d \langle QQ \rangle_\ell}{\d \log \mu} = -(2G M \omega)^2 \left[ \frac{-11 + 15\ell(1+\ell)}{(-1 + 2\ell)(1 + 2\ell)(3+2\ell)} \langle QQ \rangle_\ell \right]\,, 
\ee
where $\mu$ is the renormalization scale. As we can see this RG equation depends on the specific form of the tidal response function. Remarkably, there is another type of running which is independent of the nature of the compact object. For the $\ell=0$ sector, we find that
\be
\frac{\d \langle QQ \rangle_{\ell = 0}}{\d \log \mu} = (\text{self-induced}) - 4\pi (2GM)^3 ~.
\ee
The second term is universal for all types of compact objects as it depends solely on their masses. For black holes, we can determine various coefficients in Eq.~\eqref{eq:QQ_expand} by matching our EFT amplitude to black hole perturbation theory using the $\overline{\rm MS}$ renormalization scheme.
\begin{gather}
C_{\ell = 0, 0} = 0, \qquad C_{\ell=1,0} = 0,\\
C_{\ell=0,2} = -4\pi r_s^3 \left[ \frac{1}{4 \epsilon_{\text{UV}}} + \log(\mu r_s) + \frac{19}{12} + \gamma_E \right] ~.
\end{gather}
We find that the static tidal Love numbers for $\ell=0$ and $\ell=1$ vanish, and the dynamical tidal Love numbers receive logarithmic corrections.

In summary, the worldline EFT offers a model-independent framework for studying the tidal effects of astrophysical compact objects. Given the diverse applications discussed in this section, the study of tidal effects is poised to enter a new era of precision science.


\section[Radiation in gravitational observables]{Radiation in gravitational observables\\
\normalfont{\textit{Hofie Hannesdottir}}}
\label{sec:Hannesdottir}
\newcommand{\cloud}{\tikzset{every picture/.style={line width=0.75pt}}  
\begin{tikzpicture}[x=0.75pt,y=0.75pt,yscale=-0.5,xscale=0.5]
\draw[color=charcoal,fill=white  ,fill opacity=1 ]   (107.28,126.75) .. controls (106.64,121.92) and (108.75,117.13) .. (112.73,114.42) .. controls (116.7,111.71) and (121.85,111.56) .. (125.97,114.03) .. controls (127.43,111.21) and (130.1,109.27) .. (133.18,108.79) .. controls (136.26,108.3) and (139.38,109.33) .. (141.6,111.57) .. controls (142.84,109.02) and (145.29,107.31) .. (148.06,107.04) .. controls (150.84,106.77) and (153.55,107.98) .. (155.24,110.25) .. controls (157.49,107.55) and (161.07,106.41) .. (164.42,107.33) .. controls (167.78,108.25) and (170.32,111.06) .. (170.93,114.55) .. controls (173.69,115.31) and (175.98,117.27) .. (177.22,119.9) .. controls (178.47,122.53) and (178.53,125.58) .. (177.41,128.27) .. controls (180.12,131.88) and (180.75,136.68) .. (179.07,140.89) .. controls (177.39,145.11) and (173.65,148.09) .. (169.24,148.74) .. controls (169.21,152.69) and (167.09,156.32) .. (163.7,158.22) .. controls (160.3,160.12) and (156.16,160.01) .. (152.88,157.91) .. controls (151.48,162.65) and (147.54,166.13) .. (142.76,166.86) .. controls (137.99,167.59) and (133.23,165.43) .. (130.55,161.32) .. controls (127.26,163.34) and (123.31,163.93) .. (119.59,162.94) .. controls (115.88,161.95) and (112.71,159.46) .. (110.8,156.05) .. controls (107.44,156.45) and (104.19,154.67) .. (102.67,151.59) .. controls (101.14,148.51) and (101.66,144.79) .. (103.98,142.27) .. controls (100.98,140.47) and (99.45,136.89) .. (100.18,133.4) .. controls (100.92,129.92) and (103.76,127.31) .. (107.21,126.94) ; \draw[color=charcoal]   (103.98,142.27) .. controls (105.39,143.13) and (107.03,143.51) .. (108.66,143.38)(110.8,156.05) .. controls (111.51,155.96) and (112.2,155.79) .. (112.85,155.52)(130.55,161.32) .. controls (130.05,160.56) and (129.64,159.75) .. (129.31,158.9)(152.88,157.91) .. controls (153.13,157.05) and (153.3,156.16) .. (153.37,155.26)(169.24,148.74) .. controls (169.28,144.53) and (166.94,140.67) .. (163.23,138.83)(177.41,128.27) .. controls (176.81,129.7) and (175.89,130.97) .. (174.73,131.98)(170.93,114.55) .. controls (171.04,115.12) and (171.08,115.71) .. (171.07,116.3)(155.24,110.25) .. controls (154.68,110.92) and (154.22,111.68) .. (153.87,112.49)(141.6,111.57) .. controls (141.3,112.18) and (141.08,112.83) .. (140.93,113.5)(125.97,114.03) .. controls (126.84,114.55) and (127.65,115.18) .. (128.37,115.9)(107.28,126.75) .. controls (107.37,127.42) and (107.51,128.08) .. (107.7,128.72) ;
\end{tikzpicture}}

QFT provides the framework to compute many different observables in particle physics and gravity. In previous sections, we have already discussed one such observable, the \emph{scattering amplitude}, which encodes the probability amplitude for a set of ``in'' states to turn into a set of ``out'' states. In addition to the application mentioned in previous sections, scattering amplitudes form the basis for computations in collider-physics experiments: we send in a pair of particles (such as protons at the Large Hadron Collider), and measure the outgoing particles -- at least to the fullest extent possible with current technology.

However, many other observables are relevant in high-energy physics. In particular, recent years have seen the advent of measurements of \emph{gravitational waves}, for example in the LIGO--Virgo--Kagra experiments. The physical setup in gravitational-wave measurements is very different from the one in particle colliders. First, the measured waves originate in events occurring far away from Earth, such as black hole merging, instead of a particle collision on Earth. Second, we only have access to the signal in a small corner of the phase space, and have to sum over unobserved configurations. This section is devoted to a discussion of the gravitational waveform as measured after a scattering of two black holes and is based on~\cite{Caron-Huot:2023vxl} \footnote{Compared to~\cite{Caron-Huot:2023vxl} and the contribution~\cite{chapterCaronHuotGiroux},  we here use mostly minus signature as in the previous sections.}. For a more general discussion on different asymptotic observables and their analytic continuations, see the contribution~\cite{chapterCaronHuotGiroux}.

Before we go on, let us emphasize that this waveform is different from the ones measured so far in the LIGO--Virgo--Kagra experiments. Current measurements correspond to radiation after \emph{merging} of two heavy objects, while here we are computing the corresponding \emph{scattering} events, that is, when the black holes do not form a bound state but rather scatter off each other on a hyperbolic orbit. When, and to which extent, future gravitational-wave experiments will be able to measure the scattering waveforms is an open question. One of the challenges in such measurements is that the scattering waveforms do not have the characteristic periodic behavior of the bound waveforms: as discussed in Sec.~\ref{sec:resumming}, the period of the merging waveform gets traded for a time delay of the scattering waveform. Nevertheless, it remains an open problem whether the bound-state waveforms can generally be obtained as analytic continuations of the scattering ones, see the discussion around~\eqref{eq:taileffects} and Ref.~\cite{Adamo:2024oxy}.

\subsection{The gravitational waveform}

For a computation of the gravitational waveform, we model the black holes as heavy scalars with masses $m_i$, and assume that they scatter off each other. Following the KMOC formalism~\cite{Kosower:2018adc}, we write the incoming state as
\begin{equation}
    | \psi \rangle_{\text{in}} = \prod_{i=1}^2 \left[ \int \d \Phi(p_i) f_i(p_i) e^{i b_i \cdot p_i/ \hbar} \right] | p_1 p_2 \rangle\, ,
\end{equation}
where $b\equiv b_1-b_2$ is the impact parameter and the phase space is given by
\begin{equation}
    \d \Phi(p_i) \equiv \frac{\d^\rD p_i}{(2\pi)^\rD} 2 \pi \Theta(p^0) \delta(p_i^2-m_i^2) \,.
\end{equation}

The functions $f(p_i)$ correspond to wavepackets, and we assume they are sharply peaked around the classical momenta of the black holes, as described in~\cite{Kosower:2018adc}. Following~\cite{Elkhidir:2023dco}, we define the wavepackets as
\begin{equation}
    \phi_b (p_1,p_2)\equiv e^{i p_1 \cdot b_1 / \hbar} e^{i p_2 \cdot b_2 / \hbar} f_1(p_1) f_2(p_2) \,.
\end{equation} We have labeled the state with the two black holes using their four-momenta $p_1$ and $p_2$, but we recall that they are on-shell external states, so $p_i^2=m_i^2$. We define the state $ | p_1 p_2 \rangle$ as usual, with creation operators acting on the vacuum: 
\begin{equation}
    | p_1 p_2 \rangle = 
    a_1^\dag a_2^\dag | 0 \rangle\,.
\end{equation}

The expectation value of the curvature in gravity is given by
\begin{equation}
   R_{\mu \nu \rho \sigma} (x) = \, _{\text{in}} \,\langle \psi | S^\dag \mathbb{R}_{\mu \nu \rho \sigma} S | \psi \rangle_{\text{in}}  \,.
\end{equation}
Expanding $\mathbb{R}_{\mu \nu \rho \sigma}$ out in its Fourier modes gives~\cite{Cristofoli:2021vyo,Elkhidir:2023dco}
\begin{align}
    R_{\mu \nu \rho \sigma} (x)  & = \kappa \Re \sum_{\eta}
    \prod_{i\in \{1,2,1',2'\}} \left[ \int \d \Phi(p_i)\right]
    \phi_b (p_1,p_2) \phi_b^\ast (p_{1'},p_{2'})
    \label{eq:R}
    \\ & \times
    \int \d \Phi(k)
    k_{[\mu}\epsilon^\eta_{\nu]}(k)
    k_{[\rho}\epsilon^\eta_{\sigma]}(k)
    \langle p_1' p_2' | b_\eta (k) | p_1 p_2 \rangle \,,
    \nonumber
\end{align}
with $\kappa=\sqrt{32 \pi G}$, and $b_\eta (k)$ is the future annihilation operator for a graviton with momentum $k$ and helicity $\eta$.\footnote{The creation and annihilation operators are often referred to as $a_{\text{out}}^\dag$ and $a_{\text{out}}$ in the literature, but we refer to them as $b$'s instead to avoid clutter, as in \cite[Sec.~1]{chapterCaronHuotGiroux}. Similarly, we simply write $a$ and $a^\dag$ for the past operators, instead of $a_{\text{in}}$ and $a_{\text{in}}^\dag$.} We will henceforth suppress helicity labels for simplicity. The intuition behind this expression is that, since $\mathbb{R}_{\mu \nu \rho \sigma}$ depends linearly on the creation and annihilation operators $b^\dag (k)$ and $b (k)$, the expectation value of $b (k)$ emerges as the key dynamical quantity to compute. In this subsection, we will focus on computing the momentum-space expectation value
\begin{equation}
    {\rm Exp}_k \equiv
    \langle p_1' p_2' | b (k) | p_1 p_2 \rangle \,.
\end{equation}
The position-space waveform can then be obtained from~\eqref{eq:R}, see~\cite{Brandhuber:2023hhy,Herderschee:2023fxh,Elkhidir:2023dco,Georgoudis:2023lgf,Georgoudis:2023ozp}.

As presented in \cite[Sec.~1]{chapterCaronHuotGiroux}, the future operators in the Heisenberg picture are related to the past operators via the $S$-matrix, which acts as an evolution operator:
\begin{equation}
    b(k) = S^\dag a (k)  S \,, \qquad b^\dag (k)  = S^\dag a^\dag (k)  S \,,
\end{equation}
where we assume that $S$ is unitary, $S S^\dag = \mathbbm{1}$. 
We will save the wavepacket convolution for last, and start with computing the expression in the integral,
\begin{equation}
    {\rm Exp}_k =
    \langle p_1' p_2' | S^\dag a (k)  S | p_1 p_2 \rangle = \langle p_1' p_2' | S^\dag \sumint_X | X \rangle \langle X k | S | p_1 p_2 \rangle
\end{equation}
where we have inserted a complete set of states, $\mathbbm{1} = \sumint_X | X \rangle \langle X |$ and used that $\langle X| a (k)  = \langle X k |$. We call this expectation value the \emph{waveshape} to avoid confusion with the position-space waveform. The waveshape is a product of an $S$-matrix for $p_1 p_2 \to k X$, and a conjugated $S$-matrix for $X \to p_{1'} p_{2'}$, summed and integrated over all states $X$, which we denote pictorially as
\begin{equation}
    {\rm Exp}_k =
    \begin{gathered}
    \begin{tikzpicture}[baseline={([yshift=-45]current bounding box.center)},line width=1,photon/.style={decorate, decoration={snake, amplitude=1pt, segment length=6pt}},xscale=-1]
    \draw[line width = 1] (0,0.3) -- (-1.2,0.3) node[above] {\small$1'$};
    \draw[line width = 1] (0,-0.3) -- (-1.2,-0.3) node[below] {\small$2'$};
    \draw[line width = 1] (2,0.3) -- (3.2,0.3) node[above] {\small$1$};
    \draw[line width = 1] (2,-0.3) -- (3.2,-0.3) node[below] {\small$2$};
    \draw[photon] (2,0) -- (1,1) node[above left] {\small$k$};
    \filldraw[fill=gray!30, very thick](0,-0.3) rectangle (2,0.3);
    \draw[] (1,0) node {$X$};
    \filldraw[fill=gray!5, line width=1.2,yshift=1](0,0) circle (0.6) node {$S^\dag$};
    \filldraw[fill=gray!5, line width=1.2](2,0) circle (0.6) node {$S$};
    \draw[dashed,Orange] (1,1.1) -- (1,-1.1);
    \end{tikzpicture}
    \end{gathered}
\end{equation}
The shaded $X$ in this picture corresponds to an insertion of a complete set of on-shell states with positive energy flowing across the dashed cut.

Using the definition of the amplitude $\mathcal{M}$ through $S = \mathbbm{1} + i(2 \pi)^{\rD} \delta^{\rD}(\sum p_i) \mathcal{M}$, we can rewrite the waveshape in terms of amplitudes and cut amplitudes,
\begin{equation}
\hspace{-0.7cm}
  \stackrel{\adjustbox{valign=t}{
\begin{tikzpicture}[line width=1,photon/.style={decorate, decoration={snake, amplitude=1pt, segment length=6pt}},xscale=-1]
\draw[line width = 1] (0,0.3) -- (-1.2,0.3) node[above] {\small$1'$};
\draw[line width = 1] (0,-0.3) -- (-1.2,-0.3) node[below] {\small$2'$};
\draw[line width = 1] (2,0.3) -- (3.2,0.3) node[above] {\small$1$};
\draw[line width = 1] (2,-0.3) -- (3.2,-0.3) node[below] {\small$2$};
\draw[photon] (2,0) -- (1,1) node[above left] {\small$k$};
\filldraw[fill=gray!30, very thick](0,-0.3) rectangle (2,0.3);
\draw[] (1,0) node {$X$};
\filldraw[fill=gray!5, line width=1.2,yshift=1](0,0) circle (0.6) node {$S^\dag$};
\filldraw[fill=gray!5, line width=1.2](2,0) circle (0.6) node {$S$};
\draw[dashed,Orange] (1,1.1) -- (1,-1.1);
\end{tikzpicture}
}}{{\rm Exp}_k}
  =  \stackrel{\adjustbox{valign=t}{
\begin{tikzpicture}[line width=1, photon/.style={decorate, decoration={snake, amplitude=1pt, segment length=6pt}
	},xscale=-1]
\draw[line width = 1] (2,0.3) -- (0.8,0.3) node[above] {\small$1'$};
\draw[line width = 1] (2,-0.15) -- (0.8,-0.15) node[below] {\small$2'$};
\draw[line width = 1] (2,0.3) -- (3.2,0.3) node[above] {\small$1$};
\draw[line width = 1] (2,-0.3) -- (3.2,-0.3) node[below] {\small$2$};
\draw[photon] (2,0) -- (1,1) node[above left] {\small$k$};
\filldraw[fill=gray!5, line width=1.2](2,0) circle (0.6) node {$i \mathcal{M}$};
\draw[draw=none] (1,1.1) -- (1,-1.1);
\end{tikzpicture}
}}{i\mathcal{M}} + 
  \stackrel{\adjustbox{valign=t}{\begin{tikzpicture}[line width=1,photon/.style={decorate, decoration={snake, amplitude=1pt, segment length=6pt}
	},xscale=-1]
\draw[line width = 1] (0,0.3) -- (-1.2,0.3) node[above] {\small$1'$};
\draw[line width = 1] (0,-0.3) -- (-1.2,-0.3) node[below] {\small$2'$};
\draw[line width = 1] (2,0.3) -- (3.2,0.3) node[above] {\small$1$};
\draw[line width = 1] (2,-0.3) -- (3.2,-0.3) node[below] {\small$2$};
\draw[photon] (2,0) -- (1,1) node[above left] {\small$k$};
\filldraw[fill=gray!30, very thick](0,-0.3) rectangle (2,0.3);
\draw[] (1,0) node {$X$};
\filldraw[fill=gray!5, line width=1.2,yshift=1](0,0) circle (0.6) node {$-i\mathcal{M}^\dag$};
\filldraw[fill=gray!5, line width=1.2](2,0) circle (0.6) node {$i\mathcal{M}$};
\draw[dashed,Orange] (1,1.1) -- (1,-1.1);
\end{tikzpicture}}}{{\rm Cut}_{1'2'}\,.} 
\label{eq:Exp_k-pics}
\end{equation}
Importantly, the first term on the left-hand side is called \emph{superclassical} (or \emph{hyperclassical}), since it is proportional to a power of $\frac{1}{\hbar}$ in perturbation theory. As discussed in previous subsections, it thus does not make sense to take a classical limit $\hbar \to 0$ directly at the level of scattering amplitudes at a fixed order in perturbation theory.

Nevertheless, this power counting in $\hbar$ is entirely expected from classical physics: The exclusive amplitude in gravity is exponentially suppressed since the probability to create some fixed state is exponentially small. The amplitude therefore behaves as $\sim \e^{i \mathcal{S}/\hbar}$, where $\mathcal{S}$ is the action, and expanding out in $G$ results in inverse powers of $\hbar$. The waveshape, ${\rm Exp}_k$, is, on the other hand, a perfectly sensible classical observable since it sums over unobserved configurations. At a mathematical level, the cut term labeled ${\rm Cut}_{1'2'}$ in~\eqref{eq:Exp_k-pics} precisely works to cancel off the $\frac{1}{\hbar}$ dependence of the scattering amplitude term, rendering ${\rm Exp}_k$ well defined in the $\hbar \to 0$ limit. In addition to cancelling the superclassical contribution, the cut term contributes to the infrared divergence and the finite part of ${\rm Exp}_k$ as we discuss below.

\subsection{The classical limit}
Our strategy will be to compute $\text{Exp}_k$ in the classical limit of quantum field theory, following~\cite{Kosower:2018adc}.
The waveshape $\text{Exp}_k$ for measuring a graviton in the background of black hole scattering is a five-point process, which we label with
\begin{equation}
    \begin{gathered}
        \begin{tikzpicture}[scale=0.6,thick,
        baseline={([yshift=-0.4ex]current bounding box.center)},photon/.style={decorate, decoration={snake, amplitude=1pt, segment length=6pt}},xscale=-1]
        \draw[line width=2] (-1,1) -- (0,0) -- (1,1)  (-1,-1) -- (0,0) -- (1,-1);
        \draw[photon] (-0.5,0) -- +(-1.2,0);
        \filldraw[fill=gray!5, line width=1.2](0,0) circle (0.8);
        \node[xshift=-10] at (2,1.5) {$p_1=\bar{p}_1 + \frac{q_1}{2}$};
        \node[xshift=10] at (-2,1.5) {$p_1'=\bar{p}_1 - \frac{q_1}{2}$};
        \node[xshift=-10] at (2,-1.5) {$p_2=\bar{p}_2 + \frac{q_2}{2}$};
        \node[xshift=10] at (-2,-1.5) {$p_2'=\bar{p}_2 - \frac{q_2}{2}$};
        \node[] at (-3.5,0) {$k=q_1+q_2$};
\end{tikzpicture}
    \end{gathered}
\end{equation}
We define the barred masses as $\bar{m}_i = \bar{p}_i^2$.
The classical limit is obtained by taking the masses $\bar{m}_i$ of the black holes to be much larger than the Planck scale, and the impact parameter to be much larger than the wavelength of the graviton, 
\begin{equation}
    \bar{m}_1, \bar{m}_2 \gg M_{\text{pl}}\,, \qquad b \gg \lambda\,.
\end{equation}
In the classification of~\ref{eq:regimes}, this condition corresponds to the Eikonal regime.
This limit enforces the graviton to be soft, so
\begin{equation}
     \qquad q_1 \sim \hbar \,, \qquad q_2 \sim \hbar\,, \qquad k \sim \hbar \,.
\end{equation}
The coupling scales as $\kappa \sim \frac{1}{\sqrt{\hbar}}$, so Newton's constant scales as $G \sim \frac{1}{\hbar}$.

Note that this power counting is very different from the one used when doing computations purely within classical general relativity (GR), where $\hbar$ never makes an appearance. In GR, the perturbation series often involves a low-velocity expansion, in terms of powers of $G \sim v^2 G$, referred to as the \emph{Post-Newtonian} (PN) \emph{expansion}.
When computing classical observables from QFT computations, the perturbation series is instead organized as a series expansion in $G$, but is exact in $v$, called the \emph{Post-Minkowskian} (PM) \emph{expansion}.

\begin{center}
\begin{tabular}{ c|c|c|c|c } 
  & 0PN & 1PN & 2PN & 3PN \\ \hline 
 1PM & $G$ & $v^2 G$ & $v^4 G$ & $v^6 G$ \\ 
 2PM & $\bullet$ & $G^2$ & $v^2 G^2$ & $v^4 G^2$ \\ 
 3PM & $\bullet$ & $\bullet$ & $G^3$ & $v^2 G^3$ \\ 
\end{tabular}
\end{center}

The $0$PM expansion corresponds to flat space (no gravity), and the $0$PN expansion corresponds to Newton's law of gravity.

\subsection{One-loop contributions to the waveshape}
We now move onto analyzing the causal properties of the one-loop contributions to the waveshape. This subsection can be skipped if one wants to avoid the technical details.

Using tensor reduction and integration by parts, we can reduce the set of integrals needed for the one-loop computation in the eikonal limit into a basis of 16 master integrals, which are subtopologies of a pentagon diagram along with the permutations of where the graviton legs attach. For illustration and simplicity, we focus here on the pentagon topology itself. We  refer to Refs.~\cite{Brandhuber:2023hhy,Herderschee:2023fxh,Elkhidir:2023dco,Georgoudis:2023lgf,Caron-Huot:2023vxl} for the full computation of all master integrals.

The four topologies we consider, which we label with $A$, $B$, $C$ and $D$ are
\begin{equation}
\stackrel{\adjustbox{valign=c,scale={0.75}{0.75}}{\tikzset{every picture/.style={line width=0.75pt}} 
\begin{tikzpicture}[x=0.75pt,y=0.75pt,yscale=-1,xscale=-1]
\draw [line width=2.25]    (301,103) -- (391,103) ;
\draw [line width=2.25]    (301,173) -- (391,173) ;
\draw    (329,173) .. controls (327.35,171.32) and (327.36,169.65) .. (329.04,168) .. controls (330.71,166.34) and (330.72,164.67) .. (329.07,163) .. controls (327.42,161.32) and (327.43,159.65) .. (329.11,158) .. controls (330.78,156.34) and (330.79,154.67) .. (329.14,153) .. controls (327.49,151.32) and (327.5,149.65) .. (329.18,148) .. controls (330.85,146.34) and (330.86,144.67) .. (329.21,143) .. controls (327.56,141.32) and (327.57,139.65) .. (329.25,138) .. controls (330.93,136.35) and (330.94,134.68) .. (329.29,133) .. controls (327.64,131.33) and (327.65,129.66) .. (329.32,128) .. controls (331,126.35) and (331.01,124.68) .. (329.36,123) .. controls (327.71,121.33) and (327.72,119.66) .. (329.39,118) .. controls (331.07,116.35) and (331.08,114.68) .. (329.43,113) .. controls (327.78,111.33) and (327.79,109.66) .. (329.46,108) .. controls (331.14,106.35) and (331.15,104.68) .. (329.5,103) -- (329.5,103) -- (329.5,103) ;
\draw    (365,172) .. controls (363.32,170.35) and (363.31,168.68) .. (364.96,167) .. controls (366.61,165.33) and (366.6,163.66) .. (364.93,162) .. controls (363.25,160.35) and (363.24,158.68) .. (364.89,157) .. controls (366.54,155.33) and (366.53,153.66) .. (364.86,152) .. controls (363.18,150.35) and (363.17,148.68) .. (364.82,147) .. controls (366.47,145.32) and (366.46,143.65) .. (364.78,142) .. controls (363.11,140.34) and (363.1,138.67) .. (364.75,137) .. controls (366.4,135.32) and (366.39,133.65) .. (364.71,132) .. controls (363.03,130.35) and (363.02,128.68) .. (364.67,127) .. controls (366.32,125.33) and (366.31,123.66) .. (364.64,122) .. controls (362.96,120.35) and (362.95,118.68) .. (364.6,117) .. controls (366.25,115.33) and (366.24,113.66) .. (364.57,112) .. controls (362.89,110.35) and (362.88,108.68) .. (364.53,107) -- (364.5,103) -- (364.5,103) ;
\draw    (300,138) .. controls (301.67,136.33) and (303.33,136.33) .. (305,138) .. controls (306.67,139.67) and (308.33,139.67) .. (310,138) .. controls (311.67,136.33) and (313.33,136.33) .. (315,138) .. controls (316.67,139.67) and (318.33,139.67) .. (320,138) .. controls (321.67,136.33) and (323.33,136.33) .. (325,138) -- (329.25,138) -- (329.25,138) ;
\end{tikzpicture}
\tikzset{every picture/.style={line width=1}} }}{A}
\qquad
\stackrel{\adjustbox{valign=c,scale={0.75}{0.75}}{\tikzset{every picture/.style={line width=0.75pt}}
\begin{tikzpicture}[x=0.75pt,y=0.75pt,yscale=-1,xscale=-1]
\draw [line width=2.25]    (301,103) -- (391,103) ;
\draw [line width=2.25]    (301,173) -- (391,173) ;
\draw    (329,173) .. controls (328.27,170.76) and (329.02,169.27) .. (331.26,168.54) .. controls (333.5,167.81) and (334.25,166.32) .. (333.52,164.08) .. controls (332.79,161.84) and (333.54,160.35) .. (335.78,159.62) .. controls (338.02,158.89) and (338.78,157.4) .. (338.05,155.16) .. controls (337.32,152.92) and (338.07,151.43) .. (340.31,150.7) .. controls (342.55,149.97) and (343.3,148.48) .. (342.57,146.24) .. controls (341.84,144) and (342.59,142.51) .. (344.83,141.78) .. controls (347.07,141.05) and (347.82,139.57) .. (347.09,137.33) .. controls (346.36,135.09) and (347.11,133.6) .. (349.35,132.87) .. controls (351.59,132.14) and (352.35,130.65) .. (351.62,128.41) .. controls (350.89,126.17) and (351.64,124.68) .. (353.88,123.95) .. controls (356.12,123.22) and (356.87,121.73) .. (356.14,119.49) .. controls (355.41,117.25) and (356.16,115.76) .. (358.4,115.03) .. controls (360.64,114.3) and (361.39,112.81) .. (360.66,110.57) .. controls (359.93,108.33) and (360.68,106.84) .. (362.92,106.11) -- (364.5,103) -- (364.5,103) ;
\draw    (365,172) .. controls (362.71,171.45) and (361.83,170.03) .. (362.38,167.74) .. controls (362.93,165.45) and (362.05,164.03) .. (359.76,163.48) .. controls (357.47,162.94) and (356.59,161.52) .. (357.14,159.23) .. controls (357.69,156.94) and (356.81,155.52) .. (354.52,154.97) .. controls (352.23,154.42) and (351.35,153) .. (351.9,150.71) .. controls (352.45,148.42) and (351.57,147) .. (349.28,146.45) -- (349,146) -- (349,146) ;
\draw    (322,103) .. controls (324.29,103.53) and (325.17,104.95) .. (324.64,107.24) .. controls (324.11,109.54) and (324.99,110.96) .. (327.29,111.49) .. controls (329.58,112.02) and (330.46,113.44) .. (329.93,115.73) .. controls (329.4,118.02) and (330.29,119.44) .. (332.58,119.97) .. controls (334.87,120.51) and (335.75,121.93) .. (335.22,124.22) .. controls (334.69,126.51) and (335.58,127.93) .. (337.87,128.46) .. controls (340.16,128.99) and (341.04,130.41) .. (340.51,132.7) .. controls (339.98,135) and (340.86,136.42) .. (343.16,136.95) -- (343.5,137.5) -- (343.5,137.5) ;
\begin{scope}[xshift=-3,yshift=10]
\draw    (342,145) .. controls (340.28,146.61) and (338.61,146.56) .. (337,144.84) .. controls (335.38,143.13) and (333.71,143.08) .. (332,144.69) .. controls (330.28,146.3) and (328.62,146.25) .. (327.01,144.53) .. controls (325.39,142.82) and (323.72,142.77) .. (322.01,144.38) .. controls (320.29,145.99) and (318.62,145.94) .. (317.01,144.22) .. controls (315.4,142.5) and (313.73,142.45) .. (312.01,144.06) -- (310,144) -- (310,144) ;
\end{scope}
\end{tikzpicture}}}{B}
\qquad
\stackrel{\adjustbox{valign=c,scale={0.75}{0.75}}{\tikzset{every picture/.style={line width=0.75pt}}   
\begin{tikzpicture}[x=0.75pt,y=0.75pt,yscale=-1,xscale=-1]
\draw [line width=2.25]    (301,103) -- (391,103) ;
\draw [line width=2.25]    (301,173) -- (391,173) ;
\draw    (329,173) .. controls (328.27,170.76) and (329.02,169.27) .. (331.26,168.54) .. controls (333.5,167.81) and (334.25,166.32) .. (333.52,164.08) .. controls (332.79,161.84) and (333.54,160.35) .. (335.78,159.62) .. controls (338.02,158.89) and (338.78,157.4) .. (338.05,155.16) .. controls (337.32,152.92) and (338.07,151.43) .. (340.31,150.7) .. controls (342.55,149.97) and (343.3,148.48) .. (342.57,146.24) .. controls (341.84,144) and (342.59,142.51) .. (344.83,141.78) .. controls (347.07,141.05) and (347.82,139.57) .. (347.09,137.33) .. controls (346.36,135.09) and (347.11,133.6) .. (349.35,132.87) .. controls (351.59,132.14) and (352.35,130.65) .. (351.62,128.41) .. controls (350.89,126.17) and (351.64,124.68) .. (353.88,123.95) .. controls (356.12,123.22) and (356.87,121.73) .. (356.14,119.49) .. controls (355.41,117.25) and (356.16,115.76) .. (358.4,115.03) .. controls (360.64,114.3) and (361.39,112.81) .. (360.66,110.57) .. controls (359.93,108.33) and (360.68,106.84) .. (362.92,106.11) -- (364.5,103) -- (364.5,103) ;
\draw    (365,172) .. controls (362.71,171.45) and (361.83,170.03) .. (362.38,167.74) .. controls (362.93,165.45) and (362.05,164.03) .. (359.76,163.48) .. controls (357.47,162.94) and (356.59,161.52) .. (357.14,159.23) .. controls (357.69,156.94) and (356.81,155.52) .. (354.52,154.97) .. controls (352.23,154.42) and (351.35,153) .. (351.9,150.71) .. controls (352.45,148.42) and (351.57,147) .. (349.28,146.45) -- (349,146) -- (349,146) ;
\draw    (322,103) .. controls (324.29,103.53) and (325.17,104.95) .. (324.64,107.24) .. controls (324.11,109.54) and (324.99,110.96) .. (327.29,111.49) .. controls (329.58,112.02) and (330.46,113.44) .. (329.93,115.73) .. controls (329.4,118.02) and (330.29,119.44) .. (332.58,119.97) .. controls (334.87,120.51) and (335.75,121.93) .. (335.22,124.22) .. controls (334.69,126.51) and (335.58,127.93) .. (337.87,128.46) .. controls (340.16,128.99) and (341.04,130.41) .. (340.51,132.7) .. controls (339.98,135) and (340.86,136.42) .. (343.16,136.95) -- (343.5,137.5) -- (343.5,137.5) ;
\draw    (332.75,120.25) .. controls (331.03,121.86) and (329.36,121.81) .. (327.75,120.09) .. controls (326.13,118.38) and (324.46,118.33) .. (322.75,119.94) .. controls (321.03,121.55) and (319.37,121.5) .. (317.76,119.78) .. controls (316.14,118.07) and (314.47,118.02) .. (312.76,119.63) .. controls (311.04,121.24) and (309.37,121.19) .. (307.76,119.47) .. controls (306.15,117.75) and (304.48,117.7) .. (302.76,119.31) -- (300.75,119.25) -- (300.75,119.25) ;
\end{tikzpicture}}}{C}
\qquad
\stackrel{\adjustbox{valign=c,scale={0.75}{0.75}}{\tikzset{every picture/.style={line width=0.75pt}}      
\begin{tikzpicture}[x=0.75pt,y=0.75pt,yscale=-1,xscale=-1]
\draw [line width=2.25]    (301,103) -- (391,103) ;
\draw [line width=2.25]    (301,173) -- (391,173) ;
\draw    (329,173) .. controls (327.35,171.32) and (327.36,169.65) .. (329.04,168) .. controls (330.71,166.34) and (330.72,164.67) .. (329.07,163) .. controls (327.42,161.32) and (327.43,159.65) .. (329.11,158) .. controls (330.78,156.34) and (330.79,154.67) .. (329.14,153) .. controls (327.49,151.32) and (327.5,149.65) .. (329.18,148) .. controls (330.85,146.34) and (330.86,144.67) .. (329.21,143) .. controls (327.56,141.32) and (327.57,139.65) .. (329.25,138) .. controls (330.93,136.35) and (330.94,134.68) .. (329.29,133) .. controls (327.64,131.33) and (327.65,129.66) .. (329.32,128) .. controls (331,126.35) and (331.01,124.68) .. (329.36,123) .. controls (327.71,121.33) and (327.72,119.66) .. (329.39,118) .. controls (331.07,116.35) and (331.08,114.68) .. (329.43,113) .. controls (327.78,111.33) and (327.79,109.66) .. (329.46,108) .. controls (331.14,106.35) and (331.15,104.68) .. (329.5,103) -- (329.5,103) -- (329.5,103) ;
\draw    (365,172) .. controls (363.32,170.35) and (363.31,168.68) .. (364.96,167) .. controls (366.61,165.33) and (366.6,163.66) .. (364.93,162) .. controls (363.25,160.35) and (363.24,158.68) .. (364.89,157) .. controls (366.54,155.33) and (366.53,153.66) .. (364.86,152) .. controls (363.18,150.35) and (363.17,148.68) .. (364.82,147) .. controls (366.47,145.32) and (366.46,143.65) .. (364.78,142) .. controls (363.11,140.34) and (363.1,138.67) .. (364.75,137) .. controls (366.4,135.32) and (366.39,133.65) .. (364.71,132) .. controls (363.03,130.35) and (363.02,128.68) .. (364.67,127) .. controls (366.32,125.33) and (366.31,123.66) .. (364.64,122) .. controls (362.96,120.35) and (362.95,118.68) .. (364.6,117) .. controls (366.25,115.33) and (366.24,113.66) .. (364.57,112) .. controls (362.89,110.35) and (362.88,108.68) .. (364.53,107) -- (364.5,103) -- (364.5,103) ;
\draw    (334,138) .. controls (335.64,136.31) and (337.31,136.28) .. (339,137.92) .. controls (340.69,139.56) and (342.36,139.53) .. (344,137.84) .. controls (345.64,136.15) and (347.31,136.12) .. (349,137.76) .. controls (350.69,139.39) and (352.36,139.36) .. (354,137.67) .. controls (355.64,135.98) and (357.31,135.95) .. (359,137.59) .. controls (360.69,139.23) and (362.36,139.2) .. (364,137.51) -- (364.75,137.5) -- (364.75,137.5) ;
\draw    (295,138) .. controls (296.64,136.31) and (298.31,136.28) .. (300,137.92) .. controls (301.69,139.56) and (303.36,139.53) .. (305,137.84) .. controls (306.64,136.15) and (308.31,136.12) .. (310,137.76) .. controls (311.69,139.39) and (313.36,139.36) .. (315,137.67) .. controls (316.64,135.98) and (318.31,135.95) .. (320,137.59) .. controls (321.69,139.23) and (323.36,139.2) .. (325,137.51);
\end{tikzpicture}}}{D}
\label{eq:topologies}
\end{equation}
Using the following momentum labeling for the internal edges
\be 
\adjustbox{valign=c,scale={1}{1}}{\tikzset{every picture/.style={line width=0.75pt}} 
\begin{tikzpicture}[x=0.75pt,y=0.75pt,yscale=-1,xscale=-1]
\draw [line width=2.25]    (301,103) -- (391,103) node[midway,above]{$\ell+\Bar{p}_1+\frac{q_1}{2}$};
\draw[<-] [line width=1.25]    (341,103) -- (351,103); 
\draw [line width=2.25]    (301,173) -- (391,173) node[midway,below]{$-\ell+\Bar{p}_2+\frac{q_2}{2}$};
\draw[<-] [line width=1.25]    (341,173) -- (351,173); 
\draw[->] [line width=0.25]    (319,110) -- (319,133) node[right,midway]{$\ell+q_1$}; 
\draw[->] [line width=0.25]    (319,143) -- (319,166) node[right,midway]{$\ell-q_2$};
\draw[<-] [line width=0.25]    (372,123) -- (372,146) node[left,midway]{$\ell$}; 
\draw    (329,173) .. controls (327.35,171.32) and (327.36,169.65) .. (329.04,168) .. controls (330.71,166.34) and (330.72,164.67) .. (329.07,163) .. controls (327.42,161.32) and (327.43,159.65) .. (329.11,158) .. controls (330.78,156.34) and (330.79,154.67) .. (329.14,153) .. controls (327.49,151.32) and (327.5,149.65) .. (329.18,148) .. controls (330.85,146.34) and (330.86,144.67) .. (329.21,143) .. controls (327.56,141.32) and (327.57,139.65) .. (329.25,138) .. controls (330.93,136.35) and (330.94,134.68) .. (329.29,133) .. controls (327.64,131.33) and (327.65,129.66) .. (329.32,128) .. controls (331,126.35) and (331.01,124.68) .. (329.36,123) .. controls (327.71,121.33) and (327.72,119.66) .. (329.39,118) .. controls (331.07,116.35) and (331.08,114.68) .. (329.43,113) .. controls (327.78,111.33) and (327.79,109.66) .. (329.46,108) .. controls (331.14,106.35) and (331.15,104.68) .. (329.5,103) -- (329.5,103) -- (329.5,103);
\draw    (365,172) .. controls (363.32,170.35) and (363.31,168.68) .. (364.96,167) .. controls (366.61,165.33) and (366.6,163.66) .. (364.93,162) .. controls (363.25,160.35) and (363.24,158.68) .. (364.89,157) .. controls (366.54,155.33) and (366.53,153.66) .. (364.86,152) .. controls (363.18,150.35) and (363.17,148.68) .. (364.82,147) .. controls (366.47,145.32) and (366.46,143.65) .. (364.78,142) .. controls (363.11,140.34) and (363.1,138.67) .. (364.75,137) .. controls (366.4,135.32) and (366.39,133.65) .. (364.71,132) .. controls (363.03,130.35) and (363.02,128.68) .. (364.67,127) .. controls (366.32,125.33) and (366.31,123.66) .. (364.64,122) .. controls (362.96,120.35) and (362.95,118.68) .. (364.6,117) .. controls (366.25,115.33) and (366.24,113.66) .. (364.57,112) .. controls (362.89,110.35) and (362.88,108.68) .. (364.53,107) -- (364.5,103) -- (364.5,103) ;
\draw    (300,138) .. controls (301.67,136.33) and (303.33,136.33) .. (305,138) .. controls (306.67,139.67) and (308.33,139.67) .. (310,138) .. controls (311.67,136.33) and (313.33,136.33) .. (315,138) .. controls (316.67,139.67) and (318.33,139.67) .. (320,138) .. controls (321.67,136.33) and (323.33,136.33) .. (325,138) -- (329.25,138) -- (329.25,138) ;
\end{tikzpicture}
\tikzset{every picture/.style={line width=1}} }
\label{eq:I_labels}
\ee
the master integrals all belong to the following family of integrals,
\begin{equation} \label{eik topology}
 I_{\mathcal S_i} = \frac{\e^{\epsilon\gamma_{\text{E}}}}{\bar{m}_1 \bar{m}_2}\int \frac{\d^\rD\ell}{i\pi^{\rD/2}}
 \frac{1}{[\ell^2]^{a_1}[2\ell{\cdot}\barv_1]^{a_2} [(\ell{+}q_1)^2]^{a_3}[(\ell{-}q_2)^2]^{a_4}
 [-2\ell{\cdot}\barv_2]^{a_5}}\,.
\end{equation}
where $\mathcal{S}_i = \{a_1,a_2,a_3,a_4,a_5\}$ labels a set of indices. For the pentagon diagram itself, we take $a_i=1$ for all $i$.
To write this integral family, we have defined the velocities $\bar{v}_i = \bar{p}_i/\bar{m}_i$ and used the expansion of the propagators appearing in the diagram from~\eqref{eq:I_labels} in the eikonal limit, e.g.,
\be \label{series one propagator}
 \frac{1}{[({-}\ell{+}\bar{p}_2{+}\frac{1}{2}q_2)^2-m_2^2]}  =
 \frac{1}{\bar{m}_2} \frac{1}{[-2\ell{\cdot}\barv_2]}
- \frac{1}{\bar{m}_2^2} \frac{\ell{\cdot}(\ell-q_2)}
{[2\ell{\cdot}\barv_2]^2}+\ldots \,.
\ee
We deliberately used square brackets to emphasize that the eikonal expansion holds for any $i \varepsilon$ prescription, which can be inserted into the brackets.

To compute the waveshape $\text{Exp}_k$, we have to sum over the contributions corresponding to diagrams $A$, $B$, $C$ and $D$ from~\eqref{eq:topologies} (including the relevant $+i\varepsilon$'s in all propagators), and, additionally, the cut of diagram $D$. A cut through the massive particles in the $D$ topology allows for the graviton to be emitted before the cut, so this term must be included by~\eqref{eq:Exp_k-pics}. In the eikonal limit, the on-shell delta function of the cut propagator in topology $D$ admits the following expansion,
\begin{align}
    \delta\Big[(\ell{+}\bar{p}_2{-}\tfrac{1}{2}q_2)^2{-}m_2^2\Big] & = 
    \frac{ \delta \left(2 \ell \cdot \barv_2 \right) }{\bar{m}_2} + \frac{\ell{\cdot}(\ell{-}q_2)}{\bar{m}_2^2} \delta' \left(2 \ell \cdot \barv_2 \right) + \mathcal{O}\left(\bar{m}_2^{-3}\right)\,,
    \label{eq:delta-expansion}
\end{align}
which is the analog of~\eqref{series one propagator} for the propagators.

Looking at the contribution to the master integral~\eqref{eik topology} with $a_i=1$ for all $i$, we get the following expansions corresponding to the pentagon-topology diagrams to first subleading order in $\frac{1}{\bar{m}_2}$,
\begin{subequations}\begin{align}
{I^A_{\mathcal S_1} \equiv }
\adjustbox{valign=c,scale={0.65}{0.65}}{\tikzset{every picture/.style={line width=0.75pt}} 
\begin{tikzpicture}[x=0.75pt,y=0.75pt,yscale=-1,xscale=-1]
\draw [line width=2.25]    (301,103) -- (391,103) ;
\draw [line width=2.25]    (301,173) -- (391,173) ;
\draw    (329,173) .. controls (327.35,171.32) and (327.36,169.65) .. (329.04,168) .. controls (330.71,166.34) and (330.72,164.67) .. (329.07,163) .. controls (327.42,161.32) and (327.43,159.65) .. (329.11,158) .. controls (330.78,156.34) and (330.79,154.67) .. (329.14,153) .. controls (327.49,151.32) and (327.5,149.65) .. (329.18,148) .. controls (330.85,146.34) and (330.86,144.67) .. (329.21,143) .. controls (327.56,141.32) and (327.57,139.65) .. (329.25,138) .. controls (330.93,136.35) and (330.94,134.68) .. (329.29,133) .. controls (327.64,131.33) and (327.65,129.66) .. (329.32,128) .. controls (331,126.35) and (331.01,124.68) .. (329.36,123) .. controls (327.71,121.33) and (327.72,119.66) .. (329.39,118) .. controls (331.07,116.35) and (331.08,114.68) .. (329.43,113) .. controls (327.78,111.33) and (327.79,109.66) .. (329.46,108) .. controls (331.14,106.35) and (331.15,104.68) .. (329.5,103) -- (329.5,103) -- (329.5,103) ;
\draw    (365,172) .. controls (363.32,170.35) and (363.31,168.68) .. (364.96,167) .. controls (366.61,165.33) and (366.6,163.66) .. (364.93,162) .. controls (363.25,160.35) and (363.24,158.68) .. (364.89,157) .. controls (366.54,155.33) and (366.53,153.66) .. (364.86,152) .. controls (363.18,150.35) and (363.17,148.68) .. (364.82,147) .. controls (366.47,145.32) and (366.46,143.65) .. (364.78,142) .. controls (363.11,140.34) and (363.1,138.67) .. (364.75,137) .. controls (366.4,135.32) and (366.39,133.65) .. (364.71,132) .. controls (363.03,130.35) and (363.02,128.68) .. (364.67,127) .. controls (366.32,125.33) and (366.31,123.66) .. (364.64,122) .. controls (362.96,120.35) and (362.95,118.68) .. (364.6,117) .. controls (366.25,115.33) and (366.24,113.66) .. (364.57,112) .. controls (362.89,110.35) and (362.88,108.68) .. (364.53,107) -- (364.5,103) -- (364.5,103) ;
\draw    (300,138) .. controls (301.67,136.33) and (303.33,136.33) .. (305,138) .. controls (306.67,139.67) and (308.33,139.67) .. (310,138) .. controls (311.67,136.33) and (313.33,136.33) .. (315,138) .. controls (316.67,139.67) and (318.33,139.67) .. (320,138) .. controls (321.67,136.33) and (323.33,136.33) .. (325,138) -- (329.25,138) -- (329.25,138) ;
\end{tikzpicture}
\tikzset{every picture/.style={line width=1}} }
&  {\approx  \int \d \hat{I}
\frac{1}{(2\ell{\cdot}\barv_1 {+}i\varepsilon)({-}2\ell{\cdot}\barv_2 {+} i\varepsilon)}
\left[1 {-} \medmath{\frac{1}{\bar{m}_2}\frac{\ell{\cdot}(\ell{-}q_2)}{({-}2 \ell \cdot \barv_2 {+} i\varepsilon)}}  \right],}
\\
{I^B_{\mathcal S_1} \equiv }
\adjustbox{valign=c,scale={0.65}{0.65}}{\tikzset{every picture/.style={line width=0.75pt}}
\begin{tikzpicture}[x=0.75pt,y=0.75pt,yscale=-1,xscale=-1]
\draw [line width=2.25]    (301,103) -- (391,103) ;
\draw [line width=2.25]    (301,173) -- (391,173) ;
\draw    (329,173) .. controls (328.27,170.76) and (329.02,169.27) .. (331.26,168.54) .. controls (333.5,167.81) and (334.25,166.32) .. (333.52,164.08) .. controls (332.79,161.84) and (333.54,160.35) .. (335.78,159.62) .. controls (338.02,158.89) and (338.78,157.4) .. (338.05,155.16) .. controls (337.32,152.92) and (338.07,151.43) .. (340.31,150.7) .. controls (342.55,149.97) and (343.3,148.48) .. (342.57,146.24) .. controls (341.84,144) and (342.59,142.51) .. (344.83,141.78) .. controls (347.07,141.05) and (347.82,139.57) .. (347.09,137.33) .. controls (346.36,135.09) and (347.11,133.6) .. (349.35,132.87) .. controls (351.59,132.14) and (352.35,130.65) .. (351.62,128.41) .. controls (350.89,126.17) and (351.64,124.68) .. (353.88,123.95) .. controls (356.12,123.22) and (356.87,121.73) .. (356.14,119.49) .. controls (355.41,117.25) and (356.16,115.76) .. (358.4,115.03) .. controls (360.64,114.3) and (361.39,112.81) .. (360.66,110.57) .. controls (359.93,108.33) and (360.68,106.84) .. (362.92,106.11) -- (364.5,103) -- (364.5,103) ;
\draw    (365,172) .. controls (362.71,171.45) and (361.83,170.03) .. (362.38,167.74) .. controls (362.93,165.45) and (362.05,164.03) .. (359.76,163.48) .. controls (357.47,162.94) and (356.59,161.52) .. (357.14,159.23) .. controls (357.69,156.94) and (356.81,155.52) .. (354.52,154.97) .. controls (352.23,154.42) and (351.35,153) .. (351.9,150.71) .. controls (352.45,148.42) and (351.57,147) .. (349.28,146.45) -- (349,146) -- (349,146) ;
\draw    (322,103) .. controls (324.29,103.53) and (325.17,104.95) .. (324.64,107.24) .. controls (324.11,109.54) and (324.99,110.96) .. (327.29,111.49) .. controls (329.58,112.02) and (330.46,113.44) .. (329.93,115.73) .. controls (329.4,118.02) and (330.29,119.44) .. (332.58,119.97) .. controls (334.87,120.51) and (335.75,121.93) .. (335.22,124.22) .. controls (334.69,126.51) and (335.58,127.93) .. (337.87,128.46) .. controls (340.16,128.99) and (341.04,130.41) .. (340.51,132.7) .. controls (339.98,135) and (340.86,136.42) .. (343.16,136.95) -- (343.5,137.5) -- (343.5,137.5) ;
\begin{scope}[xshift=-3,yshift=10]
\draw    (342,145) .. controls (340.28,146.61) and (338.61,146.56) .. (337,144.84) .. controls (335.38,143.13) and (333.71,143.08) .. (332,144.69) .. controls (330.28,146.3) and (328.62,146.25) .. (327.01,144.53) .. controls (325.39,142.82) and (323.72,142.77) .. (322.01,144.38) .. controls (320.29,145.99) and (318.62,145.94) .. (317.01,144.22) .. controls (315.4,142.5) and (313.73,142.45) .. (312.01,144.06) -- (310,144) -- (310,144) ;
\end{scope}
\end{tikzpicture}}
&  {\approx \int \d \hat{I}
\frac{1}{({-}2\ell{\cdot}\barv_1 {+}i\varepsilon)({-}2\ell{\cdot}\barv_2 {+} i\varepsilon)}\!
\left[1 {-} \medmath{\frac{1}{\bar{m}_2}\frac{\ell{\cdot}(\ell{-}q_2)}{({-}2 \ell \cdot \barv_2 {+} i\varepsilon)}}  \right]\!,}
\\
{I^C_{\mathcal S_1} \equiv } \adjustbox{valign=c,scale={0.65}{0.65}}{\tikzset{every picture/.style={line width=0.75pt}}   
\begin{tikzpicture}[x=0.75pt,y=0.75pt,yscale=-1,xscale=-1]
\draw [line width=2.25]    (301,103) -- (391,103) ;
\draw [line width=2.25]    (301,173) -- (391,173) ;
\draw    (329,173) .. controls (328.27,170.76) and (329.02,169.27) .. (331.26,168.54) .. controls (333.5,167.81) and (334.25,166.32) .. (333.52,164.08) .. controls (332.79,161.84) and (333.54,160.35) .. (335.78,159.62) .. controls (338.02,158.89) and (338.78,157.4) .. (338.05,155.16) .. controls (337.32,152.92) and (338.07,151.43) .. (340.31,150.7) .. controls (342.55,149.97) and (343.3,148.48) .. (342.57,146.24) .. controls (341.84,144) and (342.59,142.51) .. (344.83,141.78) .. controls (347.07,141.05) and (347.82,139.57) .. (347.09,137.33) .. controls (346.36,135.09) and (347.11,133.6) .. (349.35,132.87) .. controls (351.59,132.14) and (352.35,130.65) .. (351.62,128.41) .. controls (350.89,126.17) and (351.64,124.68) .. (353.88,123.95) .. controls (356.12,123.22) and (356.87,121.73) .. (356.14,119.49) .. controls (355.41,117.25) and (356.16,115.76) .. (358.4,115.03) .. controls (360.64,114.3) and (361.39,112.81) .. (360.66,110.57) .. controls (359.93,108.33) and (360.68,106.84) .. (362.92,106.11) -- (364.5,103) -- (364.5,103) ;
\draw    (365,172) .. controls (362.71,171.45) and (361.83,170.03) .. (362.38,167.74) .. controls (362.93,165.45) and (362.05,164.03) .. (359.76,163.48) .. controls (357.47,162.94) and (356.59,161.52) .. (357.14,159.23) .. controls (357.69,156.94) and (356.81,155.52) .. (354.52,154.97) .. controls (352.23,154.42) and (351.35,153) .. (351.9,150.71) .. controls (352.45,148.42) and (351.57,147) .. (349.28,146.45) -- (349,146) -- (349,146) ;
\draw    (322,103) .. controls (324.29,103.53) and (325.17,104.95) .. (324.64,107.24) .. controls (324.11,109.54) and (324.99,110.96) .. (327.29,111.49) .. controls (329.58,112.02) and (330.46,113.44) .. (329.93,115.73) .. controls (329.4,118.02) and (330.29,119.44) .. (332.58,119.97) .. controls (334.87,120.51) and (335.75,121.93) .. (335.22,124.22) .. controls (334.69,126.51) and (335.58,127.93) .. (337.87,128.46) .. controls (340.16,128.99) and (341.04,130.41) .. (340.51,132.7) .. controls (339.98,135) and (340.86,136.42) .. (343.16,136.95) -- (343.5,137.5) -- (343.5,137.5) ;
\draw    (332.75,120.25) .. controls (331.03,121.86) and (329.36,121.81) .. (327.75,120.09) .. controls (326.13,118.38) and (324.46,118.33) .. (322.75,119.94) .. controls (321.03,121.55) and (319.37,121.5) .. (317.76,119.78) .. controls (316.14,118.07) and (314.47,118.02) .. (312.76,119.63) .. controls (311.04,121.24) and (309.37,121.19) .. (307.76,119.47) .. controls (306.15,117.75) and (304.48,117.7) .. (302.76,119.31) -- (300.75,119.25) -- (300.75,119.25) ;
\end{tikzpicture}}
& {\approx  \int \d \hat{I}
\frac{1}{(2\ell{\cdot}\barv_1 {+}i\varepsilon)(2\ell{\cdot}\barv_2 {+} i\varepsilon)}
\left[1 {-} \medmath{\frac{1}{\bar{m}_2}\frac{\ell{\cdot}(\ell-q_2)}{(2 \ell \cdot \barv_2+ i\varepsilon)}}  \right],}
\\
{I^D_{\mathcal S_1} \equiv }
\adjustbox{valign=c,scale={0.65}{0.65}}{\tikzset{every picture/.style={line width=0.75pt}}      
\begin{tikzpicture}[x=0.75pt,y=0.75pt,yscale=-1,xscale=-1]
\draw [line width=2.25]    (301,103) -- (391,103) ;
\draw [line width=2.25]    (301,173) -- (391,173) ;
\draw    (329,173) .. controls (327.35,171.32) and (327.36,169.65) .. (329.04,168) .. controls (330.71,166.34) and (330.72,164.67) .. (329.07,163) .. controls (327.42,161.32) and (327.43,159.65) .. (329.11,158) .. controls (330.78,156.34) and (330.79,154.67) .. (329.14,153) .. controls (327.49,151.32) and (327.5,149.65) .. (329.18,148) .. controls (330.85,146.34) and (330.86,144.67) .. (329.21,143) .. controls (327.56,141.32) and (327.57,139.65) .. (329.25,138) .. controls (330.93,136.35) and (330.94,134.68) .. (329.29,133) .. controls (327.64,131.33) and (327.65,129.66) .. (329.32,128) .. controls (331,126.35) and (331.01,124.68) .. (329.36,123) .. controls (327.71,121.33) and (327.72,119.66) .. (329.39,118) .. controls (331.07,116.35) and (331.08,114.68) .. (329.43,113) .. controls (327.78,111.33) and (327.79,109.66) .. (329.46,108) .. controls (331.14,106.35) and (331.15,104.68) .. (329.5,103) -- (329.5,103) -- (329.5,103) ;
\draw    (365,172) .. controls (363.32,170.35) and (363.31,168.68) .. (364.96,167) .. controls (366.61,165.33) and (366.6,163.66) .. (364.93,162) .. controls (363.25,160.35) and (363.24,158.68) .. (364.89,157) .. controls (366.54,155.33) and (366.53,153.66) .. (364.86,152) .. controls (363.18,150.35) and (363.17,148.68) .. (364.82,147) .. controls (366.47,145.32) and (366.46,143.65) .. (364.78,142) .. controls (363.11,140.34) and (363.1,138.67) .. (364.75,137) .. controls (366.4,135.32) and (366.39,133.65) .. (364.71,132) .. controls (363.03,130.35) and (363.02,128.68) .. (364.67,127) .. controls (366.32,125.33) and (366.31,123.66) .. (364.64,122) .. controls (362.96,120.35) and (362.95,118.68) .. (364.6,117) .. controls (366.25,115.33) and (366.24,113.66) .. (364.57,112) .. controls (362.89,110.35) and (362.88,108.68) .. (364.53,107) -- (364.5,103) -- (364.5,103) ;
\draw    (334,138) .. controls (335.64,136.31) and (337.31,136.28) .. (339,137.92) .. controls (340.69,139.56) and (342.36,139.53) .. (344,137.84) .. controls (345.64,136.15) and (347.31,136.12) .. (349,137.76) .. controls (350.69,139.39) and (352.36,139.36) .. (354,137.67) .. controls (355.64,135.98) and (357.31,135.95) .. (359,137.59) .. controls (360.69,139.23) and (362.36,139.2) .. (364,137.51) -- (364.75,137.5) -- (364.75,137.5) ;
\draw    (295,138) .. controls (296.64,136.31) and (298.31,136.28) .. (300,137.92) .. controls (301.69,139.56) and (303.36,139.53) .. (305,137.84) .. controls (306.64,136.15) and (308.31,136.12) .. (310,137.76) .. controls (311.69,139.39) and (313.36,139.36) .. (315,137.67) .. controls (316.64,135.98) and (318.31,135.95) .. (320,137.59) .. controls (321.69,139.23) and (323.36,139.2) .. (325,137.51);
\end{tikzpicture}}
&  {\approx  \int \d \hat{I}
\frac{1}{({-}2\ell{\cdot}\barv_1 {+}i\varepsilon)(2\ell{\cdot}\barv_2 {+} i\varepsilon)}
\left[1 {-} \medmath{\frac{1}{\bar{m}_2}\frac{\ell{\cdot}(\ell-q_2)}{(2 \ell \cdot \barv_2 + i\varepsilon)}} \right],} \\
{{\rm Cut} I^D_{\mathcal S_1} \equiv } \adjustbox{valign=c,scale={0.65}{0.65}}{\tikzset{every picture/.style={line width=0.75pt}}      
\begin{tikzpicture}[x=0.75pt,y=0.75pt,yscale=-1,xscale=-1]
\draw [line width=2.25]    (301,103) -- (391,103) ;
\draw [line width=2.25]    (301,173) -- (391,173) ;
\draw    (329,173) .. controls (327.35,171.32) and (327.36,169.65) .. (329.04,168) .. controls (330.71,166.34) and (330.72,164.67) .. (329.07,163) .. controls (327.42,161.32) and (327.43,159.65) .. (329.11,158) .. controls (330.78,156.34) and (330.79,154.67) .. (329.14,153) .. controls (327.49,151.32) and (327.5,149.65) .. (329.18,148) .. controls (330.85,146.34) and (330.86,144.67) .. (329.21,143) .. controls (327.56,141.32) and (327.57,139.65) .. (329.25,138) .. controls (330.93,136.35) and (330.94,134.68) .. (329.29,133) .. controls (327.64,131.33) and (327.65,129.66) .. (329.32,128) .. controls (331,126.35) and (331.01,124.68) .. (329.36,123) .. controls (327.71,121.33) and (327.72,119.66) .. (329.39,118) .. controls (331.07,116.35) and (331.08,114.68) .. (329.43,113) .. controls (327.78,111.33) and (327.79,109.66) .. (329.46,108) .. controls (331.14,106.35) and (331.15,104.68) .. (329.5,103) -- (329.5,103) -- (329.5,103) ;
\draw    (365,172) .. controls (363.32,170.35) and (363.31,168.68) .. (364.96,167) .. controls (366.61,165.33) and (366.6,163.66) .. (364.93,162) .. controls (363.25,160.35) and (363.24,158.68) .. (364.89,157) .. controls (366.54,155.33) and (366.53,153.66) .. (364.86,152) .. controls (363.18,150.35) and (363.17,148.68) .. (364.82,147) .. controls (366.47,145.32) and (366.46,143.65) .. (364.78,142) .. controls (363.11,140.34) and (363.1,138.67) .. (364.75,137) .. controls (366.4,135.32) and (366.39,133.65) .. (364.71,132) .. controls (363.03,130.35) and (363.02,128.68) .. (364.67,127) .. controls (366.32,125.33) and (366.31,123.66) .. (364.64,122) .. controls (362.96,120.35) and (362.95,118.68) .. (364.6,117) .. controls (366.25,115.33) and (366.24,113.66) .. (364.57,112) .. controls (362.89,110.35) and (362.88,108.68) .. (364.53,107) -- (364.5,103) -- (364.5,103) ;
\draw    (334,138) .. controls (335.64,136.31) and (337.31,136.28) .. (339,137.92) .. controls (340.69,139.56) and (342.36,139.53) .. (344,137.84) .. controls (345.64,136.15) and (347.31,136.12) .. (349,137.76) .. controls (350.69,139.39) and (352.36,139.36) .. (354,137.67) .. controls (355.64,135.98) and (357.31,135.95) .. (359,137.59) .. controls (360.69,139.23) and (362.36,139.2) .. (364,137.51) -- (364.75,137.5) -- (364.75,137.5) ;
\draw    (295,138) .. controls (296.64,136.31) and (298.31,136.28) .. (300,137.92) .. controls (301.69,139.56) and (303.36,139.53) .. (305,137.84) .. controls (306.64,136.15) and (308.31,136.12) .. (310,137.76) .. controls (311.69,139.39) and (313.36,139.36) .. (315,137.67) .. controls (316.64,135.98) and (318.31,135.95) .. (320,137.59) .. controls (321.69,139.23) and (323.36,139.2) .. (325,137.51); 
\draw [color=Orange,very thick, dashed]    (347,97) -- (347,180) ;
\end{tikzpicture}}
& \approx 4 \pi^2 \int \d \hat{I} \, 
    \delta(2 \ell \cdot \barv_1) \Big[ \delta (2 \ell \cdot \barv_2 ) + \medmath{\frac{\ell{\cdot}(\ell-q_2)}{\bar{m}_2}} \delta' (2 \ell \cdot \barv_2) \Big],
    \label{eq:CutGDeik}
\end{align} \label{eq:GvsGeik}\end{subequations}
where we have labelled $\mathcal{S}_1 \equiv \{1,1,1,1,1\}$, and used a short-hand notation for the phase space and the graviton propagators,
\begin{equation}
    \int \d \hat{I} \equiv
    \frac{\e^{\epsilon \gamma_{\mathrm{E}}}}{\bar{m}_1 \bar{m}_2}
    \int \frac{\d^\rD\ell}{i\pi^{\rD/2} [\ell^2]^{a_1} [(\ell{+}q_1)^2]^{a_3} [(\ell{-}q_2)^2]^{a_4}} \,.
    \label{eq:elldenominator}
\end{equation}
Using the distributional identity $\frac{1}{x+i \varepsilon} - \frac{1}{x-i\varepsilon} = -2 \pi i \delta(x)$, we see that when adding these contributions (i.e.\ computing $I^A_{\mathcal S_1}+I^B_{\mathcal S_1}+I^C_{\mathcal S_1}+I^D_{\mathcal S_1}+{\rm Cut} I^D_{\mathcal S_1}$), the leading term in $\frac{1}{\bar{m}_2}$ cancels.

To analyze the subleading terms, let us take a step back and analyze more generally different ways in which these diagrams could be added or subtracted to contribute to the waveform. First, note that the topologies $B$, $C$ and $D$ are permutations of $A$, obtained by taking $\bar{p}_i \to - \bar{p}_i$ in different ways. Generally, we can write,
\be\begin{aligned}
    \text{Exp}_k & = \sum_{i} \Big[c_i(\barp_1,\barp_2) I^A_{\mathcal{S}_i} + c_i(-\barp_1,\barp_2) I^B_{\mathcal{S}_i}  \\ & \hspace{2cm}
    + c_i(\barp_1,-\barp_2) I^C_{\mathcal{S}_i} + c_i(-\barp_1,- \barp_2) \left(  I^D_{\mathcal{S}_i} + \text{Cut} \, I^D_{\mathcal{S}_i} \right)
    \Big] \,, \label{Expk 5 terms}
\end{aligned}\ee
where each $\mathcal{S}_i$ is a set of indices, and the sum is over all master integrals needed to compute $\text{Exp}_k$. The numerator factors $c_i$ are those obtained by tensor reductions and integration by parts.
Recall that the numerators come from vertices between the heavy black holes and the gravitons,
\begin{equation}
    \begin{tikzpicture}[scale=0.6,thick,
    baseline={([yshift=-0.4ex]current bounding box.center)},photon/.style={decorate, decoration={snake, amplitude=1pt, segment length=6pt}}]
    \draw[line width=2] (-1,-1)--(1,-1);
    \draw[photon] (0,-1)--(0,0);
    \end{tikzpicture}
    \propto
    \sqrt{G} \bar{m}_i^2 \,,
    \label{eq:threeptgrav}
\end{equation}
so the leading term in the classical expansion of the numerator of the pentagon diagrams will be of order $\bar{m}_1^4\bar{m}_2^4$. The first subleading order will come from numerators that are proportional to either $\bar{m}_1^3\bar{m}_2^4$ or $\bar{m}_1^4\bar{m}_2^3$.

The numerator terms that are proportional to $\bar{m}_1^4\bar{m}_2^4$ is symmetric in $\bar{p}_i \to - \bar{p}_i$ for $i=1,2$, meaning that, according to~\eqref{Expk 5 terms}, the five terms from~\eqref{eq:GvsGeik} must be added up to form the waveshape. As noted before, we see that the first terms in the square brackets all cancel each other, i.e., the term proportional to $\bar{m}_1^4\bar{m}_2^4$ vanishes.
According to the power counting, these leading terms that cancel are precisely the superclassical contributions that scale as $\frac{1}{\hbar}$ around $\rD=4$ spacetime dimensions. However, this symmetric contribution also contributes to the next order,
\begin{equation}
        \label{Geik: sub}
\bar{m}_1^4\bar{m}_2^4 (I^A {+} I^B {+} I^C {+} I^D {+} \text{Cut} \, I^D) 
= \bar{m}_1^4\bar{m}_2^3 \int \d \hat{I} \, \delta (2 \ell \cdot \barv_1) 
\frac{4\pi i \, \ell{\cdot}(\ell-q_2)}{(2 \ell \cdot \barv_2-i\eps)^2} + \ldots,
\end{equation}
where the dots indicate the terms that are of $\mathcal{O}(\bar{m}_1^3 \bar{m}_2^4)$, as well as the terms of higher orders in $1/\bar{m}_i$, which are suppressed in the classical expansion. Similarly, the subleading numerator factors of order $\bar{m}_1^4\bar{m}_2^3$ are antisymmetric under $\bar{p}_2 \to - \bar{p}_2$, so topologies $B$ and $C$ get subtracted in their contribution to $\text{Exp}_k$. These terms, that have antisymmetric numerators, thus contribute as
\begin{equation}
\label{Geik: odd}
\bar{m}_1^4\bar{m}_2^3 (I^A {-} I^B {-} I^C {+} I^D {+} \text{Cut} \, I^D) 
=
\bar{m}_1^4\bar{m}_2^3
\int \d \hat{I} \, \delta (2 \ell \cdot \barv_1)
\frac{2\pi i}{2\ell\cdot\barv_2-i\varepsilon} + \ldots \,,
\end{equation}
where, again, the dots indicate terms of $\mathcal{O}(\bar{m}_1^3 \bar{m}_2^4)$, or higher orders in $1/\bar{m}_i$.

As a last remark about this computation, note that the role of the cut contribution $\text{Cut} I^D$ is not solely to cancel the classical term of $\text{Exp}_k$, but also to change the causal properties of the propagators in the diagrams. In particular, if we had computed the corresponding amplitude contribution $\mathcal{M}$, the propagators would have been symmetric, principal-valued propagators, see~\cite{Brandhuber:2023hhy,Herderschee:2023fxh,Elkhidir:2023dco,Georgoudis:2023lgf,Caron-Huot:2023vxl}. However, after adding $\text{Cut} I^D$, we get retarded propagators instead, as indicated with the $i\varepsilon$ prescription in~\eqref{Geik: odd} and~\eqref{Geik: sub} The physical interpretation is that these propagators enforce the causal condition of the waveshape $\text{Exp}_k$: the emitted graviton is in the future of all other particles.

\subsection{Infrared divergence and physical interpretation}
We have already seen two ways in which the cut term on the right-hand side of~\eqref{eq:Exp_k-pics} contributes to the waveshape. First, it eliminates the superclassical $\frac{1}{\hbar}$ contribution, and second, it modifies the Feynman propagators to become retarded propagators, thus enforcing that the graviton is in the future of the incoming black holes. The cut term has yet another role, however; it enforces the infrared divergence to match with the classical prediction for the Shapiro time delay of this process.

Let us first discuss what result we expect for the infrared divergence for the scattering amplitude $\mathcal{M}$ and for the waveshape $\text{Exp}_k$. It is well-known that the infrared divergence of the scattering amplitude $\mathcal{M}$ is caused by the Shapiro time delay for the graviton escaping the black hole potential, causing its wavelength to getting redshifted. We denote this Shapiro time delay as $\Delta t_{\text{obs}}$. Its value in classical GR is
\begin{align}
 \Delta t_{\rm obs}=
   2 G\, \hat{k}{\cdot}(\bar{p}_1+\bar{p}_2) \log\frac{r_{\rm obs}}{b}\,,
\label{eq:tobs}
\end{align}
where $b$ is the impact parameter, $r_{\rm obs}$ is the distance between the observation point and the black holes, and $\hat{k}$ is the unit vector in the direction of the graviton. Note that this time delay is not present in electromagnetism, since photons do not get delayed or redshifted in the presence of an electromagnetic potential. Since the time delay is logarithmically divergent as $r_{\rm obs} \to \infty$, we expect the avatar of this time delay to be an infrared divergence in the perturbation theory computation, where we are working in a framework where $r_{\rm obs} \to \infty$ from the outset. The infrared divergences can be computed using Weinberg soft factors, which gives
\begin{equation}
\label{M IR classical}
    \mathcal{M}^{(1\text{-loop})}\Big\vert_{\text{IR div}}
    =  \mathcal{M}^{({\rm tree})} \times 2i G \, k{\cdot}(\bar{p}_1+\bar{p}_2) \log\frac{\Lambda}{\mu_{\rm IR}}\,,
\end{equation}
at one-loop order,
where $\Lambda\sim b^{-1}$ is a physical momentum scale, and $\mu_{\rm IR}$ is an infrared renormalization scale. When including all-orders contributions, these infrared divergences exponentiate to recover the classical time delay.

The waveshape $\text{Exp}_k$ contains an additional source of Shapiro time delay which is not present in the exclusive amplitude $\mathcal{M}$. Recall that the waveshape is an ``in-in'' observable, see~\cite{chapterHaehlRangamani}. So it contains an additional contribution to the time delay coming from fixing the initial distance $r_{\text{in}}$ between the two black holes. This contribution also has a classical explanation. Imagine two black holes venturing into a scattering process with momenta $p_1$ and $p_2$. If they are moving at non-relativistic speeds, they will start attracting each other and scatter off each other earlier in time than expected if they had been moving on free trajectories. In fact, the time advance is logarithmic in the initial separation between them, $r_{\text{in}}$. This attraction results in a time advance of the scattering point compared with free motion. It turns out that relativistically, there is a value of $\bar{p}_1 \cdot \bar{p}_2$ at which the time advance turns into a time delay instead. In any case, the effect can easily be computed in classical GR assuming black holes moving in a Schwarzschild metric~\cite{Caron-Huot:2023vxl}. The time delay due to the fixed initial distance between the two black holes is 
\begin{equation} \label{eq:tin}
\Delta t_{\rm in} = 
   G \, \hat{k}{\cdot}(\bar{p}_1+\bar{p}_2)
   \times \frac{2-3\bar{y}^{-2}}{(1-\bar{y}^{-2})^{\frac32}}
   \log\frac{r_{\rm in}}{b} \,,
\end{equation}
where $\bar{y} =  \barv_1 \cdot \barv_2$. Note that $\Delta t_{\rm in}<0$ corresponds to a time advance.

Pictorially, we can represent the contributions from the two sources of time delay for $\text{Exp}_k$ in gravity as
\begin{equation}
\begin{gathered}
\begin{tikzpicture}[photon/.style={decorate, decoration={snake, amplitude=1pt, segment length=6pt}},scale=0.8]
\tikzset{ar/.style={decoration={
				markings,
				mark=at position 0.5 with {\arrow{latex}}},postaction={decorate}}};
\draw[lightgray, ->] (-0.5,0) -- (3,0) node[below] {$r$};
\draw[lightgray, ->] (0,-3) -- (0,4.5) node[right] {$t$};
\draw[RoyalBlue, very thick, ar] (0,-3) -- (0,1);
\draw[charcoal, dashed] (0,0) -- (3,-3);
\draw[<->] (0.1,-3) -- (2.9,-3) node[midway, above] {$r_{\rm in}$};
\draw[Maroon, very thick, ar] (3,-3) to[out=135,in=-90,distance=1cm] (0.05,1);
\draw[yellowish, very thick, ar] (0.06,1) to[out=90,in=-110,distance=1cm] (0.8,4.5);
\draw[yellowish] (-0.3,3.7) node {$X$};
\draw[yellowish] (0.29,3.7) node {$\cdots$};
\draw[RoyalBlue] (-0.3,-2.8) node {$2$};
\draw[Maroon] (3.2,-2.8) node {$1$};
\draw[yellowish, very thick, ar] (0,1) -- (0,4.5);
\draw[charcoal, dashed] (0,1) -- (2.5,3.5);
\draw[photon, thick] (0,1) to[] (2.5,4);
\draw[<->] (0.1,4) -- (2.4,4) node[midway, above] {$r_{\rm obs}$};
\draw[white, very thick, decoration={
		markings,
		mark=at position 0.5 with {\arrow[black]{latex}}},postaction={decorate}] (1.25, 2.5) -- (1.25+0.025, 2.5+0.03); 
\node[scale=0.7] at (0,1) {\cloud};
\draw[<->] (-0.45,0.05) -- (-0.45,1) node[midway,left] {$\Delta t_{\rm in}$};
\draw[<->] (2.7,3.5) -- (2.7,4) node[midway,right] {$\Delta t_{\rm obs}$};
\end{tikzpicture}
\end{gathered}
\end{equation}
Adding the two time delays, from~\eqref{eq:tobs} and~\eqref{eq:tin}, we find that the corresponding IR divergence of $\text{Exp}_k$ when computed at one loop in QFT is
\begin{equation} \label{prediction IR divergences}
   \underbracket[0.4pt]{\text{Exp}_k^{(1\text{-loop})}\Big\vert_{\rm div}
    =  \text{Exp}_k^{({\rm tree})}\times iG k{\cdot}(\bar{p}_1+\bar{p}_2)
 \left( 2+ \frac{2-3\bar{y}^{-2}}{(1-\bar{y}^{-2})^{\frac 32}}\right)
    \log\frac{\Lambda}{\mu_{\rm IR}}}
    _{\mbox{\small (classical prediction)}}\,,
\end{equation}
where, as before $\Lambda\sim b^{-1}$ is a physical scale and $\mu_{\text{IR}}$ is an infrared scale. This IR divergence was computed in~\cite{Caron-Huot:2023vxl}, and was later confirmed in explicit computations of $\text{Exp}_k$ at one loop in~\cite{Brandhuber:2023hhy,Herderschee:2023fxh,Elkhidir:2023dco,Georgoudis:2023ozp,Bohnenblust:2023qmy,Bini:2023fiz,Bini:2024rsy}.

Note that alternatively, we can compute the leading IR divergence using Wilson lines, analogous to the QED computation in \cite[Sec.~2.3]{chapterCaronHuotGiroux}. The difference between QED and gravity in this case is that here we will get an extra contribution from~\eqref{eq:tobs}, which corresponds to the time-delay that the graviton experiences when moving in the field of the two massive particles. The computation in gravity was carried out in~\cite{Caron-Huot:2023vxl}.

\section[Hierarchical 3-body problem]{Hierarchical 3-body problem\\
\normalfont{\textit{Anna M. Wolz}}}
\label{sec:Wolz}

In this section, we will complicate the already complicated story of 2-body scattering by adding an extra body. So, we will have three compact massive objects in total, interacting via gravity.

The 3-body problem is a notoriously difficult problem in astrophysics. It was approached from a scattering amplitudes and EFT perspective in Ref.~\cite{Jones:2022aji}, whose construction we will review in this section. Then, following~\cite{Solon:2024zhr}, we will motivate a version of the 3-body problem in which one of the massive objects is very far away from the other two, called the \emph{hierarchical limit}.

In this formalism, one can treat both bound and unbound states. For example, we can consider the case in which a supermassive black hole interacts with two other heavy bodies, out of which one falls in and the other flies out. This setup is relevant to many astrophysical systems. If we assume that velocities of the black holes are small compared with the speed of light, we can use the PN expansion in the bound case. Otherwise, we need the PM expansion.

Of course, 3-body systems are extremely complicated even in Newtonian gravity. Newtonian solutions to the equations of motion are unknown in most situations and otherwise difficult to obtain.
Within the framework of GR, it is difficult to even explicitly write down all the interactions, since we have intrinsic $N$-body effects.
Due to this complexity in GR, we will instead approach the 3-body problem using the tools from particle physics, similar to the ones used in the previous sections. We will be able to make headway by using approximations and treating computations perturbatively.

As we mentioned above, the main tools at our disposal are the PN expansion, which works well for bound systems that move slowly and where the bodies are far apart (equivalently, we deal with weak fields so Newton's constant $G$ is small). In this approximation, the expansion parameter is $G v^2$, where $v$ denotes the velocities of the bodies. On the other hand, PM is relevant to unbound encounters, where the bodies move fast, though they are still far apart. Then, we expand in $G$ alone. Amplitudes are most naturally suited for the PM case where we have relativistic high-energy scattering, and perturbation theory is already linked to the $G$-expansion.

The setup will be as follows. We consider $3 \to 3$ scattering of masses $m_i$, whose initial momenta are $p_i$ and the final momenta are $p_i - q_i$ for $i = 1,2,3$:
\be
\adjustbox{valign=c}{
\begin{tikzpicture}[line width=1, photon/.style={decorate, decoration={snake, amplitude=1pt, segment length=6pt}
	}]
\tikzset{ar/.style={decoration={
				markings,
				mark=at position 0.5 with {\arrow{latex}}},postaction={decorate}}};
\tikzset{ar2/.style={decoration={
				markings,
				mark=at position 0.8 with {\arrow{latex}}},postaction={decorate}}};
\draw[very thick, color=Maroon, ar] (0.8,0.3) -- (1.8,0.3);
\draw[very thick, color=yellowish, ar] (0.8,0) -- (1.8,0);
\draw[very thick, color=RoyalBlue, ar] (0.8,-0.3) -- (1.8,-0.3);
\draw[very thick, color=Maroon, ar2] (2.2,0.3) -- (3.2,0.3);
\draw[very thick, color=yellowish, ar2] (2.2,0) -- (3.2,0);
\draw[very thick, color=RoyalBlue, ar2] (2.2,-0.3) -- (3.2,-0.3);
\filldraw[fill=gray!5, line width=1.2](2,0) circle (0.6) node {$\mathcal{M}$};
\draw[] (-0.5,0) node {\textcolor{black}{$p_i, m_i$}};
\draw[] (5,0) node {\textcolor{black}{$p_i - q_i, m_i$}};
\end{tikzpicture}
}
\label{eq:3bodyamp}
\ee
The initial momentum of each body can be written as $p_i^\mu=(E_i(\vec{p}_i),\vec{p}_i)$, and the momentum transfer of each body is $q_i^\mu = (q_i^0, \vec{q}_i)$.
We can go freely between the positions of the objects $\vec{r}_i$ and the spatial momentum transfers $\vec{q}_i$ by performing a Fourier transform 
\be
\vec{r}_i \xleftrightarrow{\text{FT}} \vec{q}_i\, .
\ee
In this section, we will describe how to calculate the classical, conservative 3-body potential $V^{(3)}(\{\vec{p},\vec{q}\})$ (from now on, curly brackets denote the dependence on the momenta $\vec{p}_i$ and momentum transfers $\vec{q}_i$ for all $i=1,2,3$) in momentum space, as done in \cite{Jones:2022aji}. The Fourier transform of this function into position space $V^{(3)}(\{\vec{p},\vec{r}\})$ in the hierarchical limit is described in \cite{Solon:2024zhr}.

The way we will calculate $V^{(3)}(\{\vec{p},\vec{q}\})$ below is by taking the classical limit of the amplitude in~\eqref{eq:3bodyamp} with scalars coupled to gravity. More precisely, we will compute the amplitude from a low-energy EFT of particles interacting via the potential $V(\{ \vec{p}, \vec{q}\})$ and then formally match it onto the amplitude $\mathcal{M}$ using the Lippmann--Schwinger equation. 

Note that the 3-body problem necessarily involves potentials with two types of terms:
\be
V = \sum_{\substack{\text{pairs}\\ i,j}} V^{(2)}_{ij} + V^{(3)}\, ,
\ee
where the first piece encodes the 2-body interactions and the second one is intrinsically 3-body and it depends on all three momenta $\vec{p}_i$ and positions $\vec{r}_i$. Note that we do not have the 3-body potential $V^{(3)}$ in Newtonian dynamics.

Here we are interested in the leading $V^{(3)}$ interaction. In this case, leading order translates to $\mathcal{O}(G^2)$, i.e., the 2PM order.

\subsection{Lippmann--Schwinger equation}

The setup looks simple so let us get to it. We start with the Einstein--Hilbert action minimally coupled to scalars:
\be
S = \int \d^4 x \sqrt{-g} \left( \frac{-R}{16\pi G} + \sum_{i=1}^{3} \left( (\nabla_\mu \phi_i)^2 - m_i^2 \phi_i^2 \right) \right)\, ,
\ee
where $\phi$ is a real field. One could of course add higher-derivative corrections, but here we are working in the simplest setup so we approximate black holes as point particles. This approximation is valid when their Schwarzschild radii $R_S$ are much smaller than the distances $|\vec{r}_i - \vec{r}_j|$ between the black holes. If we wanted to include finite-size effects, charge, etc., we could use exactly the same terms as in the 2-body case.

In order to obtain the 3-body potential at 2PM, the necessary ingredients will be the tree-level matrix elements
\be
\mathcal{M}^{(2)}_G \qquad \text{and} \qquad \mathcal{M}^{(3)}_{G^2}
\ee
for the $2 \to 2$ and $3 \to 3$ cases respectively. The subscript denotes the order in $G$ at which are working in.
We need them computed in a general frame, which is a slight complication compared to the usual 2-body case which are usually calculated in the center-of-mass frame.

The classical limit for us means large impact parameter, or equivalently, small momentum transfer $q$. For the $2\to 2$ case, the leading scaling in $q$ will be $\mathcal{O}(q^{-2})$ and for $3 \to 3$, it will be $\mathcal{O}(q^{-4})$.

Using Feynman rules, the two amplitudes at tree level are simply given by
\be
\mathcal{M}^{(2)}_G = \raisebox{-1.5em}{
\begin{tikzpicture}[line width=1.1, scale=0.7]
    \coordinate (left1) at (-1,-1);
    \coordinate (left2) at (-1, 1);
    \coordinate (right1) at (1, -1);
    \coordinate (right2) at (1,1);
    \coordinate (up1) at (0,1);
    \coordinate (low1) at (0,-1);

    \draw[decorate, decoration=snake] (up1) -- (low1);
    \draw[decorate, decoration=snake] (0.15,1) -- (0.15,-1);
    
    \draw[RoyalBlue] (left1) -- (right1);
    \draw[Maroon] (left2) -- (right2);

\end{tikzpicture}
}
\ee
and 
\be
\mathcal{M}^{(3)}_{G^2} = \raisebox{-1.5em}{
\begin{tikzpicture}[line width=1.1, scale=0.7]
    \coordinate (left1) at (0,-1);
    \coordinate (left2) at (-1, 1);
    \coordinate (right1) at (3, -1);
    \coordinate (right2) at (1, 1);
    \coordinate (up1) at (0,1);
    \coordinate (low1) at (1.5,-1);

    \draw[decorate, decoration=snake] (up1) -- (low1);
    \draw[decorate, decoration=snake] (0.15,1) -- (1.65,-1);

    \draw[decorate, decoration=snake] (low1) -- (3,1);
    \draw[decorate, decoration=snake] (1.65,-1) -- (3.15,1);
    
    \draw[RoyalBlue] (left1) -- (right1);
    \draw[Maroon] (left2) -- (right2);
    \draw[yellowish] (2,1) -- (4,1);
\end{tikzpicture}
}
+
\raisebox{-1.5em}{
\begin{tikzpicture}[line width=1.1, scale=0.7]
    \coordinate (left1) at (0,-1);
    \coordinate (left2) at (-1, 1);
    \coordinate (right1) at (3, -1);
    \coordinate (right2) at (1, 1);
    \coordinate (up1) at (0,1);
    \coordinate (low1) at (1.5,0);

    \draw[decorate, decoration=snake] (up1) -- (low1);
    \draw[decorate, decoration=snake] (0.15,1) -- (1.65,0);

    \draw[decorate, decoration=snake] (low1) -- (3,1);
    \draw[decorate, decoration=snake] (1.65,0) -- (3.15,1);

    \draw[decorate, decoration=snake] (low1) -- (1.5,-1);
    \draw[decorate, decoration=snake] (1.65,0) -- (1.65,-1);
    
    \draw[RoyalBlue] (left1) -- (right1);
    \draw[Maroon] (left2) -- (right2);
    \draw[yellowish] (2,1) -- (4,1);
\end{tikzpicture}
}
+
\raisebox{-1.5em}{
\begin{tikzpicture}[line width=1.1, scale=0.7]
    \coordinate (left1) at (0,-1);
    \coordinate (left2) at (-1, 1);
    \coordinate (right1) at (3, -1);
    \coordinate (right2) at (1, 1);
    \coordinate (up1) at (0,1);
    \coordinate (low1) at (1,-1);
    \coordinate (low2) at (2,-1);

    \draw[decorate, decoration=snake] (up1) -- (low1);
    \draw[decorate, decoration=snake] (0.15,1) -- (1.15,-1);

    \draw[decorate, decoration=snake] (low2) -- (3,1);
    \draw[decorate, decoration=snake] (2.15,-1) -- (3.15,1);
    
    \draw[RoyalBlue] (left1) -- (right1);
    \draw[Maroon] (left2) -- (right2);
    \draw[yellowish] (2,1) -- (4,1);
\end{tikzpicture}
}
\ee
We are keeping only the diagrams that contribute non-trivially at the orders we require. We use the notation $M^{(N)}$ to denote the $N\to N$ amplitude multiplied by a normalization factor:
\be\label{eq:normM}
M^{(N)}=\prod_{i=1}^N\frac{1}{2\sqrt{E_i(\vec{p}_i)E_i(\vec{p}_i-\vec{q}_i)}}\mathcal{M}^{(N)}.
\ee

Next, we will write down a low-energy EFT of relativistic particles. It is governed by the Hamiltonian
\be
H = \sum_{i=1}^{3} \sqrt{\vec{p}_i^2 + m_i^2} + V(\{\vec{p},\vec{r}\})\, .
\ee
As mentioned above, the potential is related via a Fourier transform to $V(\{\vec{p},\vec{q}\})$.
The Lippmann--Schwinger equation can formally be expressed as:
\be\label{eq:LS2}
T(\{\vec{p},\vec{q}\}) \stackrel{!}{=} - V(\{\vec{p}, \vec{q}\}) - \int_\vec{k} \frac{V(\{\vec{p}, \vec{p}-\vec{k}\}) \,V(\{\vec{k}, \vec{k}-\vec{p}+\vec{q}\})}{\sum_{i=1}^{3} [E_i(\vec{p}) - E_i(\vec{k})] + i \varepsilon} + \ldots
\ee
In this equation, $T$ denotes the scattering amplitudes, i.e.\ the matrix elements of the interacting part of the S-matrix (as in $S = \mathbbm{1} + i \hat{T}$). We have written the iteration terms and stopped at the quadratic order in $V$ since this will be sufficient for our purposes. Note that $\overset{!}{=}$ denotes that the potential is defined only up to terms that vanish on-shell, which are the gauge/coordinate artifacts.

We are going to think of this equation in terms of scattering amplitudes. It is actually valid for any $N$-body scattering, but here we will apply it only to $N=2,3$ specifically. The potentials $V^{(2)}$ and $V^{(3)}$ are like the Wilson coefficients in the EFT. We will use Feynman diagrams to compute $T$ and then match on to the right hand-side of the above equation. Note that $T$ contains the fully connected term, but also partially disconnected pieces when $N \geq 3$.

\subsection{Solving the \texorpdfstring{$2$}{2}-body potential}

Let us first deal with the $2$-body case, which we have already encountered in Sec.~\ref{sec:Correia}. At $\mathcal{O}(G)$, we have the tree-level contribution
\be\label{eq:V2G}
\adjustbox{valign=c}{
\begin{tikzpicture}[line width=1, photon/.style={decorate, decoration={snake, amplitude=1pt, segment length=6pt}
	}]
\draw[very thick, color=Maroon] (0.8,0.3) -- (1.8,0.3);
\draw[very thick, color=RoyalBlue] (0.8,-0.3) -- (1.8,-0.3);
\draw[very thick, color=Maroon] (2.2,0.3) -- (3.2,0.3);
\draw[very thick, color=RoyalBlue] (2.2,-0.3) -- (3.2,-0.3);
\filldraw[fill=gray!5, line width=1.2](2,0) circle (0.6) node {$V^{(2)}_G$};
\end{tikzpicture}
}
\quad\stackrel{!}{=}\quad - \quad
\adjustbox{valign=c}{
\begin{tikzpicture}[line width=1, photon/.style={decorate, decoration={snake, amplitude=1pt, segment length=6pt}
	}]
\draw[very thick, color=Maroon] (0.8,0.3) -- (1.8,0.3);
\draw[very thick, color=RoyalBlue] (0.8,-0.3) -- (1.8,-0.3);
\draw[very thick, color=Maroon] (2.2,0.3) -- (3.2,0.3);
\draw[very thick, color=RoyalBlue] (2.2,-0.3) -- (3.2,-0.3);
\filldraw[fill=gray!5, line width=1.2](2,0) circle (0.6) node {$M^{(2)}_G$};
\end{tikzpicture}
}
\ee
The notation $V^{(2)}_G$ denotes the 2-body potential at $\mathcal{O}(G)$. This is simply the Newtonian potential, computed with the $t$-channel exchange of a graviton.

At the next order, $\mathcal{O}(G^2)$, we encounter the contribution from Feynman diagrams of $\mathcal{M}$ at this order, but also the iteration of the $V^{(2)}$ terms we computed above:
\be\label{eq:V2G2}
\adjustbox{valign=c}{
\begin{tikzpicture}[line width=1, photon/.style={decorate, decoration={snake, amplitude=1pt, segment length=6pt}
	}]
\draw[very thick, color=Maroon] (0.8,0.3) -- (1.8,0.3);
\draw[very thick, color=RoyalBlue] (0.8,-0.3) -- (1.8,-0.3);
\draw[very thick, color=Maroon] (2.2,0.3) -- (3.2,0.3);
\draw[very thick, color=RoyalBlue] (2.2,-0.3) -- (3.2,-0.3);
\filldraw[fill=gray!5, line width=1.2](2,0) circle (0.6) node {$V^{(2)}_{G^2}$};
\end{tikzpicture}
}
\quad\stackrel{!}{=}\quad - \quad
\adjustbox{valign=c}{
\begin{tikzpicture}[line width=1, photon/.style={decorate, decoration={snake, amplitude=1pt, segment length=6pt}
	}]
\draw[very thick, color=Maroon] (0.8,0.3) -- (1.8,0.3);
\draw[very thick, color=RoyalBlue] (0.8,-0.3) -- (1.8,-0.3);
\draw[very thick, color=Maroon] (2.2,0.3) -- (3.2,0.3);
\draw[very thick, color=RoyalBlue] (2.2,-0.3) -- (3.2,-0.3);
\filldraw[fill=gray!5, line width=1.2](2,0) circle (0.6) node {$M^{(2)}_{G^2}$};
\end{tikzpicture}
}
\quad-\quad
\adjustbox{valign=c}{
\begin{tikzpicture}[line width=1, photon/.style={decorate, decoration={snake, amplitude=1pt, segment length=6pt}
	}]
\draw[very thick, color=Maroon] (0.8,0.3) -- (1.8,0.3);
\draw[very thick, color=RoyalBlue] (0.8,-0.3) -- (1.8,-0.3);
\draw[very thick, color=Maroon] (2.2,0.3) -- (4.7,0.3);
\draw[very thick, color=RoyalBlue] (2.2,-0.3) -- (4.7,-0.3);
\filldraw[fill=gray!5, line width=1.2](2,0) circle (0.6) node {$V^{(2)}_{G}$};
\filldraw[fill=gray!5, line width=1.2](3.5,0) circle (0.6) node {$V^{(2)}_{G}$};
\end{tikzpicture}
}
\ee
The notation of two lines between $V_G^{(2)}$ in the last term represents the integration over the appropriate phase space of the intermediate particles (as in the last term of \eqref{eq:LS2}).
This procedure allows us to determine $V^{(2)}_{G^2}$ diagrammatically, which is all we will need.

\subsection{Solving the \texorpdfstring{$3$}{3}-body potential}

Let us move on to the 3-body case. Now, the $T$-matrix contains contributions from both $3 \to 3$ and $2 \to 2$ amplitudes. Diagrammatically, the expansion of \eqref{eq:LS2} looks as follows:

\begin{align}\label{eq:V3G2}
 \Bigg(&
\adjustbox{valign=c}{
\begin{tikzpicture}[line width=1, photon/.style={decorate, decoration={snake, amplitude=1pt, segment length=6pt}
	}]
\draw[very thick, color=Maroon] (0.8,0.3) -- (1.8,0.3);
\draw[very thick, color=yellowish] (0.8,0) -- (1.8,0);
\draw[very thick, color=RoyalBlue] (0.8,-0.3) -- (1.8,-0.3);
\draw[very thick, color=Maroon] (2.2,0.3) -- (3.2,0.3);
\draw[very thick, color=yellowish] (2.2,0) -- (3.2,0);
\draw[very thick, color=RoyalBlue] (2.2,-0.3) -- (3.2,-0.3);
\filldraw[fill=gray!5, line width=1.2](2,0) circle (0.6) node {$M^{(3)}_{G^2}$};
\end{tikzpicture}
}
+
\adjustbox{valign=c}{
\begin{tikzpicture}[line width=1, photon/.style={decorate, decoration={snake, amplitude=1pt, segment length=6pt}
	}]
\draw[very thick, color=Maroon] (0.8,0.25) -- (1.8,0.25);
\draw[very thick, color=yellowish] (0.8,-0.7) -- (3.2,-0.7);
\draw[very thick, color=RoyalBlue] (0.8,-0.25) -- (1.8,-0.25);
\draw[very thick, color=Maroon] (2.2,0.25) -- (3.2,0.25);
\draw[very thick, color=RoyalBlue] (2.2,-0.25) -- (3.2,-0.25);
\filldraw[fill=gray!5, line width=1.2](2,0) circle (0.5) node {$M^{(2)}_{G^2}$};
\end{tikzpicture}
}
+ \text{perm.}
\Bigg)\nonumber\\
\stackrel{!}{=} & - \Bigg(
\adjustbox{valign=c}{
\begin{tikzpicture}[line width=1, photon/.style={decorate, decoration={snake, amplitude=1pt, segment length=6pt}
	}]
\draw[very thick, color=Maroon] (0.8,0.3) -- (1.8,0.3);
\draw[very thick, color=yellowish] (0.8,0) -- (1.8,0);
\draw[very thick, color=RoyalBlue] (0.8,-0.3) -- (1.8,-0.3);
\draw[very thick, color=Maroon] (2.2,0.3) -- (3.2,0.3);
\draw[very thick, color=yellowish] (2.2,0) -- (3.2,0);
\draw[very thick, color=RoyalBlue] (2.2,-0.3) -- (3.2,-0.3);
\filldraw[fill=gray!5, line width=1.2](2,0) circle (0.6) node {$V^{(3)}_{G^2}$};
\end{tikzpicture}
}
+
\adjustbox{valign=c}{
\begin{tikzpicture}[line width=1, photon/.style={decorate, decoration={snake, amplitude=1pt, segment length=6pt}
	}]
\draw[very thick, color=Maroon] (0.8,0.25) -- (1.8,0.25);
\draw[very thick, color=yellowish] (0.8,-0.7) -- (3.2,-0.7);
\draw[very thick, color=RoyalBlue] (0.8,-0.25) -- (1.8,-0.25);
\draw[very thick, color=Maroon] (2.2,0.25) -- (3.2,0.25);
\draw[very thick, color=RoyalBlue] (2.2,-0.25) -- (3.2,-0.25);
\filldraw[fill=gray!5, line width=1.2](2,0) circle (0.5) node {$V^{(2)}_{G^2}$};
\end{tikzpicture}
}
+ \text{perm.}
\Bigg)\\
& - \Bigg(
\adjustbox{valign=c}{
\begin{tikzpicture}[line width=1, photon/.style={decorate, decoration={snake, amplitude=1pt, segment length=6pt}
	}]
\draw[very thick, color=Maroon] (0.8,0.25) -- (4.7,0.25);
\draw[very thick, color=yellowish] (0.8,-0.7) -- (4.7,-0.7);
\draw[very thick, color=RoyalBlue] (0.8,-0.25) -- (1.8,-0.25);
\draw[very thick, color=Maroon] (2.2,0.25) -- (3.2,0.25);
\draw[very thick, color=RoyalBlue] (2.2,-0.25) -- (4.7,-0.25);
\filldraw[fill=gray!5, line width=1.2](2,0) circle (0.5) node {$V^{(2)}_{G}$};
\filldraw[fill=gray!5, line width=1.2](3.5,0) circle (0.5) node {$V^{(2)}_{G}$};
\end{tikzpicture}
}
+
\adjustbox{valign=c}{
\begin{tikzpicture}[line width=1, photon/.style={decorate, decoration={snake, amplitude=1pt, segment length=6pt}
	}]
\draw[very thick, color=Maroon] (0.8,0.25) -- (4.7,0.25);
\draw[very thick, color=yellowish] (0.8,-0.7) -- (4.7,-0.7);
\draw[very thick, color=RoyalBlue] (0.8,-0.25) -- (1.8,-0.25);
\draw[very thick, color=Maroon] (2.2,0.25) -- (3.2,0.25);
\draw[very thick, color=RoyalBlue] (2.2,-0.25) -- (4.7,-0.25);
\filldraw[fill=gray!5, line width=1.2](2,0) circle (0.5) node {$V^{(2)}_{G}$};
\filldraw[fill=gray!5, line width=1.2](3.5,-0.5) circle (0.5) node {$V^{(2)}_{G}$};
\end{tikzpicture}
}
+ \text{perm.}
\Bigg).\nonumber
\end{align}
The second line corresponds to the terms linear in $V$ in the Lippmann--Schwinger equation and the third line involves the quadratic terms.
Our goal is to isolate the piece $V^{(3)}_{G^2}$. We plug in \eqref{eq:V2G2} for $V_{G^2}^{(2)}$ with the spectator particle (in yellow). This cancels with the $M_{G^2}^{(2)}$ piece on the LHS and the ``loop" iteration piece on the LHS of the third line.
All of the disconnected pieces therefore cancel. Solving for $V_{G^2}^{(3)}$, the final expression reads diagrammatically
\be
\adjustbox{valign=c}{
\begin{tikzpicture}[line width=1, photon/.style={decorate, decoration={snake, amplitude=1pt, segment length=6pt}
	}]
\draw[very thick, color=Maroon] (0.8,0.3) -- (1.8,0.3);
\draw[very thick, color=yellowish] (0.8,0) -- (1.8,0);
\draw[very thick, color=RoyalBlue] (0.8,-0.3) -- (1.8,-0.3);
\draw[very thick, color=Maroon] (2.2,0.3) -- (3.2,0.3);
\draw[very thick, color=yellowish] (2.2,0) -- (3.2,0);
\draw[very thick, color=RoyalBlue] (2.2,-0.3) -- (3.2,-0.3);
\filldraw[fill=gray!5, line width=1.2](2,0) circle (0.6) node {$V^{(3)}_{G^2}$};
\end{tikzpicture}
}
=
-\adjustbox{valign=c}{
\begin{tikzpicture}[line width=1, photon/.style={decorate, decoration={snake, amplitude=1pt, segment length=6pt}
	}]
\draw[very thick, color=Maroon] (0.8,0.3) -- (1.8,0.3);
\draw[very thick, color=yellowish] (0.8,0) -- (1.8,0);
\draw[very thick, color=RoyalBlue] (0.8,-0.3) -- (1.8,-0.3);
\draw[very thick, color=Maroon] (2.2,0.3) -- (3.2,0.3);
\draw[very thick, color=yellowish] (2.2,0) -- (3.2,0);
\draw[very thick, color=RoyalBlue] (2.2,-0.3) -- (3.2,-0.3);
\filldraw[fill=gray!5, line width=1.2](2,0) circle (0.6) node {$M^{(3)}_{G^2}$};
\end{tikzpicture}
}
- 
\adjustbox{valign=c}{
\begin{tikzpicture}[line width=1, photon/.style={decorate, decoration={snake, amplitude=1pt, segment length=6pt}
	}]
\draw[very thick, color=Maroon] (0.8,0.25) -- (4.7,0.25);
\draw[very thick, color=yellowish] (0.8,-0.7) -- (4.7,-0.7);
\draw[very thick, color=RoyalBlue] (0.8,-0.25) -- (1.8,-0.25);
\draw[very thick, color=Maroon] (2.2,0.25) -- (3.2,0.25);
\draw[very thick, color=RoyalBlue] (2.2,-0.25) -- (4.7,-0.25);
\filldraw[fill=gray!5, line width=1.2](2,0) circle (0.5) node {$M^{(2)}_{G}$};
\filldraw[fill=gray!5, line width=1.2](3.5,-0.5) circle (0.5) node {$M^{(2)}_{G}$};
\end{tikzpicture}
} - \text{perm.}
\ee

We now have all the ingredients to compute the effective 3-body potential. After the dust settles, the final answer is:
\be\label{eq:V-ijk}
V^{(3)}_{G^2} \overset{!}{=} \sum_{(i,j,k)\in S_3} \Bigg(- \frac{\mathcal{M}_{G^2,ijk}^{(3)}}{8 E_i E_j E_k} - \frac{V_{G,ij}^{(2)}(\vec{p}_i, \vec{p}_j; \vec{q}_i)\, V_{G,kj}^{(2)} (\vec{p}_k,\vec{p}_j + \vec{q}_i ; \vec{q}_k) }{ E_j + q_i^0 - \sqrt{(\vec{p}_j + \vec{q}_i)^2 + m_j^2}}\Bigg)\, ,
\ee
where $S_3$ is the set of distinct permutations of the particle labels $(1,2,3)$ and $\mathcal{M}^{(3)}$ is the fully connected matrix element for the $3 \to 3$ scattering process.

Let us do some cross checks on this result. There are a couple of criteria that a classical potential should satisfy. First, it should behave as $\mathcal{O}(q^{-4})$. Naively, it might appear from~\eqref{eq:V-ijk} that the three-body potential behaves as $\mathcal{O}(q^{-5})$, which would lead to a superclassical $\sim 1/\hbar$ scaling. Second, the potential should have no matter singularities, i.e., only the graviton poles should survive. Let us illustrate these problems at the level of the individual diagram, say:
\be
\raisebox{-1.5em}{
\begin{tikzpicture}[line width=1.1, scale=0.7]
    \coordinate (left1) at (0,-1);
    \coordinate (left2) at (-1, 1);
    \coordinate (right1) at (3, -1);
    \coordinate (right2) at (1, 1);
    \coordinate (up1) at (0,1);
    \coordinate (low1) at (1,-1);
    \coordinate (low2) at (2,-1);

    \draw[decorate, decoration=snake] (up1) -- (low1);
    \draw[decorate, decoration=snake] (0.15,1) -- (1.15,-1);

    \draw[decorate, decoration=snake] (low2) -- (3,1);
    \draw[decorate, decoration=snake] (2.15,-1) -- (3.15,1);
    
    \draw[RoyalBlue] (left1) -- (right1);
    \draw[Maroon] (left2) -- (right2);
    \draw[yellowish] (2,1) -- (4,1);
\end{tikzpicture}
}
\ee
One can indeed show that it scales as $\mathcal{O}(q^{-5})$. Likewise, it contains an explicit matter pole (associated to the blue propagator) of the form
\be
\frac{1}{(p_i+q_j)^2 - m_i^2}\, .
\ee

After some work, one can show that these two problems go away once we sum over all the diagrams and iteration terms. In other words, \eqref{eq:V-ijk} goes as $\mathcal{O}(q^{-4})$ and does not have any matter poles, although it is not manifest at the level of the formula \eqref{eq:V-ijk}.

These cross-checks give us confidence that \eqref{eq:V-ijk} is the correct 3-body potential at the leading order. Note that it is supposed to be valid for both unbound and bound orbits. The authors of \cite{Jones:2022aji} checked the result for $V^{(3)}_{G^2}(\{\vec{p},\vec{q}\})$ with \cite{Loebbert:2020aos}, who computed the 3-body effective potential using other techniques.

\newpage
\bibliographystyle{jhep}
\bibliography{references}

\end{document}